\documentclass[11pt]{article}
\usepackage{amsmath}
\usepackage{amssymb}
\usepackage{amsthm,amsxtra}
\newlength{\mytopmargin}
\newlength{\myleftmargin}
\newtheorem{thm}{Theorem}
\newtheorem{cor}[thm]{Corollary}
\newtheorem{lemma}{Lemma}
\newtheorem{prop}[lemma]{Proposition}
\begin{document}
\vspace{4cm}
\noindent
{\bf Correlations for superpositions and decimations of Laguerre and
Jacobi orthogonal matrix ensembles with a parameter}

\vspace{5mm}
\noindent
Peter J.~Forrester${}^*$ and Eric M.~Rains${}^\dagger$

\noindent
${}^*$Department of Mathematics and Statistics,
University of Melbourne, \\
Victoria 3010, Australia ; 
${}^\dagger$AT\&T Research, Florham Park, NJ 07932, USA	 \\
(Present address:
Center for Communications
Research, Princeton, NJ 08540, USA) 

\small
\begin{quote}
A superposition of a matrix ensemble refers to the ensemble constructed
from two independent copies of the original, while a decimation refers to
the formation of a new ensemble by observing only every second eigenvalue.
In the cases of the classical matrix ensembles with orthogonal symmetry,
it is known that forming superpositions and decimations gives rise to
classical matrix ensembles with unitary and symplectic symmetry. The
basic identities expressing these facts can be extended to include a
parameter, which in turn provides us with  probability density
functions which we take as the definition of special parameter
dependent matrix ensembles. The parameter dependent ensembles 
relating to superpositions interpolate
between superimposed orthogonal ensembles and a 
unitary ensemble, 
while the parameter dependent ensembles
relating to decimations interpolate between 
an orthogonal ensemble with an even number of eigenvalues
and a symplectic ensemble of half the number
of eigenvalues.
 By the construction of new families of
biorthogonal and skew orthogonal polynomials, we are able to compute
the corresponding correlation functions, both in the finite system and
in various scaled limits. Specializing back to the cases of orthogonal
and symplectic symmetry, we find that our results imply  different
functional forms to those known previously.
\end{quote}

\section{Introduction}
Dyson \cite{Dy62} introduced three ensembles of random unitary matrices ---
the circular orthogonal ensemble (COE), circular unitary ensemble (CUE)
and circular symplectic ensemble (CSE). The corresponding joint eigenvalue
probability density functions (PDFs) were calculated to be
\begin{equation}\label{0.1}
{1 \over C} \prod_{1 \le j < k \le n} |e^{i \theta_j} -
e^{i \theta_k} |^\beta
\end{equation}
where $C$ is the normalization (throughout the symbol $C$ will be used to
denote {\it some} normalization) and $\beta = 1$ for the COE, $\beta = 2$
for the CUE and $\beta = 4$ for the CSE. In the theory of the COE, a
technique known as integration over alternate variables has a special place.
This technique draws one naturally to study statistical properties of
every second eigenvalue (parity respecting correlations) as well as
the statistical properties of the complete COE sequence (parity blind
correlations). Thus for matrices from the ensemble COE${}_{2n}$
($2n \times 2n$ members of the COE)
Mehta and Dyson \cite{MD63} considered the statistical properties of
every second eigenvalue (referred to as an alternating sequence) 
by integrating out the complementary alternating sequence.
With the  resulting distribution denoted alt(COE${}_{2n}$)
they showed 
\begin{equation}\label{2.1}
{\rm  alt(COE}_{2n}) = \rm {CSE}_n,
\end{equation}
where on the RHS we mean the joint eigenvalue
distribution of the ensemble
CSE${}_n$ (below we will state similar equations
with this convention without further comment). 

Prior to the work of Mehta and Dyson, Dyson \cite{Dy62a} was led to
conjecture that superimposing two independent eigenvalue sequences from
the COE and integrating out every second eigenvalue leaves an
eigenvalue sequence with the same joint distribution as the CUE, and thus
\begin{equation}\label{1.1}
{\rm  alt(COE}_{n} \cup {\rm COE}_n) = \rm {CUE}_n.
\end{equation}
This was subsequently proved by Gunson \cite{Gu62}. The results
(\ref{2.1}) and (\ref{1.1}) together imply that the physically
important gap probability, that is the probability that the interval
$[-s,s]$ is free of eigenvalues, is inter-related for the three
ensembles COE, CUE and CSE. As formulas for this quantity were known
for the COE and CUE in terms of the Fredholm determinant of the
integral operator on $[-s,s]$ with kernel $\sin \pi (x-y)/ \pi
(x-y)$, the inter-relationships imply a formula for the gap
probability in the CSE in terms of the same Fredholm determinant
\cite{MD63}.

Some years after the pioneering work of Dyson and Mehta,
Baik and Rains \cite{BR01a} were led to study the distribution of
every second row of random Young tableaux specified according to
some specific probability measures. In particular, a probability
measure was identified for which the distribution of the even
numbered rows is independent of a parameter $\alpha$ occurring
in the measure. In the continuum limit, in
which the integer valued row lengths go over to continuous valued
variables, the PDF of the $2n$ row lengths is specified by
\begin{equation}\label{3.1}
{1 \over C} \prod_{j=1}^{2n} e^{-x_j/2} \prod_{j=1}^n e^{A(x_{2j-1} - x_{2j})/2}\prod_{1 \le j < k \le 2n}(x_j - x_k)
\end{equation}
where
\begin{equation}\label{3.1a}
x_1 > x_2 > \cdots > x_{2n} \ge 0,
\end{equation}
$C$ is the normalization and $A$ is the analogue of the
parameter $\alpha$. 
For (\ref{3.1}) to be normalizable we must have $A < 1$.
In the case $A =0$ this coincides with 
the joint distribution of the eigenvalues in the matrix ensemble LOE
--- the Laguerre orthogonal ensemble with parameter $a=0$ (for
general parameter $a$ the Laguerre weight is $x^{a} e^{-x/2}$). We recall
the LOE distribution of $2n$ variables with parameter $a=0$ is
realized by the eigenvalues of Wishart matrices $X^T X$ where the real
matrix $X$ has dimension $(2n+1) \times 2n$ and independent elements,
identically distributed with the standard normal distribution
N$[0,1]$. Let us denote the ensemble corresponding to (\ref{3.1}) by
LOE${}_{2n}^A$. With ME denoting a general matrix ensemble
and the operation even(ME) denoting the distribution
of the even labelled coordinates with the ordering (\ref{3.1a}), and thus
the operation of integrating out the odd labelled coordinates, the result
of Baik and Rains gives 
\begin{equation}\label{3.1b}
{\rm even}({\rm LOE}_{2n}^A) = {\rm LSE}_n
\end{equation}
where the PDF for the
LSE$_{n}$ is the $A \to - \infty$ limit of the PDF (\ref{3.1}),
and is thus given by 
\begin{equation}\label{3.2}
{1 \over C} \prod_{j=1}^n e^{-x_j} \prod_{1 \le j < k \le n} 
(x_j - x_k)^4,
\end{equation}
after re-labelling coordinates $\{x_{2j-1}, x_{2j} \} \mapsto x_j$
$(j=1,\dots,n)$ (see (\ref{3.12a})).
As a matrix ensemble the LSE refers to the Laguerre
symplectic ensemble with parameter $a=0$. It can be realized as a
$2n \times 2n$ antisymmetric matrix, in which the elements are
pure imaginary numbers with each $2 \times 2$ block having a real
quaternion structure. Such matrices are equivalent to block
matrices of the form
$$
\left [ \begin{array}{cc} 0_{n \times n} & X_{n \times n} \\
 X_{n \times n}^\dagger &  0_{n \times n} \end{array} \right ]
$$
where $X$ is an antisymmetric complex matrix.
As is evident from the transition from (\ref{3.1})
to (\ref{3.2}), the corresponding eigenvalue spectrum is doubly
degenerate.

Implicit in \cite{BR01a} is the matrix ensemble corresponding to the
eigenvalue PDF
\begin{equation}\label{4.1}
{1 \over C} \prod_{j=1}^{2n} e^{-x_j/2}
\prod_{j=1}^n e^{A(x_{2j-1} - x_{2j})/2}
\prod_{1 \le j < k \le n} (x_{2j-1} - x_{2k-1})(x_{2k} - x_{2j}),
\end{equation}
where the ordering (\ref{3.1a}) is assumed, 
and as with (\ref{3.1}) we must have $A < 1$ for the
PDF to be normalizable. When $A=0$ the matrix ensemble
with this eigenvalue PDF is 
the superimposed ensemble
$$
{\rm LOE}_{n} \cup {\rm LOE}_{n}
$$
(two independent copies of the LOE). The $A \to - \infty$ limit of
(\ref{4.1}) gives the LUE, parameter $a=0$, joint  eigenvalue distribution
\begin{equation}\label{4.1a}
{1 \over C} \prod_{j=1}^n e^{-x_j} \prod_{1 \le j < k \le n}
(x_k - x_j)^2,
\end{equation}
where as in (\ref{3.2}) we have re-labelled the coordinates
$\{x_{2j-1}, x_{2j} \} \mapsto x_j$ ($j=1,\dots,n$). As a matrix
ensemble the LUE refers to the Laguerre unitary ensemble, which
can be realized by matrices of the form $X^\dagger X$ where
$X$ is a $n \times n$ matrix with independent,
identically distributed complex Gaussian entries.
Analogous to
the derivation of (\ref{3.1}), one can use arguments based on the
underlying combinatorial model to show that 
\begin{equation}\label{4.2}
{\rm even}(({\rm LOE}_n \cup {\rm LOE}_n)^A) = {\rm LUE}_n
\end{equation}
independent of the parameter value $A$.

Inspired by the results (\ref{3.1b}) and (\ref{4.2}), Forrester
and Rains \cite{FR01} considered general matrix ensembles with orthogonal
symmetry OE${}_n(f)$ corresponding to the joint distribution
\begin{equation}\label{4.3}
{1 \over C} \prod_{j=1}^n f(x_j) \prod_{1 \le j < k \le n}
(x_j - x_k).
\end{equation}
They sought to classify all differentiable weight functions $f$ in
(\ref{4.3}) such that
\begin{equation}\label{5.1}
{\rm even}({\rm OE}_n(f) \cup {\rm OE}_n(f) ) = {\rm UE}_n(g)
\end{equation}
where UE${}_n(g)$ is the matrix ensemble with unitary symmetry corresponding to 
the joint distribution
$$
{1 \over C} \prod_{j=1}^n g(x_j) \prod_{1 \le j < k \le n}
(x_j - x_k)^2.
$$
Up to linear fractional transformations only two pairs of weights
$(f,g)$ were found to possess this property:
\begin{equation}\label{5.2}
(f,g) = \left \{ \begin{array}{ll}
(e^{-x/2}, e^{-x}), & \quad x>0 \\
((1-x)^{(a-1)/2}, (1-x)^a), & \quad 0 < x < 1. \end{array} \right.
\end{equation}
The first of these is the $a=0$ Laguerre weight in (\ref{4.1}),
while the second is an example of the Jacobi weight $x^b(1-x)^a$ with
parameter $b=0$. Furthermore, it was proved \cite{FR01} that the
statement (\ref{5.1}) is equivalent to the statement
\begin{equation}\label{1.13a}
{\rm even(OE}{}_{2n}(f)) = {\rm SE}_n((g/f)^2)
\end{equation}
where SE${}_n(h)$ denotes the matrix ensemble with symplectic symmetry
corresponding to the joint distribution
\begin{equation}\label{hs}
{1 \over C} \prod_{j=1}^n h(x_j) \prod_{1 \le j < k \le n}
(x_j - x_k)^4.
\end{equation}

Our interests in this work are parameter dependent generalizations of the
orthogonal ensembles, and superimposed orthogonal ensembles, specified
by the weights $f$ in (\ref{5.2}). For the weight $f(x) = e^{-x/2}$,
$(x>0)$, these are precisely the ensembles 
with PDFs (\ref{3.1}) and (\ref{4.1}).
For the weight $f(x) = (1-x)^{(a-1)/2}$, $0 < x < 1$, the parameter
dependent ensembles are specified by the PDFs
\begin{equation}\label{6.1}
{1 \over C} \prod_{j=1}^{2n} (1 - x_j)^{(a-1)/2}
\prod_{l=1}^n \Big ( {1 - x_{2l-1} \over 1 - x_{2l}} \Big )^{-A/2}
\prod_{1 \le j < k \le 2n} (x_j - x_k)
\end{equation}
and
\begin{equation}\label{6.2}
{1 \over C} \prod_{j=1}^{2n} (1 - x_j)^{(a-1)/2}
\prod_{l=1}^n \Big ( {1 - x_{2l-1} \over 1 - x_{2l}} \Big )^{-A/2}
\prod_{1 \le j < k \le n} (x_{2j-1} - x_{2k-1})(x_{2j} - x_{2k})
\end{equation}
where, for convenience, we shift the origin and scale the variable $x$
relative to that used in (\ref{5.2}) so that
$$
-1 < x_j < 1 \qquad (j=1,\dots, 2n).
$$
For (\ref{6.1}) and (\ref{6.2}) to be normalizable we must have $A < a+1$.
We remark that after shifting the origin back to that used in (\ref{5.2})
by $x_j \mapsto {1 \over 2}(x_j + 1)$, then scaling the variables and
parameters
$$
x_j \mapsto x_j/L, \quad a \mapsto L, \quad A \mapsto LA
$$
and taking the limit $L \to \infty$, (\ref{6.1}) and (\ref{6.2})
reduce to (\ref{3.1}) and (\ref{3.2}) respectively.

We note that with respect to each even labelled coordinate the
parameter $A$ in (\ref{6.1}) and (\ref{6.2}) can be viewed as a
change of parameter $a \mapsto a + A$
in the Jacobi weight.
Let us make this change 
in (\ref{5.2}) and
form the identities (\ref{5.1}) and (\ref{1.13a}). If we then divide
both sides of the resulting identities by $\prod_{l=1}^{2n}
(1-x_l)^{A/2}$ so that in each case the RHS is independent of $A$,
we deduce that
\begin{eqnarray}
{\rm even}\Big ( {\rm OE}_n(f_{\rm o}, f_{\rm e}) \cup
{\rm OE}_n(f_{\rm o}, f_{\rm e}) \Big ) = {\rm UE}_n(g), \label{1m} \\
{\rm even}\Big ( {\rm OE}_{2n}(f_{\rm o}, f_{\rm e}) \Big ) =
{\rm SE}_n \Big ( (g/f_{\rm e} |_{A = 0})^2 \Big ),
\label{1m'}
\end{eqnarray}
with
\begin{equation}\label{1ma}
(f_{\rm o}, f_{\rm e}, g) = \Big (
(1-x)^{(a-A-1)/2}, (1-x)^{(a+A-1)/2}, (1-x)^a \Big ), \qquad
-1 < x < 1.
\end{equation}
In (\ref{1m}), assuming the ordering $x_1 > x_2  > \cdots 
> x_{2n}$, ${\rm OE}_n(f_{\rm o}, f_{\rm e}) \cup
{\rm OE}_n(f_{\rm o}, f_{\rm e})$ refers to the ensemble with PDF
\begin{equation}\label{1.20a}
\prod_{j=1}^n f_{\rm o}(x_{2j-1}) f_{\rm e}(x_{2j})
\prod_{1 \le j < k \le n} (x_{2j} - x_{2k}) (x_{2j-1} - x_{2k-1}),
\end{equation}
while in (\ref{1m'}) ${\rm OE}_{2n}(f_{\rm o}, f_{\rm e})$ refers to
the ensemble with PDF
\begin{equation}\label{1.20b}
\prod_{j=1}^n f_{\rm o}(x_{2j-1}) f_{\rm e}(x_{2j})
\prod_{1 \le j < k \le 2n} (x_j - x_k).
\end{equation}
An analogous argument starting with the Laguerre case  of (\ref{5.2}),
modified so that $x$ therein is replaced by $(1+A)x$, can be used to
deduce (\ref{3.1}) and (\ref{4.2}).

The parameter dependent PDFs (\ref{3.1}), (\ref{4.1}), (\ref{6.1})
and (\ref{6.2}) each have at least two interpretations in distinct
applied settings. One, already made explicit \cite{BR01a,Ba02} in the case
of (\ref{3.1}), is as the continuum limit of certain measures on
partitions. These measures in turn are intimately related to increasing
subsequence problems \cite{BDJ99,BO01,BR01},  growth models \cite{Jo99a,Jo02}
and non-intersecting lattice paths \cite{Kr95,GOV98,Fo01,HI02}.
The other is as the eigenvalue PDF for certain computable
ensembles of random matrices. 
Development of these settings will be undertaken in a separate
publication \cite{FR02b}. Here
we seek the evaluation of the multi-point correlation functions, both
parity aware and parity blind (recall the terminology from the first
paragraph) associated with the probability densities (\ref{3.1}),
(\ref{4.1}), (\ref{6.1}) and (\ref{6.2}). 
Let us first revise the definition of a multi-point correlation function.
Consider a general PDF
\begin{equation}\label{6.0.1}
p(x_1,\dots,x_n;y_1,\dots,y_n)
\end{equation}
which is symmetric in $\{x_j\}_{j=1,\dots,n}$ and 
$\{y_j\}_{j=1,\dots,n}$ with the support of $p$ on some region $I$
of the real line for each coordinate. This can be thought of as the
PDF for a two species system of particles free to move on $I$.
The $(k_1,k_2)$-point correlation function for $k_1$ particles of
species $x$ and $k_2$ particles of species $y$ is defined as
\begin{eqnarray*}
&&{\rho_{(k_1,k_2)}(x_1,\dots,x_{k_1}; y_1,\dots,y_{k_2})
= n(n-1) \cdots (n-k_1+1)n (n-1) \cdots (n - k_2 + 1)} \\
&& \times
\prod_{l=k_1+1}^n \prod_{l'=k_2+1}^n\int_I dx_l \int_I dy_{l'} \,
p(x_1,\dots, x_n; y_1, \dots, y_n).
\end{eqnarray*}
All the PDFs (\ref{3.1}),
(\ref{4.1}), (\ref{6.1}) and (\ref{6.2}) can be written in the form
(\ref{6.0.1}) with $x_1,\dots,x_n$ denoting the odd labelled
particles in the ordering (\ref{3.1a}), and
$y_1,\dots,y_n$ denoting the even labelled particles in the ordering
(\ref{3.1a}).

In the parity aware cases, it follows from  general formulas due to Rains 
\cite{Ra00} that for the PDFs (\ref{4.1}) and (\ref{6.1}) the
$(k_1,k_2)$-point correlation functions have the determinant structure
\begin{eqnarray}\label{1.17}
\lefteqn{\rho_{(k_1,k_2)}(x_1,\dots,x_{k_1}; y_1,\dots,y_{k_2})} \nonumber \\
&&= \det \left [ \begin{array}{cc}
[ K_{\rm oo}(x_j, x_l) ]_{j,l=1,\dots,k_1} &
[ K_{\rm oe}(x_j, y_l) ]_{j =1,\dots,k_1 \atop l=1,\dots,k_2} \\
{}[ K_{\rm eo}(y_j, x_l) ]_{j =1,\dots,k_2 \atop l=1,\dots,k_1} &
{}[ K_{\rm ee}(y_j, y_l) ]_{j,l=1,\dots,k_2} \end{array}
\right ]
\end{eqnarray}
for certain functions $K_{\rm oo}$, $K_{\rm oe}$, $K_{\rm eo}$ and
$K_{\rm ee}$. The latter are defined in terms of arbitrary polynomials
$p_j(y)$, $Q_j(x)$ of degree $j$,
as well as the inverse of the matrix
\begin{equation}\label{6.1.3}
\Big [ \int_0^\infty dy \, e^{-(1+A)y/2} p_j(y)
\int_y^\infty dx \, e^{-(1-A)x/2} Q_k(x) \Big ]_{j,k=0,\dots,n-1}
\end{equation}
in the case of (\ref{3.1}), and the inverse of the matrix
\begin{equation}\label{6.1.4}
\Big [ \int_{-1}^1 dy \, (1-y)^{(a-A-1)/2} p_j(y)
\int_y^1 dx \, (1-x)^{(a+A-1)/2} Q_k(x) \Big ]_{j,k=0,\dots,n-1}
\end{equation}
in the case of (\ref{6.1}). Also, a general formula of
\cite{Ra00} gives that for the PDFs (\ref{3.1}) and (\ref{6.2})
the $(k_1,k_2)$-point correlation functions have the quaternion
determinant (the definition of a quaternion
determinant is revised in Section \ref{s3}) structure
\begin{eqnarray}\label{6.2.1}
\lefteqn{\rho_{(k_1,k_2)}(x_1,\dots,x_{k_1}; y_1,\dots,y_{k_2})} \nonumber \\
&& = {\rm qdet} \left [ \begin{array}{cc}
[ f_{\rm oo}(x_j, x_l) ]_{j,l=1,\dots,k_1} &
[ f_{\rm oe}(x_j, y_l) ]_{j =1,\dots,k_1 \atop l=1,\dots,k_2} \\
{}[ f_{\rm eo}(y_j, x_l) ]_{j =1,\dots,k_2 \atop l=1,\dots,k_1} &
[f_{\rm ee}(y_j, y_l) ]_{j,l=1,\dots,k_2} \end{array}
\right ]
\end{eqnarray}
where the $f_{s_1 s_2}$ are $2 \times 2$ matrices with elements defined
in terms of arbitrary $j$th degree polynomials $R_j(x)$,
together with quantities which differ in their specification depending
on whether one is considering (\ref{3.1}) or (\ref{6.2}). For
(\ref{3.1}) these quantities are
\begin{equation}\label{1.23c}
\Phi_j^{\rm e}(x) :=  \int_x^\infty  e^{-t/2}
e^{A(t - x)/2} R_j(t) \, dt
\end{equation}
and the inverse of the antisymmetric matrix
$$
\Big [ \int_0^\infty e^{-x/2} R_j(x) \Phi_k^{\rm e}(x) \, dx
\Big ]_{j,k=0,\dots, 2n-1},
$$
while for (\ref{6.2}) they are
\begin{equation}\label{1.23d}
\Phi_j^{\rm e}(x) :=  \int_{x}^1 (1-t)^{(a-1)/2} 
\Big ( {1 - t \over 1 - x} \Big )^{-A/2 } R_j(t) \, dt
\end{equation}
and the inverse of the antisymmetric matrix
$$
\Big [ \int_{-1}^1 (1-x)^{(a-1)/2} R_j(x) \Phi_k^{\rm e}(x) \, dx 
\Big ]_{j,k=0,\dots, 2n-1}.
$$

Regarding the parity blind correlations,
$\rho_k(x_1,\dots,x_k)$ (here the symbol $x_j$ is used in its original
sense of (\ref{3.1}), (\ref{4.1}), (\ref{6.1}) and (\ref{6.2}))
we note that because  (\ref{4.1}) and (\ref{6.2}) do not vanish
when $x_{2j-1} = x_{2j}$, $\rho_k$ does not vanish at coincident points
and so cannot have a simple determinental form analogous to (\ref{6.1}).
However (\ref{3.1}) and (\ref{6.1}) can both be written in a form
involving a Pfaffian factor,
\begin{equation}\label{6.3.1}
\prod_{j=1}^{2n} w(x_j) \prod_{1 \le j < l \le 2n}
(x_j - x_l) \, {\rm Pf}[\epsilon(x_j, x_l]_{j,l=1,\dots,2n},
\end{equation}
for which the general
structure of the $k$-point correlation is known \cite{MM91} in terms
of a quaternion determinant
\begin{equation}\label{6.3.2}
\rho_k(x_1,\dots,x_k) = {\rm qdet} [f(x_j, x_l) ]_{j,l=1,\dots,k}
\end{equation}
where $f$ is a $2 \times 2$ matrix defined in terms of the same
quantities as those specifying (\ref{6.2.1}).

We find that in the case of the parameter dependent Laguerre ensembles
(\ref{3.1}) and (\ref{4.1}), the correlations can be written in terms of
\begin{equation}\label{6.3.3}
K^L_n(x,y) := e^{-(x+y)/2} \sum_{l=0}^{n-1}
{1 \over h_l^L} L_l(x) L_l(y),  
\end{equation}
where $ L_l(x)$ denotes the Laguerre polynomial of degree $l$ and parameter
$a=0$, with the orthogonality property
\begin{equation}\label{6.3.4}
\int_0^\infty e^{-t} L_j(t) L_k(t) \, dt = h_j^L \delta_{j,k}, \quad
h_j^L = 1.
\end{equation}
The function $K^L_n$ is familiar as determining the correlation function of
the Laguerre unitary ensemble with parameter $a=0$. The latter has the
eigenvalue PDF (\ref{4.1a}). Explicitly, the $k$-point correlation function
is given in terms of (\ref{6.3.3}) by
\begin{equation}\label{1.31'}
\rho_k(x_1,\dots,x_k) = \det \Big [ K_n^L(x_j,x_l) \Big ]_{j,l=1,\dots,k}.
\end{equation}
Similarly, in the case of the parameter dependent Jacobi ensembles
(\ref{6.1}) and (\ref{6.2}), the correlations can be written in terms
of
\begin{equation}\label{1.25}
K^J_n(x,y) := (1-x)^{(a-1)/2} (1 - y)^{(a-1)/2}
\sum_{l=0}^{n-1} {1 \over h_l^J}
P_l^{(a,0)}(x) P_l^{(a,0)}(y)
\end{equation}
where $P_l^{(a,0)}(x)$ denotes the Jacobi polynomial of degree $l$ and
parameter $b=0$, with the orthogonality property
\begin{equation}\label{6.4.2}
\int_{-1}^1 (1 - t)^a P_j^{(a,0)}(t) P_k^{(a,0)}(t) \, dt =
h_j^J \delta_{j,k}, \quad h_j^J = {2^{a+1} \over 2j + a+1}.
\end{equation}
We remark that
\begin{equation}\label{1.33'}
\rho_k(x_1,\dots,x_k) =
\det \Big [ (1 - x_l) {K}^J_n(x_j, x_l) \Big ]_{j,l=1,\dots,k}
\end{equation}
is the $k$-point correlation function for the eigenvalue PDF
$$
{1 \over C} \prod_{l=1}^n (1 - x_l)^a \prod_{1 \le j < k \le n}
(x_k - x_j)^2, \qquad |x_l|< 1,
$$
which corresponds to the Jacobi unitary ensemble with parameter $b=0$.

Let us now turn to the plan of the paper. 
In Section 2 we take up the problem
of the explicit computation of the entries of (\ref{6.1}), and in
Section 3 we compute the explicit form of the entries in (\ref{6.2.1})
and (\ref{6.3.2}). Various scaled limits of the correlations are
computed in Section 4, while we finish in Section 5 by relating the
correlation functions found in this study to correlation and
distribution functions known from previous studies.

\section{Correlations for superimposed orthogonal ensembles with a parameter}
\setcounter{equation}{0}
Consider a PDF of the form
\begin{equation}\label{8.0}
\prod_{j=1}^n w_{\rm o}(x_j) w_{\rm e}(y_j) \prod_{1 \le j < k \le n}
(x_j - x_k) (y_j - y_k) \det [ \kappa(x_j,y_k) ]_{j,k=1,\dots,n},
\end{equation}
where $x_j, y_j \in \mathbb R$ (although the support of $w_{\rm o}, 
w_{\rm e}$ may be some subset of $\mathbb R$). 
A general formula of \cite{Ra00} gives that the $(k_1,k_2)$-point
correlation function is given by (\ref{1.17}) with 
\begin{eqnarray}\label{8.1}
K_{\rm oo}(x,x') & = &
\sum_{j,k=0}^{n-1}  w_{\rm o}(x)
Q_j(x)  \, M_{jk}^{-t} \, \int_{-\infty}^{\infty}  
\kappa(x',u)w_{\rm e}(u) p_{k}(u) \, du 
\nonumber \\
K_{\rm oe}(x,y) & = & \sum_{j,k=0}^{n-1}  w_{\rm o}(x)
Q_j(x)  \, M_{jk}^{-t} \,
w_{\rm e}(y) p_{k}(y)
\nonumber \\
K_{\rm eo}(y,x) & = & - \kappa(x,y) +
\sum_{j,k=0}^{n-1}  \Big ( \int_{-\infty}^\infty \kappa(u,y)
w_{\rm e}(u) Q_{j}(u) \, du \Big )\, M_{jk}^{-t} \, \Big (
\int_{-\infty}^\infty \kappa(x,v)  w_{\rm o}(v) p_{k}(v) \Big )
\, dv \nonumber \\
K_{\rm ee}(y,y')  & = & \sum_{j,k=0}^{n-1} 
\Big ( \int_{-\infty}^\infty \kappa(v,y) w_{\rm o}(v) Q_{j}(v) \, dv \Big )
 \, M_{jk}^{-t} \, 
w_{\rm e}(y')p_k(y')
\end{eqnarray}
where ${}^{-t}$ denotes the operation of taking the transpose of the inverse, 
$p_j(y)$ and
$Q_j(x)$ are as in (\ref{6.1.3}) and $[M_{jk}]$ is the matrix with
entries
\begin{equation}\label{8.2}
M_{jk} = \int_{-\infty}^\infty dx \, w_{\rm o}(x) Q_j(x)
\int_{-\infty}^\infty dy \, \kappa(x,y) w_{\rm e}(y) p_k(y).
\end{equation}
This result is relevant to (\ref{4.1}) and (\ref{6.1}) because both these
PDFs can be written in the form (\ref{8.0}) with
\begin{equation}\label{9.1}
\kappa(x,y) = \left \{ \begin{array}{ll} e^{A(x-y)/2}\chi_{x > y}, &
\quad {\rm Laguerre \, case} \\[.1cm]
\displaystyle\Big ( {1 - x \over 
1 - y} \Big )^{-A/2} \chi_{x > y}, &
\quad {\rm Jacobi \, case} \end{array} \right.
\end{equation}
where $ \chi_T=1$ if $T$ is true and $0$ otherwise, a fact which can
be seen by making note of the following determinant identity.

\begin{lemma}
For the orderings
\begin{equation}\label{9.2}
x_1 > \cdots > x_n, \qquad y_1 > \cdots > y_n
\end{equation}
we have
\begin{equation}\label{9.3}
\det[ \chi_{x_j - y_k > 0} ]_{j,k=1,\dots,n} =
\chi_{x_1 > y_1 > \cdots > x_n > y_n}.
\end{equation}
\end{lemma}

\noindent Proof. \quad For the ordering $x_1 >  y_1 > \cdots > x_n > y_n$
the determinant is triangular with $1$'s down the diagonal, so
(\ref{9.3}) is correct in this case. All other orderings must have at
least two $x$'s (or two $y$'s) in succession. The corresponding rows
(or columns) in the determinant will then be equal so the determinant
vanishes. \hfill $\square$ 

Thus with $\kappa(x,y)$ given by (\ref{9.1}), we can substitute
(\ref{9.3}) times
$$
\prod_{j=1}^n e^{A(x_{j} - y_{j})/2} \quad
{\rm Laguerre \, case,} \qquad
\prod_{j=1}^n \Big ( {1 - x_j \over 1 - y_j} \Big )^{-A/2} \quad
{\rm Jacobi \, case}
$$
for the determinant in (\ref{8.0}), and
\begin{equation}\label{eod}
w_{\rm o}(x) = w_{\rm e}(x) =
\left \{ \begin{array}{ll} e^{-x/2}\, (x>0), &
\quad {\rm Laguerre \, case}, \\[.1cm]
(1-x)^{(a-1)/2} \, (-1<x<1), &
\quad {\rm Jacobi \, case} \end{array} \right.
\end{equation}
provided the ordering
(\ref{9.2}) is assumed. On this latter point, (\ref{8.0}) is a symmetric
function of the $x$'s and $y$'s (separately), so the ordering constraint
(\ref{9.2}) is in fact irrelevant.

For general PDFs (\ref{8.0}) there is of course no explicit formula
available for the inverse of $[M_{jk}]$ which is required in (\ref{8.1}).
However, for the particular PDFs (\ref{4.1}) and (\ref{6.1}),
and thus $\kappa(x,y)$ given by (\ref{9.1}), this problem
can be overcome by choosing $\{ p_j(y) \}$ and $\{ Q_k(x) \}$ to have
the biorthogonality property
\begin{equation}\label{10.1}
\int_{-\infty}^\infty dy \, w_{\rm e}(y) p_j(y) \int_y^\infty dx \,
w_{\rm o}(x) Q_k^L(x) = n_j \delta_{j,k}.
\end{equation}
These polynomials are given simply in terms of the orthogonal
polynomials $\{L_j(x)\}$ in the Laguerre case and
$\{P_j^{(a,0)}(x)\}$ in the Jacobi case.

\begin{prop}
The sets of polynomials $\{L_j(x)\}_{j=0,1,\dots}$ and
$\{Q_j^L(x) \}_{j=0,1,\dots}$ where
\begin{equation}\label{2.21} 
Q_j^L(x)  
 =  {2 \over A - 1} e^{(1-A)x/2} {d \over dx} \Big (
e^{-(1-A)x/2} L_j(x) \Big )
\end{equation}
have the biorthogonality property
\begin{equation}\label{14.2}
\int_0^\infty dt \, e^{-(1+A)t/2} L_p(t) \int_t^\infty dx \,
e^{-(1-A)x/2} Q_q^J(x) = {\cal N}_p^L \delta_{p,q}, \quad
{\cal N}_p^L := - {2 \over A - 1},
\end{equation}
while the sets of polynomials $\{P_j^{(a,0)}(x)\}_{j=0,1,\dots}$
and $\{Q_j^J(x) \}_{j=0,1,\dots}$ where
\begin{equation}\label{2.23}
Q_j^J(x)  =  
   -{(1-x)^{-(a-A-1)/2} \over j + (a-A+1)/2} {d \over dx} \Big (
(1-x)^{(a-A+1)/2} P_j^{(a,0)}(x) \Big )
\end{equation}
have the biorthogonality property
\begin{eqnarray}\label{2.24}
&&  \int_{-1}^1 dt \, (1-t)^{(a+A-1)/2} P_p(t)
\int_t^1dx \, (1-x)^{(a-A-1)/2} Q_q(x) = {\cal N}_p^J \delta_{p,q} \nonumber \\
&&  
{\cal N}_p^J := {2^{a+1} \over (a+1+2p) (p + (a-A+1)/2)}.
\end{eqnarray} 
\end{prop}

\noindent
Proof. \quad Substituting (\ref{2.21}) in the LHS of (\ref{14.2}) and
making use of (\ref{6.3.4}) gives the RHS of (\ref{14.2}). Thus it only
remains to check that $Q_j^L(x)$ is indeed a polynomial of degree $j$.
This latter point follows by inspection of (\ref{2.21}). The
verification of (\ref{2.24}) is done analogously. \hfill $\square$

The biorthogonality properties (\ref{14.2}) and (\ref{2.24}) allow
$M_{jk}^{-t}$ in (\ref{8.1}) to be replaced by ${\cal N}_j^{-1} \delta_{j,k}$.
The double sums in (\ref{8.1}) then collapse to single sums. Furthermore,
taking note of the values of the normalizations from (\ref{6.3.4}) and
(\ref{6.4.2}) we see that (\ref{2.21}) and (\ref{2.23}) can be
rewritten
\begin{eqnarray}
{Q_j^L(x) \over {\cal N}_j^L} & = & - {1 \over h_j^L}
e^{(1-A)x/2} {d \over dx} \Big ( e^{-(1-A)x/2} L_j(x) \Big )
\label{2.25} \\
{Q_j^J(x) \over {\cal N}_j^J} & = & - {1 \over h_j^J}
(1-x)^{-(a-A-1)/2} {d \over dx} \Big (
(1-x)^{(a-A+1)/2} P_j^{(a,0)}(x) \Big ). \label{2.26}
\end{eqnarray}
It thus follows that the quantities in (\ref{8.1}) can then be
expressed simply in terms of the functions $K^L_n$ and $K^J_n$
introduced in (\ref{6.3.3}) and (\ref{1.25}).

\begin{prop}
The $(k_1,k_2)$ point parity respecting correlation for the PDF
(\ref{4.1}) is given by (\ref{1.17}) with
\begin{eqnarray}\label{KL}
K_{\rm oo}^L(x,x') & = & - e^{-A(x-x')/2} {\partial \over \partial x}
\Big \{ e^{Ax/2} \int_0^{x'} e^{-Au/2} K^L_n(x,u) \, du \Big \} \nonumber \\
K_{\rm oe}^L(x,y) & = & - e^{-Ax/2} {\partial \over \partial x}
\Big \{ e^{Ax/2}  K^L_n(x,y) \Big \} \nonumber \\
K_{\rm eo}^L(y,x) & = & - e^{A(x-y)/2} \chi_{x>y} +
e^{Ax/2} \int_0^x e^{-Av/2} K^L_n(v,y) \, dv  \nonumber \\
K_{\rm ee}^L(y,y') & = & K^L_n(y,y') 
\end{eqnarray}
(we have appended the superscripts $L$ on the LHS as notation for the
Laguerre case (\ref{4.1})). Similarly, the
$(k_1,k_2)$ point parity respecting correlation for the PDF (\ref{6.1})
--- the Jacobi case to be denoted by appending a superscript $J$ ---
is given by (\ref{1.17}) with
\begin{eqnarray}\label{KJ}
K_{\rm oo}^J(x,x') & = & -  \Big ( {1 - x \over 1 - x'} \Big )^{A/2}
 {\partial \over \partial x}
\Big \{ (1-x)^{1-A/2}
 \int_{-1}^{x'} (1-u)^{A/2} K^J_n(x,u) \, du \Big \} \nonumber \\
K_{\rm oe}^J(x,y) & = & - (1-x)^{A/2} {\partial \over \partial x}
\Big \{ (1-x)^{1-A/2}  K^J_n(x,y) \Big \} \nonumber \\
K_{\rm eo}^J(y,x) & = & - \Big ( {1 - x \over 1 - y} \Big )^{-A/2}
\chi_{x>y} + (1-y)
(1-x)^{-A/2} \int_{-1}^x (1-v)^{A/2} K^J_n(v,y) \, dv  \nonumber \\
K_{\rm ee}^J(y,y') & = & (1-y) K^J_n(y,y').
\end{eqnarray}
\end{prop}

We see from (\ref{KL}) that $K_{\rm ee}^L(x,y)$ coincides with
$K^L_n(x,y)$ which we know from (\ref{1.31'})
determines  the $k$-point correlation for the LUE. This property
of the even-even correlations is equivalent to the statement
(\ref{4.2}). Similarly, the final formula in (\ref{KJ}) implies that
the $k$-point correlation for the even-even correlations in the
Jacobi case coincides with the $k$-point correlation for the JUE with
parameter value $(a,b) \mapsto (a,0)$ (recall (\ref{1.33'})).
This result is equivalent to the statement (\ref{1m}).

\section{Correlations for decimated orthogonal ensembles with a 
parameter}\label{s3}
\setcounter{equation}{0}
\subsection{Quaternion determinant formulas}
To calculate the parity respecting correlations for (\ref{3.1}) and
(\ref{6.1}), one notes that analogous to (\ref{8.0}) they have the
general structure
\begin{equation}\label{8.0.1}
\prod_{j=1}^n w_{\rm o}(x_{2j-1}) w_{\rm e}(x_{2j})
\prod_{1 \le j < k \le 2n} (x_j - x_k)
\det [ \kappa(x_{2j-1}, x_{2k} ]_{j,k=1,\dots,n}.
\end{equation}
Explicitly, we choose $\kappa(x,y)$ as in (\ref{9.1}), $w_{\rm o}(x)$
and $w_{\rm e}(x)$ as in (\ref{eod}), and make use of (\ref{9.3}).
The significance of this is that for general PDFs of the form
(\ref{8.0.1}), a result of Rains \cite{Ra00} gives that the
parity respecting correlations have the form (\ref{6.2.1}), and
further specifies the elements of the matrix therein. Before
stating the latter, we remark that the qdet operation in
(\ref{6.2.1}) is well defined on matrices $A$ with the self dual
property
\begin{equation}\label{3.1ad}
A = Z_{2n}^{-1} A^T Z_{2n}, \qquad
Z_{2n} := {\bf 1}_{n} \otimes \left [ \begin{array}{cc} 0 & -1 \\
1 & 0 \end{array} \right ].
\end{equation}
This is equivalent to requiring that $A Z_{2n}$ be antisymmetric. On
the latter class of matrices the Pfaffian operation is well defined, and
we have in fact that \cite{Dy70}
\begin{equation}\label{3.2'}
{\rm qdet} \, A = {\rm Pf}( A Z_{2n} ),
\end{equation}
which for our present purposes can be taken as the definition of
qdet (in fact the results of \cite{Ra00} are written in terms of
Pfaffians).

Let us introduce arbitrary polynomials $R_j(x)$ of degree $j$, 
and let us follow \cite{Ra00} and introduce the
notation
\begin{equation}\label{3.17a}
(\kappa \cdot f)(x) = \int_{-\infty}^\infty w(y) \kappa(x,y) f(y)
\, dy
\end{equation}
where we have set 
\begin{equation}\label{ww}
w_{\rm o}(x) = w_{\rm e}(x) = w(x)
\end{equation}
(this can always be accomplished by changing the definition
of $\kappa(x,y)$), as well as
the $2n \times 2n$ antisymmetric matrix with elements
\begin{equation}\label{mjk}
M_{jk} = \int_{-\infty}^\infty dx \, w(x)
 \int_{-\infty}^\infty dy \, w(y) \Big ( R_j(x) R_k(y) -
R_k(x) R_j(y) \Big ) \kappa(x,y),
\end{equation}
where the $R_j(x)$ are the arbitrary $j$th degree polynomials introduced 
below (\ref{6.2.1}).
Then according to \cite{Ra00} the parity respecting
correlations for the PDF (\ref{8.0.1}) are given by (\ref{6.2.1})
with the elements of the matrices therein specified by
\begin{eqnarray}\label{fs}
f_{\rm oo}(x,x') & = & \left [ \begin{array}{cc}
 \sum_{j,k=0}^{2n-1} w(x) R_j(x) M_{jk}^{-t} (\kappa \cdot R_k)(x') &
-\sum_{j,k=0}^{2n-1} w(x) R_j(x) M_{jk}^{-t}  w(x') R_k(x') \\
 \sum_{j,k=0}^{2n-1}  (\kappa \cdot R_j)(x)  M_{jk}^{-t}
 (\kappa \cdot R_k)(x') & -\sum_{j,k=0}^{2n-1}  (\kappa \cdot R_j)(x)  
M_{jk}^{-t}   w(x') R_k(x') \end{array} \right ] \nonumber \\
f_{\rm oe}(x,y)  & = & \left [ \begin{array}{cc}
 \sum_{j,k=0}^{2n-1} w(x) R_j(x) M_{jk}^{-t} w(y) R_k(y) &
-\sum_{j,k=0}^{2n-1} w(x) R_j(x) M_{jk}^{-t} (\kappa^t \cdot R_k)(y) \\
 \sum_{j,k=0}^{2n-1}  (\kappa \cdot R_j)(x)  M_{jk}^{-t}  w(y) R_k(y) &
-\kappa(x,y) -  \sum_{j,k=0}^{2n-1}  (\kappa \cdot R_j)(x)  M_{jk}^{-t}
(\kappa^t \cdot R_k)(y) \end{array} \right ] \nonumber \\
f_{\rm eo}(y,x) & = & \left [ \begin{array}{cc}
-\kappa(x,y) +  \sum_{j,k=0}^{2n-1}  (\kappa^t \cdot R_j)(y)  M_{jk}^{-t}
(\kappa \cdot R_k)(x) &
- \sum_{j,k=0}^{2n-1}  (\kappa^t \cdot R_j(y))  M_{jk}^{-t}  w(x) R_k(x) \\
  \sum_{j,k=0}^{2n-1} w(y) R_j(y) M_{jk}^{-t} (\kappa \cdot R_k)(x) &
-\sum_{j,k=0}^{2n-1} w(y) R_j(y) M_{jk}^{-t}  w(x) R_k(x) \end{array} \right ] 
\nonumber \\ 
f_{\rm ee}(y,y') & = & \left [ \begin{array}{cc}
\sum_{j,k=0}^{2n-1}  (\kappa^t \cdot R_j)(y)  M_{jk}^{-t}  w(y') R_k(y') &
-\sum_{j,k=0}^{2n-1}  (\kappa^t \cdot R_j)(y)  M_{jk}^{-t}
(\kappa^t \cdot R_k)(y') \\
 \sum_{j,k=0}^{2n-1} w(y) R_j(y) M_{jk}^{-t}  w(y') R_k(y') &
-\sum_{j,k=0}^{2n-1} w(y) R_j(y) M_{jk}^{-t} (\kappa^t \cdot R_k)(y')
 \end{array} \right ]
\end{eqnarray}

Let us now consider the parity blind correlations. For this purpose we
write the PDFs (\ref{3.1}) and (\ref{6.1}) in the form
\begin{equation}\label{3.a}
\prod_{j=1}^{2n} w(x_{j}) \prod_{1 \le j < k \le 2n} (x_j - x_k) \,
{\rm Pf} \, [ \epsilon(x_j,x_k)]_{j,k=1,\dots,2n}
\end{equation}
where $w(x)$ is given by (\ref{ww}) with the substitution (\ref{eod}),
and
\begin{equation}\label{3.b}
\epsilon(x,y) = \left \{ \begin{array}{ll} e^{A|x-y|/2}{\rm sgn}(x-y), &
\quad {\rm Laguerre \, case} \\[.1cm]
\displaystyle \Big ( {1 - x \over 
1 - y} \Big )^{-A\, {\rm sgn}(x-y)/2} 
{\rm sgn}(x-y), &
\quad {\rm Jacobi \, case} \end{array} \right.
\end{equation}
In (\ref{3.b}) the notation sgn$(x)$ denotes the sign of $x$ and is thus
equal to 1 for $x>0$, to 0 for $x=0$ and to $-1$ for $x<0$. The equality
of (\ref{3.a}) with (\ref{3.1}) and (\ref{6.1}) follows from the
identity
\begin{equation}\label{3.c}
{\rm Pf} \, \Big [ \Big ( {f(x_j) \over f(x_k)} \Big )^{{\rm sgn}(x_j -
x_k)} {\rm sgn}(x_j - x_k) \Big ]_{j,k=1,\dots,2n} =
\prod_{j=1}^n  {f(x_{Q(2j-1)}) \over f(x_{Q(2j)})} \,
\varepsilon(Q),
\end{equation}
where
$$
x_{Q(2j-1)} > x_{Q(2j)}, \quad Q(2j) > Q(2j-1), \: (j=1,\dots,n)
$$
and $\varepsilon(Q)$ denotes the signature of the permutation $Q$.
The identity (\ref{3.c}) can be seen to follow from the definition
$$
{\rm Pf} \, A = \sum\nolimits_{P(2l) > P(2l-1)}^* \varepsilon(P)
\prod_{l=1}^n  a_{P(2l),P(2l-1)},
$$
valid for any $2n \times 2n$ antisymmetric matrix $A$, where the 
${}^*$ denotes
that only permutations which give rise to a unique product of the
$a_{jk}$'s are to be included. There are $(2n-1)!!$ such permutations,
one of which for the Pfaffian (\ref{3.c})
contributes the term on the RHS of (\ref{3.c}). All other
permutations give a contribution which cancels in pairs, so the Pfaffian
(\ref{3.c}) is in fact equal to this single term.

The quaternion determinant formula (\ref{6.3.2}) for  PDFs of the form
(\ref{3.a}) has been given by Frahm and Pichard \cite{FP95}. In
\cite{FP95} the matrix elements of the $2 \times 2$ matrix $f$ are
given in terms of skew orthogonal polynomials. In keeping with
(\ref{fs}), we prefer to follow \cite{Ra00} and state the form
of the matrix elements which involves arbitrary polynomials of degree
$k$, $R_k(x)$, as well as
\begin{equation}\label{3.8a}
(\epsilon \cdot R_k)(x) := \int_{-\infty}^\infty w(y) \epsilon(x,y)
R_k(y) \, dy
\end{equation}
(c.f.~(\ref{3.17a})) and the inverse of the antisymmetric matrix
\begin{equation}\label{ma}
\Big [ \int_{-\infty}^\infty dx \, w(x)
 \int_{-\infty}^\infty dy \, w(y) \Big ( R_j(x) R_k(y) -
R_k(x) R_j(y) \Big ) \epsilon(x,y) \Big ].
\end{equation}
Comparing the definition (\ref{3.b}) of $\epsilon(x,y)$ with the definition
(\ref{9.1}) of $\kappa(x,y)$ we see that the integral in (\ref{ma})
is unchanged if
$\epsilon(x,y)$ is replaced by $\kappa(x,y)$, and is thus equal to
$M_{jk}$ as defined by (\ref{mjk}). With this understood, reading
off from \cite{Ra00} we have
\begin{equation}\label{fxa}
f(x,y) 
 =  \left [ \begin{array}{cc}
 \sum_{j,k=0}^{2n-1} w(x) R_j(x) M_{jk}^{-t} (\epsilon \cdot R_k)(y) &
-\sum_{j,k=0}^{2n-1} w(x) R_j(x) M_{jk}^{-t}  w(y) R_k(y) \\
-\epsilon(x,y)+ \sum_{j,k=0}^{2n-1}  (\epsilon \cdot R_j)(x)  M_{jk}^{-t}
 (\epsilon \cdot R_k)(y) & -\sum_{j,k=0}^{2n-1}  (\epsilon \cdot R_j)(x)
M_{jk}^{-t}   w(y) R_k(y) \end{array} \right ].
\end{equation}

\subsection{Skew orthogonal polynomials}
Our ability to obtain closed form expressions for (\ref{fs}) in the
sense of (\ref{KL}) and (\ref{KJ}) relies on the construction
of the polynomials $\{ R_j(x) \}$ so that they exhibit the skew
orthogonality property
\begin{equation}\label{sk}
\langle R_{2j}, R_{2k} \rangle^A = 
\langle R_{2j+1}, R_{2k+1} \rangle^A = 0, \qquad
\langle R_{2j}, R_{2k+1} \rangle^A = r_j \delta_{j,k}
\end{equation}
where $\langle \, , \, \rangle^A$ denotes the skew inner product
\begin{equation}\label{fs1}
\langle f, g \rangle^A = \left \{
\begin{array}{ll} \int_0^\infty dy \, e^{-y/2} \int_y^\infty dx \,
e^{-x/2} e^{A(x-y)/2} (f(y) g(x) - g(y) f(x)), & \: {\rm Laguerre} \\[.1cm]
\int_{-1}^1 dy \, (1-y)^{(a-1)/2} \int_y^1dx \, (1-x)^{(a-1)/2}
\Big ( {\displaystyle 1 - x \over \displaystyle 1 - y} \Big )^{-A/2}
(f(y) g(x) - g(y) f(x)), & \: {\rm Jacobi} \end{array} \right.
\end{equation}
In the case $A=0$ the skew orthogonal polynomials in both the Laguerre
and Jacobi cases are known in terms of an explicit series of
Laguerre and Jacobi orthogonal polynomials respectively \cite{NW92},
as well as a more compact form involving derivatives and integrals of
these bases \cite{AFNV99}. 
Analogous explicit forms of the skew orthogonal polynomials
are available in the limit $A \to - \infty$, when the PDFs
(\ref{2.1}) and (\ref{6.1}) tend to certain PDFs of the form
(\ref{hs}) corresponding to a symplectic symmetry. Regarding this latter
point note that 
integration by parts shows
\begin{eqnarray}\label{3.12a}
&&\lim_{A \to - \infty} \Big ( {A \over 2} \Big )^{2n}
\int_{X_L} dx_1 \cdots dx_{2n} \, \prod_{l=1}^n a_l(x_{2l})
e^{-(1+A) x_{2l-1}/2} e^{-(1-A)x_{2l}/2}
\prod_{1 \le j < k \le 2n} (x_j - x_k) \nonumber \\
&& \quad =
\int_{\tilde{X}_L} dx_2 dx_4 \cdots dx_{2n} \,
\prod_{l=1}^n a_l(x_{2l}) e^{-x_{2l}} \prod_{1 \le j < k \le n}
(x_{2j} - x_{2k})^4,
\end{eqnarray}
where the $a_l$ are arbitrary, $X_L$ is the integration region (\ref{3.1a}),
and $\tilde{X}_L$ the integration region $x_2 > x_4 > \cdots > x_{2n}
\ge 0$, and
\begin{eqnarray}\label{3.12b}
&&\lim_{A \to - \infty} \Big ( {A \over 2} \Big )^{2n}
\int_{X_J} dx_1 \cdots dx_{2n} \, \prod_{l=1}^n a_l(x_{2l})
(1-x_{2l-1})^{(a-A-1)/2} (1-x_{2l})^{(a+A-1)/2}
\prod_{1 \le j < k \le 2n} (x_j - x_k) \nonumber \\
&& \quad =
\int_{\tilde{X}_J} dx_2 dx_4 \cdots dx_{2n} \,
\prod_{l=1}^n a_l(x_{2l}) (1-x_{2l})^{a+1} \prod_{1 \le j < k \le n}
(x_{2j} - x_{2k})^4,
\end{eqnarray}
where the $a_l$ are arbitrary, and
$X_J$, $\tilde{X}_J$ are the integration regions
$$
1 > x_1 > x_2 > \cdots > x_{2n} > -1, \qquad
1 > x_2 > x_4 >  \cdots > x_{2n} > -1
$$
respectively. Thus in the limit $A \to - \infty$ (\ref{3.1}) reduces to the LSE
with parameter $a=0$, while (\ref{6.1}) reduces to the JSE with parameters
$a \mapsto a+1$ and $b=0$. We see from (\ref{fs1}) that in this limit
the skew inner product takes the form
$$
\lim_{A \to - \infty} \Big ( {A \over 2} \Big )^{2}
\langle f, g \rangle^A =
\left \{ \begin{array}{ll}
 \int_0^\infty dy \, e^{-y/2} \Big ( f(y) {d \over dy} (e^{-y/2} g(y)) -
(f \leftrightarrow g) \Big ), 
& \: {\rm Laguerre} \\[.1cm]
\int_{-1}^1 dy \, (1-y)^{(a+1)/2} \Big ( f(y) {d \over dy}
((1-y)^{(a+1)/2} g(y)  ) - (f \leftrightarrow g) \Big ), & 
\: {\rm Jacobi} \end{array} \right.
$$

In keeping  with the known results in the case $A=0$ and $A \to - \infty$,
we find that the skew inner product admits skew orthogonal polynomials with
compact expressions in terms of classical Laguerre and
Jacobi polynomials.
Their derivation relies on a special
integration formula for the Laguerre polynomial $L_k(x)$ and
Jacobi polynomial $P_k^{(a,0)}(x)$, and the latter in turn rely on
differentiation formulas for the same polynomials.

\begin{lemma}
We have
\begin{eqnarray}
{d \over dt} L_p(t) & = & - \sum_{l=0}^{p-1} L_l(t), \label{11.1} \\
(1-t) {d \over dt} P_p^{(a,0)}(t) & = & -p P_p^{(a,0)}(t) +
\sum_{l=0}^{p-1}(2l+1+a) (-1)^{p-1-l} P_l^{(a,0)}(t). \label{11.2}
\end{eqnarray}
\end{lemma}

\noindent
Proof. \quad The formula (\ref{11.1}) is an immediate consequence of the
well known formula 
$$
{d \over dt} \Big ( L_n^a(t) - L_{n+1}^a(t) \Big ) = L_n^a(t)
$$
in the case $a=0$. To derive (\ref{11.2}), we make use of the general
formula 
$$
{d \over dt} P_p^{(a,b)}(t) = {1 \over 2} (p+a+b+1) P_{p-1}^{(a+1,b+1)}(t)
$$
in the case $b=0$, then the general formula
$$
(n + {1 \over 2} (a+b) + 1 ) (1 - t) P_n^{(a+1,b)}(t) =
(n+a+1) P_n^{(a,b)}(t) - (n+1)  P_{n+1}^{(a,b)}(t)
$$
again in the case $b=0$, to deduce that
$$
(1-t) {d \over dt} P_p^{(a,0)}(t) =
{p+1 + a \over 2p + 1 + a} \Big \{ (p+a) P_{p-1}^{(a,1)}(t) -
p P_{p}^{(a,1)}(t) \Big \}.
$$
The formula (\ref{11.2}) now follows from repeated use of the general
formula
$$
 P_n^{(a,b)}(t) = {1 \over n + a + b} \Big (
(2n+a+b) P_n^{(a,b-1)}(t) - (n+a) P_{n-1}^{(a,b)}(t) \Big )
$$ 
in the case $b=1$. \hfill $\square$

\begin{prop}\label{p10}
We have
\begin{equation}\label{10.2}
e^{(1-A)t/2} \int_t^\infty e^{-(1-A)x/2} L_k(x) \, dx =
\sum_{p=0}^k c_{kp}^L L_p(t)
\end{equation}
where
\begin{equation}\label{10.3}
c_{kk}^L = {2 \over 1 - A}, \qquad
c_{kp}^L = (-1)^{p-k}
 {4 \over (1-A)^2} \Big ( {1 - A \over 1 + A} \Big )^{p+1-k},
\: \: (p=0,\dots,k-1),
\end{equation}
and
\begin{equation}\label{10.4}
(1 - t)^{-(a-A+1)/2} \int_t^1(1-x)^{(a-A-1)/2} P_k^{(a,0)}(x) \, dx =
\sum_{p=0}^k c_{kp}^J P_p^{(a,0)}(t)
\end{equation}
where
\begin{equation}\label{10.5}
c_{kk}^J = {1 \over k + (a-A+1)/2}, \qquad
c_{kp}^J = (2p+1+a) A_p B_k, \: \: (p=0,\dots,k-1),
\end{equation}
\begin{equation}\label{10.6}
A_p := {\Gamma(p+(a-A+1)/2)) \over \Gamma(p+(a+A+3)/2))}, \qquad
B_k := {\Gamma(k+(a+A+1)/2)) \over \Gamma(k+(a-A+3)/2))}.
\end{equation}
\end{prop}

\noindent
Proof. \quad Consider first (\ref{10.2}). Multiplying both sides by
$e^{-(1-A)t/2}$ and differentiating, making use of
(\ref{11.1}), gives
$$
- L_k(t) = - {(1-A) \over 2} \sum_{p=0}^k c_{kp}^L L_k(t) -
\sum_{p=0}^k c_{kp}^L \sum_{l=0}^{p-1} L_l(t).
$$
Equating coefficients of $L_k(t)$ gives $c_{kk}^L$ as stated in (\ref{10.3}).
Equating coefficients of $L_p(t)$ $(p<k)$ gives
\begin{equation}\label{12.1}
0 = - {(1-A) \over 2} c_{kp}^L - \sum_{j=p+1}^k c_{kj}^L \qquad
(p<k).
\end{equation}
Replacing $p$ by $p-1$ and subtracting shows
\begin{equation}\label{12.2}
\delta_{k,p} = -{(1-A) \over 2} c_{k \, p-1}^L - {(1+A) \over 2}
c_{kp}^L \qquad (p \le k)
\end{equation}
Solving this for $c_{kp}^L$, $p=k-1,k-2, \dots, 0$ in order, making use
of the known value of $c_{pp}^L$, completes the derivation of
(\ref{10.3}).

Consider now (\ref{10.4}). Multiplying both sides by $(1-t)^{(a-A+1)/2}$
and differentiating, making use of (\ref{11.2}),  gives
\begin{eqnarray*}
- P_k^{(a,0)}(t) & = & \sum_{p=0}^k - \Big ( p + {1 \over 2} (a-A+1) \Big )
c_{kp}^J P_p^{(a,0)}(t) \nonumber \\
&& + \sum_{p=0}^k \sum_{j=0}^{p-1} c_{kp}^J (2j+1+a)
(-1)^{p-1-j} P_j^{(a,0)}(t).
\end{eqnarray*}
Equating coefficients of $P_p^{(a,0)}(t)$ gives
\begin{equation}\label{13.1}
- \delta_{p,k} = - \Big ( p + {1 \over 2} (a-A+1) \Big ) c_{kp}^J +
\sum_{l=p+1}^k c_{kl}^J (2p+1+a) (-1)^{l-1-p},
\end{equation}
and replacing $p$ by $p-1$, multiplying by $(2p+1+a)/(2p-1+a)$, then adding
to the original we obtain
\begin{equation}\label{13.2}
- \delta_{p,k} = \Big ( p + {1 \over 2} (a+A+1) \Big ) c_{kp}^J -
{(p+ (a-A-1)/2)(2p+1+a) \over 2p-1+a} c_{k \, p-1}^J.
\end{equation}
It follows from (\ref{13.1}) with $p=k$ that the value of $c_{kk}^J$ is as
stated in (\ref{10.5}). From knowledge of $c_{kk}^J$ we can use (\ref{13.2})
with $p=k-1,k-2,\dots,0$ in order to deduce the formula for
$c_{kp}^J$, $p<k$ in (\ref{10.5}).
\hfill $\square$

An immediate corollary of Proposition \ref{p10} combined with the
orthogonalities (\ref{6.3.4}) and (\ref{6.4.2}) is the following
integration formulas, which have direct use in the determination of the
sought skew orthogonal polynomials.
\begin{cor}\label{cr1}
We have
\begin{equation}\label{13.3}
\int_0^\infty dt \, e^{-(1+A)t/2} L_j(t) \int_t^\infty dx \,
e^{-(1-A)x/2} L_k(x) \: = \:
\left \{ \begin{array}{ll}0, & j > k \\
c_{kj}^L, & j \le k \end{array} \right.
\end{equation}
and
\begin{eqnarray}\label{13.4}&& \quad
\int_{-1}^1 dt \, (1-t)^{(a-A-1)/2} P_j^{(a,0)}(t)
\int_t^1 dx \, (1-x)^{(a+A-1)/2} P_k^{(a,0)}(x) \nonumber \\
&& \qquad \qquad =
\left \{ \begin{array}{ll}0, & j > k \\[.1cm]
\displaystyle{{2^{a+1} \over 2j+a+1}} c_{kj}^J, & j \le k \end{array} \right.
\end{eqnarray}
\end{cor}

The result of Corollary \ref{cr1} implies that in both the Laguerre
and Jacobi cases, we have identified a family of polynomials
$\{p_j(x)\}$ ($p_j(x) = L_j(x)$ in the Laguerre case, and
$p_j(x) = P_j^{(a,0)}(x)$ in the Jacobi case) such that
\begin{equation}\label{pA}
\langle p_j, p_k \rangle^A = \left \{ \begin{array}{ll} a_j b_k, &
j < k \\ 0, & j = k, \\ - a_j b_k, & j > k \end{array} \right.
\end{equation}
for certain $a_j$, $b_k$. We can use this special structure to construct
the corresponding skew orthogonal polynomials as series in
$\{p_j(x)\}$,
\begin{equation}\label{pAR}
R_l(x) = \sum_{j=0}^l \alpha_{lj} p_j(x), \qquad \alpha_{ll} = 1.
\end{equation}
This is equivalent to finding a lower triangular matrix
$T = [\alpha_{jk}]_{j,k=0,1,\dots,2n-1}$ with 1's down the diagonal such
that
\begin{equation}\label{3.36'}
T  \, [\alpha_{jk}] \, T^t =
\left [ \begin{array}{ccccccc} 0 & r_0 &  &  & & & \\
-r_0 & 0 & &  & & & \\ & & 0 & r_1  & & & \\
& & -r_1 & 0 & & & \\
& & & & \ddots & & \\
& & & & & 0 & r_{n-1} \\
& & & & & -r_{n-1} & 0 \end{array} \right ]
\end{equation}
where on the RHS all elements not explicitly shown are zero. An explicit
solution to this problem is given by the following result.

\begin{prop}\label{pr1a}
Let $p_j(x)$ $(j=0,1,\dots,)$ denote a polynomial of degree $j$, and
suppose the value of the skew product $\langle p_j, p_k \rangle$
factorizes as specified by (\ref{pA}). Then for $l$ even, with
\begin{equation}\label{leven}
\alpha_{l \, 2j+1} = - {\prod_{\mu = j+1}^{l/2 - 1} a_{2 \mu + 1} \over
\prod_{\mu = j+1}^{l/2 - 1} a_{2 \mu}} 
{\prod_{\mu = j+1}^{l/2} b_{2 \mu} \over
\prod_{\mu = j+1}^{l/2} b_{2 \mu-1}}, \quad
\alpha_{l \, 2j} =  {\prod_{\mu = j}^{l/2 - 1} a_{2 \mu + 1} \over
\prod_{\mu = j}^{l/2 - 1} a_{2 \mu}} 
{\prod_{\mu = j+1}^{l/2} b_{2 \mu} \over
\prod_{\mu = j+1}^{l/2} b_{2 \mu-1}}, \: j \le l/2-1
\end{equation} 
and for $l$ odd, with 
\begin{equation}\label{lodd}
\alpha_{l \, l-1} = - {b_l \over b_{l-1}}, \qquad \alpha_{lj} = 0, \:
j \le l-2
\end{equation}
the polynomials (\ref{pAR}) exhibit the skew orthogonality property
(\ref{sk}). The corresponding normalization is given by
\begin{equation}\label{norm}
r_{(l-1)/2} = a_{l-1} b_l.
\end{equation} 
\end{prop}

\noindent
Proof. \quad Suppose first that $l$ is even, and consider
$$
\langle p_j, R_l \rangle^A, \qquad j \le l.
$$
Since $p_j$ can be written in terms of $\{R_k\}_{k=0,\dots,l}$, it follows from
(\ref{sk}) that
\begin{equation}\label{pRa}
\langle p_j, R_l \rangle^A = 0.
\end{equation}
But on the other hand, it follows from (\ref{pAR}) and (\ref{pA}) that
\begin{equation}\label{pRb}
\langle p_j, R_l \rangle^A = a_j \sum_{\mu = j+1}^l \alpha_{l\mu} b_\mu
- b_j \sum_{\mu =0}^{j-1} \alpha_{l \mu} a_\mu.
\end{equation}
Equating (\ref{pRa}) and (\ref{pRb}), and calling the resulting equation
$C_j$ we see that forming
$$
- {1 \over a_j} C_j + {1 \over a_{j - 1}} C_{j-1}
$$
gives the equation
\begin{equation}\label{pRc}
 \alpha_{l j} b_j - {b_{j-1} \over a_{j-1}} \sum_{\mu = 0}^{j-2}
\alpha_{l \mu} a_\mu + {b_j \over a_j} \sum_{\mu = 0}^{j-1}
\alpha_{l \mu} a_\mu = 0,
\end{equation}
valid for $j=1,2,\dots$. Substituting $j=1,2,\dots$ in order in 
(\ref{pRc}), we deduce that
\begin{eqnarray}\label{pRd}
\alpha_{l \, 2k+1} & = & - {a_{2k} \over a_{2k+1}} \alpha_{l \, 2k}
\nonumber \\
\alpha_{l \, 2k+2} & = & - {b_{2k+1} \over b_{2k+2}} \alpha_{l \, 2k+1}
\end{eqnarray}
valid for $k=0,1,\dots, l/2 - 1$. Recalling the normalization $\alpha_{ll}
=1$ we see that the recurrences reproduce (\ref{leven}).

Consider now the case $l$ odd. Then
$$
\langle p_{l-1}, R_l \rangle^A = r_{(l-1)/2},
$$
and it follows from this and (\ref{pRb}) with $j=l-1$ that
\begin{equation}\label{s1}
r_{(l-1)/2} = \Big ( a_{l-1} B_l - B_{l-1} \sum_{\mu = 0}^{l-2}
 \alpha_{l \mu} a_\mu \Big ).
\end{equation}
Making use of (\ref{pRc}), which remains valid for $j < l-1$, we thus have
\begin{equation}\label{s2}
- {r_{(l-1)/2} \over a_{l-1}} =
\alpha_{l \, l-1} b_{l-1} - {b_{l-2} \over a_{l-2}}
\sum_{\mu = 0}^{l-3} \alpha_{l \mu} a_\mu +
{b_{l-1} \over a_{l-1}} \sum_{\mu = 0}^{l-2} \alpha_{l \mu} a_\mu.
\end{equation}
The fact that (\ref{pRc}) remains valid for $j < l-1$ means that the
equations (\ref{pRd}) are again valid, this time for $2k+1 \le l-2$ in the
first equation and $2k+2 \le l-3$ in the second equation. In particular,
since we are assuming $l$ is odd, it follows from the first equation in
(\ref{pRd}) that
$$
\sum_{\mu = 0}^m \alpha_{l \mu} a_\mu = 0, \qquad
m=1,3,\dots, l-2
$$
and thus (\ref{s1}) and (\ref{s2}) simplify to (\ref{norm}) and
\begin{equation}\label{ss1}
- {r_{(l-1)/2} \over a_{l-1}} = \alpha_{l \, l-1} b_{l-1} -
{b_{l-2} \over a_{l-2}} \alpha_{l \, l-3} a_{l-3}
\end{equation}
respectively. Substituting (\ref{norm}) in (\ref{ss1}), we see that choosing
$\alpha_{l \, l-3} = 0$ implies the value of $\alpha_{l \, l-1}$ given in
(\ref{lodd}). The final equation in (\ref{lodd}), $\alpha_{lj} = 0$,
$j \le l-2$, follows from having chosen $\alpha_{l \, l-3} = 0$ in
(\ref{ss1}) and the recurrences (\ref{pRd}). \hfill $\square$

Examination of the above proof shows that for $l$ odd the value of
$\alpha_{l \, l-1}$ is in fact completely arbitrary. This is because
the skew orthogonal polynomials as specified by (\ref{sk}) are not
unique. For a given family of polynomials $\{R_j(x)\}_{j=0,1,\dots}$
satisfying (\ref{sk}), the family with
$$
R_{2j+1}(x) \mapsto R_{2j+1}(x) + \gamma_j R_{2j}(x),
$$
$\gamma_j$ arbitrary, also satisfy (\ref{sk}). This non-uniqueness
underlies the arbitrariness of $\alpha_{l \, l-1}$; the choice made in
(\ref{lodd}) leads to the simplest result in that with this choice we
then have $\alpha_{lj} = 0$ for all $j \le l-2$.

Inserting the explicit value of $a_jb_k$ in (\ref{pA}) from
Corollary \ref{cr1}, we get from Proposition \ref{pr1a} the following
explicit formulas for the skew orthogonal polynomials in the Laguerre
and Jacobi cases.

\begin{cor}\label{cr2}
The polynomials
\begin{eqnarray}\label{rL}
R_{2l-1}^{(L)}(x) & = & L_{2l-1}(x) - {A + 1 \over A - 1} L_{2l-2}(x) 
\nonumber \\
R_{2l}^{(L)}(x) & = & \sum_{j=0}^l L_{2j}(x) - {A+1 \over A - 1}
\sum_{j=0}^{l-1} L_{2j+1}(x)
\end{eqnarray}
are skew orthogonal with respect to the skew inner product (\ref{fs1})
in the Laguerre case, while the polynomials
\begin{eqnarray}\label{rJ}
R_{2l-1}^{(J)}(x) & = & P_{2l-1}^{(a,0)}(x) - 
{2l + (a-A-3)/2  \over 2l + (a-A-1)/2 } P_{2l-2}^{(a,0)}(x)
\nonumber \\
R_{2l}^{(J)}(x) & = & P_{2l}^{(a,0)}(x) +
\sum_{j=0}^{2l-1} (-1)^j { j + (a- (-1)^j A + 1)/2 \over
2l + (a - A + 1)/2 } P_j^{(a,0)}(x)
\end{eqnarray}
are skew orthogonal with respect to the skew inner product (\ref{fs1})
in the Jacobi case. The corresponding normalizations are
\begin{equation}\label{rLJ}
r_l^{(L)} = - {4 \over (1 - A)^2}, \qquad
r_l^{(J)} = {2^{a+1} \over (2l+ (a-A+3)/2) (2l+(a-A+1)/2)}.
\end{equation}
\end{cor}

Even though the matrix elements (\ref{fs}) for
the correlations (\ref{6.2.1}) explicitly depend on $\{R_j(x)\}$,
we will not directly make use of the formulas (\ref{rL}) and
(\ref{rJ}) in our subsequent simplification of (\ref{fs}).
Rather we will make  use of these formulas to  evaluate the
indefinite integral 
\begin{equation}\label{sd}
\int_y^\infty w(x) \kappa(x,y) R_k(x) \, dx,
\end{equation}
which will then be used in (\ref{fs}).
With $R_k(x)$ replaced by $p_k(x)$, this integral is given
by (\ref{10.2}) in the Laguerre case, and (\ref{10.4}) in the Jacobi case.
Using the notation of the RHS of (\ref{pA}), and introducing the
additional symbol $\tilde{a}_k$, these results can be combined into
the single formula
\begin{equation}\label{sd1}
{1 \over \tilde{w}(y)}
\int_y^\infty w(x) \kappa(x,y) p_k(x) \, dx =
{ \tilde{a}_k \over h_k } p_k(y) +
\sum_{j=0}^{k-1} {a_j b_k \over h_j } p_j(y)
\end{equation}
where
$$
\tilde{w}(x) = \left \{ \begin{array}{ll} w(x), & \: {\rm Laguerre} \\
(1-x) w(x), & \: {\rm Jacobi} \end{array} \right. \qquad
h_j := \int_{-\infty}^\infty w(x)\tilde{w}(x) (p_j(x))^2 \, dx.
$$
We can use (\ref{sd1}) together with the result of Proposition \ref{pr1a}
to evaluate (\ref{sd}).

\begin{prop}
Let $\{R_j(x)\}$ be given by (\ref{pAR}), with the $\alpha_{lj}$ therein
specified by Proposition \ref{pr1a}. Furthermore, assume the integral
evaluation (\ref{sd1}). Then
\begin{equation}\label{3.51}
{1 \over \tilde{w}(y) }
\int_y^\infty w(x) \kappa(x,y) R_k(x) \, dx = \sum_{l=0}^k u_{kl} p_l(y)
\end{equation}
where for $k$ even
\begin{equation}\label{ue}
u_{kl} = \left \{ \begin{array}{ll}{\displaystyle
{\alpha_{kl} \tilde{a}_l \over h_l} +
{a_l \alpha_{k \, l+1} b_{l+1} \over h_l}}, & l \: {\rm odd} \\[.1cm]
{\displaystyle
{\alpha_{kl} \tilde{a}_l \over h_l}}, & l \: {\rm even}
\end{array} \right.
\end{equation}
while for $k$ odd
\begin{equation}\label{uo}
u_{kl} = \left \{ \begin{array}{ll}{\displaystyle
{\tilde{a}_k \over  h_k}}, & \: l=k \\[.1cm] {\displaystyle
{\tilde{a}_k \alpha_{k \, k-1} \over h_k}
+ {a_{k-1} b_k\over h_k}} , & \: l=k -1 \\
0, & \: {\rm otherwise} \end{array} \right.
\end{equation}
\end{prop}

\noindent
Proof. \quad Substituting (\ref{pAR}) in (\ref{sd}) and making use of
(\ref{sd1}) gives
$$
{1 \over \tilde{w}(y)}
\int_y^\infty w(x) \kappa(x,y) R_k(x) \, dx =
\sum_{j=0}^k {\alpha_{kj} \tilde{a}_j \over h_j} p_j(y)
+ \sum_{j=0}^k \alpha_{kj} \sum_{\mu=0}^{j-1}
{a_\mu b_j \over h_\mu} p_\mu(y).
$$
The coefficient of $p_l(y)$ in the above expression is
\begin{equation}\label{ce}
{ \alpha_{kl} \tilde{a}_l \over h_l} +
{ a_l \over h_l} \sum_{j=l+1}^k \alpha_{kj} b_j.
\end{equation}
But for $k$ even, the second formula in (\ref{pRd}) shows that we get
cancellation in pairs in the above summation, and (\ref{ue}) results.
For $k$ odd, we see from (\ref{lodd}) that the summation in
(\ref{ce}) vanishes for $l < k-1$, and that so too does the first term.
The terms which remain give (\ref{uo}).
\hfill $\square$

We now substitute the particular values of the quantities
$\tilde{a}_l, b_j, a_k$ implied by (\ref{10.3}) and (\ref{10.4}),
together with the explicit formulas for $\alpha_{kl}$ implied by
(\ref{rL}) and (\ref{rJ}), and the normalizations (\ref{6.3.4}) and
(\ref{6.4.2}), in (\ref{ue}) and (\ref{uo}). This shows that in both
the Laguerre and Jacobi cases the coefficients $u_{kl}$, up to a sign,
are independent of $l$. Furthermore, in the case $k$ even we can
identify the resulting series as a linear combination of
$$
{1 \over  \tilde{w}(y)}
\int_y^\infty w(x) p_k(x) \, dx \qquad {\rm and} \qquad p_k(x),
$$ 
or alternatively as a linear combination of
$$
{1 \over  \tilde{w}(y)}
\int_y^\infty w(x) p_{k+1}(x)  \, dx \qquad {\rm and} \qquad p_{k+1}(x).
$$

\begin{cor}\label{cr4}
The polynomials (\ref{rL}) have the properties that
\begin{equation}\label{rL1}
e^{y/2} \int_y^\infty e^{-t/2} e^{A(t-y)/2} R_{2k+1}^{(L)}(t) \, dt =
{2 \over 1 - A} \Big ( - L_{2k+1}(y) + L_{2k}(y) \Big ) 
\end{equation}
and
\begin{eqnarray}\label{rL2}
e^{y/2} \int_y^\infty e^{-t/2} e^{A(t-y)/2} R_{2k}^{(L)}(t) \, dt
& = & {2 \over 1 - A} \Big ( \sum_{j=0}^k L_{2j}(y) -
\sum_{j=1}^k L_{2j-1}(y) \Big ) \nonumber \\
& = & {1 \over 1 - A} L_{2k}(y) + {e^{y/2} \over 2 (1 - A)}
\int_y^\infty e^{-s/2} L_{2k}(s) \, ds \nonumber \\
& = & {1 \over 1 - A} L_{2k+1}(y) - {e^{y/2} \over 2 (1 - A)}
\int_y^\infty e^{-s/2} L_{2k+1}(s) \, ds
\end{eqnarray}
while the polynomials (\ref{rJ}) have the properties that
\begin{equation}\label{rJ1}
(1-y)^{-(a+1)/2} \int_y^1 (1-t)^{(a-1)/2} \Big ( {1-t \over 1 - y}
\Big )^{-A/2} R_{2k+1}^{(J)}(t) \, dt =
{1 \over 2k + (a-A+3)/2} \Big (
P_{2k+1}^{(a,0)}(y) + P_{2k}^{(a,0)}(y) \Big )
\end{equation}
and
\begin{eqnarray}\label{rJ2}
\lefteqn{
(1-y)^{-(a+1)/2} \int_y^1 (1-t)^{(a-1)/2} \Big ( {1-t \over 1 - y}
\Big )^{-A/2} R_{2k}^{(J)}(t)  \, dt} \nonumber \\&&
= {1 \over 2k + (a-A+1)/2} \sum_{l=0}^{2k} P_l^{(a,0)}(y) 
\nonumber \\&&
= {1 \over 2k + (a-A+1)/2} \Big (
{1 \over 2} P_{2k}^{(a,0)}(y) +
{2k + (a+1)/2 \over 2 (1-y)^{(a+1)/2}}
\int_y^1 (1 - t)^{(a-1)/2} P_{2k}^{(a,0)}(t) \, dt \Big ) \nonumber \\&&
= {1 \over 2k + (a-A+1)/2} \Big (
- {1 \over 2} P_{2k+1}^{(a,0)}(y) +
{2k + 1 + (a+1)/2 \over 2 (1-y)^{(a+1)/2}}
\int_y^1 (1 - t)^{(a-1)/2} P_{2k+1}^{(a,0)}(t) \, dt \Big )
\end{eqnarray}
\end{cor}

\noindent
Proof. \quad The series expansions follow immediately upon making the
stated substitutions. To obtain the integral formulas, we substitute
for the summations in (\ref{rL2}) and (\ref{rJ2}) according to their
value implied by (\ref{10.2}) and (\ref{10.4}) respectively.
\hfill $\square$

We remark that the series of Laguerre and Jacobi
polynomials in (\ref{rL1})--(\ref{rJ2})
can each, according to the results (\ref{rL}) and (\ref{rJ}),
be identified with $R_j(y) \Big |_{A \to - \infty}$ for $j=2k+1$
or $j=2k$ as appropriate.

\subsection{Summation formulas --- the even-even block}\label{see}
The skew orthogonality property (\ref{sk}) implies the matrix 
$M_{jk}$ as specified
by (\ref{mjk}) is equal to $-1$ times the RHS of (\ref{3.36'}). Thus
\begin{equation}\label{mf}
M_{jk}^{-t} = \left \{ \begin{array}{ll} 0, & \: (j,k) \ne (2l, 2l+1)
\: {\rm or} \: (2l+1,2l) \\
- r_l^{-1}, & \: (j,k) = (2l, 2l+1) \: \: (l=0,\dots,n-1) \\
r_l^{-1}, & \: (j,k) = (2l+1,2l) \: \: (l=0,\dots,n-1)
\end{array} \right.
\end{equation}
and so the double summations in (\ref{fs}) all collapse to single summations.
In particular, with the entry in row $s$, column $s'$ of the matrix
$f_{ab}$ $(a,b= \, {\rm e} \, {\rm or} \, {\rm o})$ denoted $f_{ab}^{ss'}$,
we have that
\begin{equation}\label{mf0}
f_{\rm ee}^{12}(y,y') =  \sum_{j=0}^{n-1}
{1 \over r_j} \Big ( \Phi_{2j}^{\rm e}(y) \Phi_{2j+1}^{\rm e}(y') -
\Phi_{2j}^{\rm e}(y') \Phi_{2j+1}^{\rm e}(y) \Big ),
\end{equation}
where $\Phi_j^{\rm e}$ is defined by (\ref{1.23c}) and (\ref{1.23d})
in the Laguerre and Jacobi cases respectively. The latter indefinite
integrals are precisely those occurring in Corollary \ref{cr4}. 
Using this result, $f_{\rm ee}^{12}$ can be expressed in terms of the
functions $K^L_{2n}$ in the Laguerre case and $K^J_{2n}$ in the Jacobi case.

\begin{prop}
In the Laguerre case
\begin{equation}\label{mf1}
f_{\rm ee}^{12}(y,y') = {1 \over 4} \Big ( \int_{y}^\infty
K_{2n}^L(y',t) \, dt - \int_{y'}^\infty K_{2n}^L(y,t) \, dt \Big )
\end{equation}
while in the Jacobi case
\begin{equation}\label{mf2}
f_{\rm ee}^{12}(y,y') = {1 \over 4} \Big ( (1-y')
 \int_{y}^1 K_{2n}^J(y',t) \, dt - (1-y)
\int_{y'}^1 K_{2n}^J(y,t) \, dt \Big ).
\end{equation}
\end{prop}

\noindent
Proof. \quad Consider the Jacobi case. Substituting for $r_j^{(J)}$
using (\ref{rLJ}) and substituting for $\Phi_{2k+1}(y')$ using
(\ref{rJ1}) we see that
\begin{eqnarray*}\lefteqn{
f_{\rm ee}^{12}(y,y') =  {1 \over 2^{a+1}} (1-y)^{(a+1)/2} 
(1-y')^{(a+1)/2}} \\&& \times
\sum_{j=0}^{n-1} (2j+(a-A+1)/2) \Big (
P_{2j+1}^{(a,0)}(y') \Phi_{2j}(y) + P_{2j}^{(a,0)}(y') \Phi_{2j}(y)
\Big ) - (y \leftrightarrow y').
\end{eqnarray*}
In the product $P_{2j+1}^{(a,0)}(y')  \Phi_{2j}(y)$ we
substitute for $(1-y)^{(a+1)/2}  \Phi_{2j}(y)$ using the third 
equality in (\ref{rJ2}), while in the product 
$P_{2j+1}^{(a,0)}(y') \Phi_{2j}(y)$ we substitute for
$(1-y)^{(a+1)/2} \Phi_{2j}(y)$ using the second equality in (\ref{rJ2}).
Straightforward simplification and comparison with (\ref{1.25}), taking note
of (\ref{6.4.2}),
gives (\ref{mf2}). The Laguerre case is similar. \hfill $\square$

The single sum form of the other matrix elements  of $f_{\rm ee}$ are
$$
f_{\rm ee}^{11}(y,y') = f_{\rm ee}^{22}(y',y) =
-\sum_{j=0}^{n-1} {w(y') \over r_j} \Big ( \Phi_{2j}^{\rm e}(y) R_{2j+1}(y')
-  \Phi_{2j+1}^{\rm e}(y) R_{2j}(y') \Big ),
$$
\begin{equation}\label{mf3}
f_{\rm ee}^{21}(y,y') = -\sum_{j=0}^{n-1} {w(y) w(y')  \over r_j}
\Big ( R_{2j}(y) R_{2j+1}(y')
-  R_{2j+1}(y) R_{2j}(y') \Big ).
\end{equation}
Noting from (\ref{1.23c}) that in the Laguerre case
$$
{A \over 2} \Phi_j^{\rm e}(x) + {d \over dx}  \Phi_j^{\rm e}(x) =
- e^{-x/2} R_j(x)
$$
and from (\ref{1.23d}) that in the Jacobi case
$$
{A \over 2(1-x)}  \Phi_j^{\rm e}(x) + {d \over dx}  \Phi_j^{\rm e}(x) =
- (1-x)^{(a-1)/2} R_j(x),
$$
we see that all the quantities $f_{\rm ee}^{11}$, $f_{\rm ee}^{22}$
and $f_{\rm ee}^{21}$ can be expressed in terms of $f_{\rm ee}^{12}$ 
and thus $K_{2n}^L$ and $K_{2n}^J$.

\begin{prop}\label{pch}
In the Laguerre case
\begin{equation}\label{ch.1}
f_{\rm ee}^{11}(y,y') = f_{\rm ee}^{22}(y',y) =
\Big ( {A \over 2} + {\partial \over \partial y'} \Big ) 
f_{\rm ee}^{12}(y,y'), \quad
f_{\rm ee}^{21}(y,y') = - 
\Big ( {A \over 2} + {\partial \over \partial y} \Big )
\Big ( {A \over 2} + {\partial \over \partial y'} \Big )
f_{\rm ee}^{12}(y,y')
\end{equation}
while in the Jacobi case
$$
f_{\rm ee}^{11}(y,y') = f_{\rm ee}^{22}(y',y) =
\Big ( {A \over 2(1-y')} + {\partial \over \partial y'} \Big ) 
f_{\rm ee}^{12}(y,y'),
$$
\begin{equation}\label{ch.2} 
f_{\rm ee}^{21}(y,y') = - 
\Big ( {A \over 2(1-y)} + {\partial \over \partial y} \Big )
\Big ( {A \over 2(1-y')} + {\partial \over \partial y'} \Big )
f_{\rm ee}^{12}(y,y').
\end{equation}
\end{prop}

The $(0,k)$-point parity aware correlation, or equivalently the
$k$-point correlation for the even labelled coordinates, is
according to (\ref{1.17}) equal to qdet$[f_{\rm ee}(y_j,y_l)]_{j,l=1,
\dots,k}$. By performing elementary row and column operations, making
sure to conserve the self dual structure (\ref{3.1ad}), we see from
Proposition \ref{pch} that in both the Laguerre and Jacobi cases all terms
dependent on the parameter $A$ can be eliminated, leaving as the final
expression
\begin{equation}\label{mf5}
\rho_{(0,k)}(y_1,\dots,y_k) = {\rm qdet}
\left [ \begin{array}{cc} \displaystyle{{\partial \over \partial y_l} 
f_{\rm ee}^{12}(y_j,y_l)}
& f_{\rm ee}^{12}(y_j,y_l) \\[.1cm]
\displaystyle{- {\partial^2 \over \partial y_j\partial y_l} 
f_{\rm ee}^{12}(y_j,y_l)}
& \displaystyle{{\partial \over \partial y_j} f_{\rm ee}^{12}(y_l,y_j)}
\end{array} \right ]_{j,l=1,\dots,k}.
\end{equation}
The independence on $A$ is required by the identity (\ref{1m'}),
valid for the Laguerre and Jacobi weights in (\ref{1ma}). 
Moreover, this $k$-point correlation must agree with the $k$-point
correlation for the LSE with parameter $a=0$ in the Laguerre case,
and for the JSE with parameters $a \mapsto a+1$, $b=0$ in the
Jacobi case (see (\ref{nd5}) and Section \ref{s5.2}). 

\subsection{Summation formulas --- the parity blind case}\label{s3.4}

The matrix element (\ref{mf3}) is fundamental with respect to all other
matrix elements in (\ref{fs}) and (\ref{fxa}) in that each of the latter
can be constructed from (\ref{mf3}) by integration. The evaluation
(\ref{mf1}) in the Laguerre case and (\ref{ch.2}) in the Jacobi case
of (\ref{mf3}) then allows us to express all matrix elements in
(\ref{fs}) and (\ref{fxa}) in terms of $K^L_{2n}$ and $K^J_{2n}$. In this
subsection we will undertake this program for the matrix elements of
(\ref{fxa}). Formulas for the matrix elements of the blocks
$f_{\rm eo}$, $f_{\rm oe}$ and $f_{\rm oo}$ in (\ref{3.12b}),
obtained using knowledge of the evaluation of the matrix elements
for the block $f_{\rm ee}$ given in subsection \ref{see}
and the evaluation of the matrix elements of (\ref{fxa}) to be given
in this subsection, will be presented in the next subsection.

Recalling the definition of $f_{\rm ee}^{21}(y,y')$ from (\ref{fs}),
and the definition (\ref{3.8a}) of $(\epsilon \cdot R_k)$, we see that
the matrix elements of $f(x,y)$ in (\ref{fxa}) can be written in
terms of $f_{\rm ee}^{21}$ according to
\begin{eqnarray}
f^{11}(x,y) & = & f^{22}(y,x) \: = \: 
\int_{-\infty}^\infty  \epsilon(y,t) f_{\rm ee}^{21}(x,t) \, dt
\label{pr1} \\
f^{12}(x,y) & = & - f_{\rm ee}^{21}(x,y) \label{pr2} \\
f^{21}(x,y)  & = & -\epsilon(x,y) + \int_{-\infty}^\infty ds \, 
 \epsilon(x,s)
\int_{-\infty}^\infty dt \,  \epsilon(y,t)  f_{\rm ee}^{21}(s,t)
\label{pr3}
\end{eqnarray}
These formulas can be made more explicit. For this purpose use will be made
of the following formulas.

\begin{lemma}
We have
\begin{equation}\label{le1}
- \int_y^\infty K_{2n}^L(0,u) \, du = \int_0^\infty K_{2n}^L(y,u) \, du
\end{equation}
and
\begin{equation}\label{le2}
- \int_y^1 K_{2n}^J(-1,u) \, du = {(1-y) \over 2}
\int_{-1}^1 K_{2n}^J(y,u) \, du
\end{equation}
\end{lemma}

\noindent
Proof. \quad It follows from (\ref{10.2}) with $t=0$, $A=0$, and the
evaluation formula
\begin{equation}\label{le3}
L_p(0) = 1
\end{equation}
that
$$
\int_0^\infty e^{-t/2} L_p(t) \, dt = 2 (-1)^p.
$$
Recalling the definition (\ref{6.3.3}) of $K_n^L$ we thus have
\begin{equation}\label{le4}
\int_0^\infty K_{2n}^L(y,u) \, du = 2 e^{-y/2}
\sum_{p=0}^{2n-1} (-1)^p L_p(y).
\end{equation}
Also, from (\ref{10.2}) with $A=0$ we have
\begin{equation}\label{le4'}
e^{t/2} \int_t^\infty e^{-x/2} L_k(x) \, dx =
2 L_k(t) + 4 (-1)^k \sum_{p=0}^{k-1} (-1)^p L_p(t).
\end{equation}
It follows from this integration formula and (\ref{le3}) that the LHS of
(\ref{le1}) reduces to the RHS of (\ref{le4}).

To establish (\ref{le2}), we note that it follows from (\ref{10.4})
with $t=0$, $A=0$, and the evaluation formula
\begin{equation}\label{le5}
P_l^{(a,0)}(0) = (-1)^l
\end{equation}
that
\begin{equation}\label{le5a}
\int_{-1}^1 (1-x)^{(a-1)/2} P_k^{(a,0)}(x) \, dx = {2^{(a+1)/2} \over
k + (a+1)/2}.
\end{equation}
It follows from this and (\ref{1.25}) that
\begin{equation}\label{le6}
\int_{-1}^1 K_{2n}^J(t,y) \, dt = 2^{-(a-1)/2} (1-y)^{(a-1)/2}
\sum_{j=0}^{2n-1} P_j^{(a,0)}(y).
\end{equation}
Regarding the LHS of (\ref{le2}), we note from (\ref{10.4}) with 
$A=0$ that
\begin{equation}\label{le5b}
(1-t)^{-(a+1)/2} \int_t^1 (1-x)^{(a-1)/2} P_k^{(a,0)}(x) \, dx =
{1 \over k + (a+1)/2} \Big \{ P_k(t) + 2
\sum_{l=0}^{k-1} P_l(t) \Big \}.
\end{equation}
This integration formula together with (\ref{le5}) shows that the LHS
of (\ref{le2}) reduces to the RHS of (\ref{le6}).
\hfill  $\square$

\begin{prop}\label{p10a}
In the Laguerre case
\begin{eqnarray}
f^{\rm 22}(x,y) & = & {1 \over 2} K_{2n}^L(x,y) - {1 \over 2}
{\partial \over \partial y} \int_0^x e^{A(x-t)/2}
K_{2n}^L(t,y) \, dt \nonumber \\
&& - {A \over 4}  \int_0^x dt \, e^{A(x-t)/2}
\int_y^\infty du \, {\partial \over \partial t} K_{2n}^L(u,t)
+ {A \over 4} e^{Ax/2} \int_0^\infty K_{2n}^L(y,u) \, du
\label{lc1} \\
f^{\rm 12}(x,y) & = & {1 \over 4}\Big ( {A \over 2} +
{\partial \over \partial x} \Big ) \Big (
{A \over 2} +
{\partial \over \partial y} \Big ) \Big \{
\int_{x}^\infty K_{2n}^L(y,t) \, dt  \nonumber \\
&&-\int_{y}^\infty K_{2n}^L(x,t) \, dt \Big \}
\label{lc2} \\
f^{\rm 21}(x,y) & = & - e^{A|x-y|/2} {\rm sgn}(x-y) -
\Big \{ \int_0^y e^{A(y-t)/2} K_{2n}^L(x,t) \, dt \nonumber \\
&& - \int_0^x e^{A(x-t)/2} K_{2n}^L(y,t) \, dt \Big \}
\label{lc3}
\end{eqnarray}
while in the Jacobi case
\begin{eqnarray}
f^{\rm 22}(x,y) & = & {1 \over 2} (1-x) K_{2n}^J(x,y) - {1 \over 2}
\Big \{ (1-y){\partial \over \partial y} -1 \Big \}
 \int_{-1}^x \Big ( {1-t \over 1 - x} \Big )^{A/2}  
K_{2n}^J(t,y) \, dt \nonumber \\
&& + {A \over 4 (1-y)}  \int_{-1}^x dt \, 
\Big ( {1-t \over 1 - x} \Big )^{A/2} 
\int_y^1 du \,\Big \{ 1 - (1-t)
{\partial \over \partial t} \Big \} K_{2n}^J(u,t) \nonumber \\ &&
+ {A \over 4 (1-y)} 
\Big ( {2 \over 1 - x} \Big )^{A/2}
 \int_{-1}^1 K_{2n}^J(y,u) \, du
\label{jc1} \\
f^{\rm 12}(x,y) & = & {1 \over 4}\Big ( {A \over 2 (1-x)} +
{\partial \over \partial x} \Big ) \Big (
{A \over 2 (1-y)} +
{\partial \over \partial y} \Big ) \Big \{
(1-y) \int_{x}^1  K_{2n}^J(y,t) \, dt  \nonumber \\
&&-(1-x) \int_{y}^1 K_{2n}^J(x,t) \, dt \Big \}
\label{jc2} \\
f^{\rm 21}(x,y) & = & - \Big ( {1-x \over 1 - y} \Big )^{-A
{\rm sgn}(x-y)/2}
 {\rm sgn}(x-y) -
\Big \{ (1-x)
\int_{-1}^y  \Big ( {1-t \over 1 - y} \Big )^{A/2}
 K_{2n}^J(x,t) \, dt \nonumber \\
&& - (1-y)
\int_{-1}^x \Big ( {1-t \over 1 - x} \Big )^{A/2}
 K_{2n}^J(y,t) \, dt \Big \}
\label{jc3}
\end{eqnarray}
\end{prop}

\noindent
Proof. \quad Consider first the Laguerre case. The formula (\ref{lc2})
follows immediately from (\ref{pr2}), upon substituting 
(\ref{mf1}) in the second formula of (\ref{ch.1}). In preparation for
deriving (\ref{lc1}), we note that the last substitution, after 
computation of the corresponding derivatives where possible, yields
\begin{eqnarray}\label{suf}
f_{\rm ee}^{21}(y,y') & = &  {1 \over 4} \Big ( {A \over 2}
\Big )^2 \left \{ \int_{y'}^\infty K_{2n}^L(y,t) \, dt +
\int_{y}^\infty K_{2n}^L(y',t) \, dt \right \} \nonumber  \\
&&+ {A \over 8}  \left \{ \int_{y'}^\infty
{\partial \over \partial y} K_{2n}^L(y,t) \, dt -
\int_{y}^\infty
{\partial \over \partial y'} K_{2n}^L(y',t) \, dt
\right \} \nonumber  \\ 
&& + {1 \over 4}  \left \{ {\partial \over \partial y} -
{\partial \over \partial y'} \right \}  K_{2n}^L(y,y').
\end{eqnarray}
Thus there are three distinct terms which must be substituted in
(\ref{pr1}).

Substituting the $A$ independent term from (\ref{suf}) in (\ref{pr1})
and integrating by parts gives the contribution
\begin{eqnarray}\label{suf1}
&&{1 \over 4} \Big \{
- e^{Ax/2} K_{2n}^L(0,y) + 2 K_{2n}^L(x,y) + {A \over 2}
\int_0^\infty e^{A|x-t|/2} K_{2n}^L(t,y) \, dt \nonumber \\ 
&& \quad
-2 {\partial \over \partial y} \int_0^x e^{A(x-t)/2} K_{2n}^L(t,y) \, dt +
{\partial \over \partial y} \int_0^\infty
e^{A|x-t|/2}  K_{2n}^L(t,y) \, dt \Big \}.
\end{eqnarray}
The expression obtained by substituting the term proportional to $A$
from (\ref{suf}) in (\ref{pr1}) can be simplified by integrating by
parts immediately in one term, while in the other first making
use of the identity
\begin{equation}\label{na}
- {A \over 2} {\rm sgn} (x-t) e^{A|x-t|/2} =
{d \over dt} e^{A|x-t|/2}
\end{equation}
and then integrating by parts. Doing this allows the contribution to be written
\begin{eqnarray}\label{suf2}
&&{1 \over 4} e^{Ax/2}
\int_0^\infty  {\partial \over \partial y} 
 K_{2n}^L(t,y) \, dt - {1 \over 4} \int_0^\infty e^{A|x-t|/2}
{\partial \over \partial y}  K_{2n}^L(t,y) \, dt\nonumber \\
&& - {A \over 4} \int_0^x dt \, e^{A|x-t|/2}
\int_y^\infty du \, {\partial \over \partial t}   K_{2n}^L(u,t) 
\nonumber \\
&& + {A \over 8} \int_0^\infty dt \, e^{A|x-t|/2}
\int_y^\infty du \, {\partial \over \partial t} K_{2n}^L(u,t).
\end{eqnarray}
Simplifying the contribution to (\ref{pr1}) from the term proportional
to $A^2$ from (\ref{suf}) in the same way gives for the final term
\begin{eqnarray}\label{suf3}
&&{A \over 8} \Big \{
- e^{Ax/2} \int_y^\infty K_{2n}^L(0,u) \, du -
\int_0^\infty dt \,  e^{A|x-t|/2}
\int_y^\infty du \, {\partial \over \partial t} K_{2n}^L(u,t) 
\nonumber \\ 
&& \quad + e^{Ax/2} \int_0^\infty K_{2n}^L(y,u) \, du -
\int_0^\infty e^{A|x-t|/2} K_{2n}^L(y,t) \, dt \Big \}
\end{eqnarray}
Adding together (\ref{suf1})--(\ref{suf3}) and simplifying using (\ref{le1})
and the formula
\begin{equation}\label{3.87'}
K_{2n}^L(0,u) = \int_0^\infty {\partial \over \partial y} 
K_{2n}^L(y,u) \, du,
\end{equation}
which follows from (\ref{le1}) by differentiation, we obtain (\ref{lc1}).
To derive (\ref{lc3}), we first note from (\ref{pr1}) that (\ref{pr3})
can be rewritten
$$
f^{21}(x,y) = \epsilon(x,y) + \int_{-\infty}^\infty ds \, 
\epsilon(x,s) f^{11}(s,y),
$$
and then substitute (\ref{lc1}) in this formula. Simplification along the
same lines as that detailed above gives (\ref{lc3}).

Next we turn our attention to the Jacobi case. Substituting (\ref{mf2})
in (\ref{ch.2}), then substituting the result in (\ref{pr2}) gives
(\ref{jc2}). The former substitution, with the derivatives computed where
possible, yields
\begin{eqnarray}\label{3.73'}
f_{\rm ee}^{21}(y,y') & = &  {A^2 \over 16}  \left \{
 {1 \over 1 - y'}  \int_{y'}^1 K_{2n}^J(y,t) \, dt -
{1 \over 1 - y} \int_{y}^1 K_{2n}^J(y',t) \, dt \right \} \nonumber  \\
&&+ {A \over 8}   \left \{ {1 \over 1 - y'}  \int_{y'}^1
\Big ( 1 - (1-y){\partial \over \partial y} \Big )
K_{2n}^J(y,t) \, dt -
{1 \over 1 - y} \int_{y}^1
\Big ( 1 - (1-y'){\partial \over \partial y'} \Big )
K_{2n}^J(y',t) \, dt
\right \} \nonumber  \\
&& - {1 \over 4}  \left \{ \Big ( (1-y){\partial \over \partial y}
-1 \Big ) - \Big ( (1-y')
{\partial \over \partial y'} - 1 \Big ) \right \}  K_{2n}^J(y,y').
\end{eqnarray}
This has an analogous structure to (\ref{suf}), consisting of three
distinct terms which must be substituted in (\ref{pr1}).

Substituting the $A$ dependent term from (\ref{3.73'}) in (\ref{pr1})
and integrating by parts gives the contribution
\begin{eqnarray}\label{nap1}
&&{1 \over 2} \Big ( {2 \over 1 - x} \Big )^{A/2} K_{2n}^J(-1,y) 
+ {1 \over 2} \int_{-1}^x \Big ( (1-t){\partial \over \partial t}
-1 \Big ) \Big ( {1 - t \over 1 - x} \Big )^{A/2}
K_{2n}^J(t,y) \, dt  \nonumber \\&&  + {A \over 8}
\int_{-1}^1 \Big ( {1 - t \over 1 - x} \Big )^{A {\rm sgn}(x-t)/2}
 K_{2n}^J(t,y) \, dt \nonumber \\ 
&&-{1 \over 2}
\Big ( (1-y) {\partial \over \partial y} - 1 \Big ) 
\int_{-1}^x \Big ( { 1 - t \over 1 - x} \Big )^{A/2}
K_{2n}^J(t,y) \, dt  \nonumber \\ && + {1 \over 4} 
\Big ( (1-y) {\partial \over \partial y} - 1 \Big ) 
\int_{-1}^1  \Big ( { 1 - t \over 1 - x} \Big )^{A
\rm{sgn}(x-t)/2}
K_{2n}^J(t,y) \, dt .
\end{eqnarray}
For the term proportional to $A$ in (\ref{3.73'}), as well as using
direct integration by parts to simplify its contribution to (\ref{pr1})
in one of the terms, we make use of the identity
\begin{equation}\label{nap}
- {A \over 2} {\rm sgn}(x-t) {1 \over 1 - t} (1-t)^{A {\rm sgn}(x-t)/2}
= {d \over dt} (1-t)^{-A {\rm sgn}(x-t)/2}
\end{equation}
(c.f.~(\ref{na})) and then integrate by parts in the other. This
contribution is then found to equal
\begin{eqnarray}\label{nap2}
&& - {1 \over 4} 
\Big ( {2 \over 1 - x} \Big )^{A/2}
 \int_{-1}^x \Big (1 - (1-y){\partial \over \partial y}
 \Big ) K_{2n}^J(t,y) \, dt \nonumber \\&& + {1 \over 4}
\int_{-1}^1 \Big ( {1 - t \over 1 - x} \Big )^{A {\rm sgn}(x-t)/2}
\Big (1 - (1-y){\partial \over \partial y}
 \Big ) K_{2n}^J(t,y) \, dt \nonumber \\ &&+
{A \over 4} {1 \over 1 - y} \int_{-1}^x \Big ( {1-t \over 1-x} \Big )^{A/2}
\Big ( \int_y^1 \Big ( 1 - (1-t){\partial \over \partial t}
 \Big ) K_{2n}^J(t,u) \, du \Big ) dt \nonumber  \\&&
- {A \over 8} {1 \over 1 - y} \int_{-1}^1 \Big ( {1-t \over 1-x} 
\Big )^{A{\rm sgn}(x-t)/2}
\Big ( \int_y^1 \Big ( 1 - (1-t){\partial \over \partial t}
 \Big ) K_{2n}^J(t,u) \, du \Big ) dt 
\end{eqnarray}
Use of (\ref{nap}) and integration by parts shows that the contribution
to (\ref{pr1}) from the term proportional to $A^2$ in (\ref{3.73'}) can
be written
\begin{eqnarray}\label{nap3}
&&- {A \over 4} {1 \over 1 - y}
\Big ( {2 \over 1 - x} \Big )^{A/2} \int_y^1 K_{2n}^J(-1,u) \, du
 \nonumber \\ &&+
{A \over 8} {1 \over 1 - y}
\int_{-1}^1 \Big ( {1-t \over 1-x}
\Big )^{A{\rm sgn}(x-t)/2} 
\Big ( \int_y^1 \Big ( 1 - (1-t){\partial \over \partial t}
\Big ) K_{2n}^J(t,u) \, du \Big ) dt \nonumber  \\
&&+ {A \over 8} \Big ( {2 \over 1 - x} \Big )^{A/2}
\int_{-1}^1 K(y,t) \, dt -{A \over 8}
\int_{-1}^1 
\Big ( {1 - t \over 1 - x} \Big )^{A {\rm sgn}(x-t)/2} K_{2n}^J(y,t) \, dt.
\end{eqnarray}
Adding together (\ref{nap1}), (\ref{nap2}) and (\ref{nap3}), and making use
of (\ref{le2}) gives (\ref{jc1}), but with the term
\begin{equation}\label{nap3a}
{1 \over 2} (1-x) K_{2n}^J(x,y)
\end{equation}
replaced by
\begin{equation}\label{nap3b}
\Big ( {2 \over 1 - x} \Big )^{A/2} K_{2n}^J(-1,y) +
{1 \over 2} \int_{-1}^x
\Big ( 1 - (1-t){\partial \over \partial t}
\Big ) \Big ( {1-t \over 1-x}
\Big )^{A/2}  K_{2n}^J(t,y) \, dt.
\end{equation}
In fact (\ref{nap3a}) and (\ref{nap3b}) are equal, as is deduced by integration
by parts in the latter. \hfill $\square$

In the orthogonal symmetry limit, $A=0$, the results of Proposition
\ref{p10a} can be written
\begin{eqnarray}\label{fsub}
f^{22}(x,y) & = & \left \{ \begin{array}{ll}\displaystyle{
{1 \over 2} K_{2n}^L(x,y) - {1 \over 2} {\partial \over \partial y}
\int_0^x  K_{2n}^L(t,y) \, dt}, & {\rm Laguerre}  \\[.1cm]
\displaystyle{
{1 \over 2} (1-x)K_{2n}^J(x,y) - {1 \over 2} \Big (
(1-y){\partial \over \partial y} - 1 \Big )
\int_{-1}^x  K_{2n}^J(t,y) \, dt}, & {\rm Jacobi} \end{array} \right.
\\
2 f^{12}(x,y) & = & {\partial \over \partial x} f^{22}(x,y) 
\label{fsub2} \\
{1 \over 2} f^{21}(x,y)  & = & -{1 \over 2} {\rm sgn}(x-y) -
\int_x^y f^{22}(x,t) \, dt. \label{fsub3}
\end{eqnarray}
The form of $f^{22}(x,y)$
given here
is different to the form known from earlier literature (see
Section \ref{s4.1}). 

It is also of interest to specialize Proposition \ref{p10a} in the
symplectic symmetry limit, $A \to - \infty$. Integration by parts shows
\begin{eqnarray}\label{fsubu}
f^{22}(x,y) & \sim & \left \{ \begin{array}{ll}\displaystyle{
{1 \over 2} K_{2n}^L(x,y) + {1 \over 2} {\partial \over \partial x}
\int_y^\infty  K_{2n}^L(t,x) \, dt}, & {\rm Laguerre}  \\[.1cm]
\displaystyle{
{1 \over 2} (1-x)K_{2n}^J(x,y) + {1 \over 2} {1 - x \over 1 - y}
\Big (
(1-x){\partial \over \partial x} - 1 \Big )
\int_{y}^1  K_{2n}^J(t,x) \, dt}, & {\rm Jacobi} \end{array} \right.
\\
{8 \over A^2} f^{12}(x,y)  & \sim & \left \{ \begin{array}{ll}\displaystyle{
-\int_y^x f^{22}(t,y) \, dt}, & {\rm Laguerre} \\[.1cm]
\displaystyle{-{1 \over 1 - x}
\int_y^x {f^{22}(t,y) \over (1-t)} \, dt}, &
{\rm Jacobi}  \end{array} \right. \label{fsubua}
\\
{A^2 \over 8} f^{21}(x,y)  & \sim &
\left \{ \begin{array}{ll}\displaystyle{ {\partial \over \partial y}
f^{22}(x,y)}, &  {\rm Laguerre}  \\[.1cm]
\displaystyle{(1-y) {\partial \over \partial y} 
(1-y) f^{22}(x,y)}, &  {\rm
Jacobi} \end{array} \right.. \label{fsubub}
\end{eqnarray} 
Equivalently, these formulas can be written
\begin{eqnarray}
{A^2 \over 8} f^{21}(x,y)  & \sim &
\left \{ \begin{array}{ll}\displaystyle{
{1 \over 2} \Big ( \int_x^\infty K_{2n}^L(y,t) \, dt -
\int_y^\infty  K_{2n}^L(x,t) \, dt \Big )}, &  {\rm Laguerre}  \\[.1cm]
\displaystyle{
{1 \over 2} \Big ( {1 \over 1 - x} \int_x^1 K_{2n}^J(y,t) \, dt -
{1 \over 1 - y} \int_y^1 K_{2n}^J(x,t) \, dt \Big )}, &
{\rm Jacobi}  \end{array} \right. \label{nd1} \\
f^{22}(x,y) & \sim & \left \{ \begin{array}{ll}\displaystyle{
{\partial \over \partial x} {A^2 \over 8} f^{21}(y,x)}
, &  {\rm Laguerre}  \\[.1cm]
\displaystyle{(1-x) {\partial \over \partial x} (1-x) 
{A^2 \over 8} f^{21}(y,x)}, & {\rm Jacobi}  \end{array} \right. \label{nd2} \\
{8 \over A^2} f^{12}(x,y)  & \sim & \left \{ \begin{array}{ll}\displaystyle{
- {\partial \over \partial y \partial x} {A^2 \over 8} f^{21}(x,y)},
&  {\rm Laguerre}  \\[.1cm]
\displaystyle{-(1-x)(1-y) {\partial^2 \over \partial x \partial y}
(1-x)(1-y) {A^2 \over 8} f^{21}(y,x)}, & {\rm Jacobi}  \end{array} \right.
\label{nd3}
\end{eqnarray}
Defining
\begin{equation}\label{nd4}
\tilde{f}^{21}(x,y) =
\left \{ \begin{array}{ll}\displaystyle{ \lim_{A \to - \infty} 
{A^2 \over 8} f^{21}(x,y)}, &  {\rm Laguerre}  \\[.1cm]
\displaystyle{(1-x)(1-y) 
\lim_{A \to - \infty} {A^2 \over 8}  f^{21}(x,y)}, &  {\rm Jacobi}
\end{array} \right.
\end{equation}
it follows from (\ref{nd1})--(\ref{nd3}), the first equality in
(\ref{pr1}) and the quaternion determinant formula (\ref{6.3.2}), that
the $k$-point correlation is identical in structure to
(\ref{mf5}), but with $f^{12}$ replaced by $\tilde{f}^{21}$.
Furthermore, comparing (\ref{nd1}) and (\ref{nd4}) with
(\ref{mf1}), (\ref{mf2}) we see that
\begin{equation}\label{nd5}
\tilde{f}^{21}(x,y) = 2 f_{\rm ee}^{12}(x,y).
\end{equation}
This is consistent with the fact that both the even-even correlation
(\ref{mf5}) and the $A \to - \infty$ limit of the parity blind
correlations coincide with the correlations for the corresponding symplectic
ensemble (the factor of 2 in (\ref{nd5}) is due to the double
degeneracy inherent in the $A \to - \infty$ limit).

\subsection{Summation formulas --- the odd-even and even-odd blocks}
The blocks $f_{\rm eo}$ and $f_{\rm oe}$ in (\ref{fs}) are duals in
the sense of (\ref{3.1ad}), and thus
$$
f_{\rm eo}(y,x) = \left [ \begin{array}{cc}
f_{\rm oe}^{22}(x,y) & - f_{\rm oe}^{12}(x,y) \\
- f_{\rm oe}^{21}(x,y) & f_{\rm oe}^{11}(x,y) \end{array} \right ].
$$
Consequently, it suffices to consider one of these blocks,
$f_{\rm oe}$ say. Now,
analogous to the formulas (\ref{pr1})--(\ref{pr3}) expressing the 
elements of the matrix $f(x,y)$ in terms of $f_{\rm ee}^{21}$, we can
express the elements of $f_{\rm oe}(x,y)$ in terms of $f_{\rm ee}^{21}$.
Thus we see from (\ref{fs}) that
\begin{eqnarray}
f_{\rm oe}^{11}(x,y) & = & f_{\rm ee}^{21}(x,y) \nonumber \\
f_{\rm oe}^{12}(x,y) & = & - \int_y^\infty  \kappa(t,y) 
f_{\rm ee}^{21}(x,t) \, dt\: = \:
- f_{\rm ee}^{11}(y,x) \nonumber \\
f_{\rm oe}^{21}(x,y)  & = & \int_{-\infty}^x  \kappa(x,t)
f_{\rm ee}^{21}(t,y) \, dt \nonumber \\ 
f_{\rm oe}^{22}(x,y)  & = & - \kappa(x,y) -
 \int_{-\infty}^x dt \,  \kappa(x,t) \int_y^\infty ds \,
\kappa(s,y) f_{\rm ee}^{21}(t,s).
\end{eqnarray}
The last two formulas can be rewritten to read
\begin{eqnarray}\label{3.97a}
f_{\rm oe}^{21}(x,y)  & = & - f^{22}(x,y) + f_{\rm ee}^{11}(x,y)
\nonumber \\
f_{\rm oe}^{21}(x,y)  & = & - \kappa(x,y) - \int_y^\infty 
\kappa(s,y) f^{22}(x,s) \, ds + f_{\rm ee}^{12}(x,y).
\label{3.97b}
\end{eqnarray}
The only quantity in these formulas
which is not known explicitly from our study of the even-even blocks and
the parity blind case 
is the integral in (\ref{3.97b}). Its simplified form is readily computed.

\begin{prop}\label{p11a}
In the Laguerre case
\begin{eqnarray}
\int_y^\infty \kappa(s,y) f^{22}(x,s) \, ds & = &
{1 \over 2} \int_0^x e^{A(x-t)/2} K_{2n}^L(t,y) \, dt
+ {1 \over 2} \int_0^x dt \, e^{A(x-t)/2} \int_y^\infty du \,
{\partial \over \partial t}  K_{2n}^L(u,t) \nonumber \\&&
+  {1 \over 2} e^{Ax/2} \int_y^\infty K_{2n}^L(0,t) \, dt
\end{eqnarray}
while in the Jacobi case
\begin{eqnarray}
\lefteqn{\int_y^\infty \kappa(s,y) f^{22}(x,s) \, ds =
{1 \over 2}(1-y) \int_{-1}^x \Big ( {1 - t \over 1 - x} \Big )^{A/2}
K_{2n}^J(t,y) \, dt} \nonumber \\
&&  - {1 \over 2}
\int_{-1}^x dt \, \Big ( {1 - t \over 1 - x} \Big )^{A/2}
\int_y^1 du \, \Big ( 1 - (1-t) {\partial \over \partial t} \Big )
K_{2n}^J(t,u) +  \Big ( {2 \over 1 - x} \Big )^{A/2}
\int_y^1 K_{2n}^J(-1,t) \, dt
\end{eqnarray}
\end{prop}

\noindent
Proof. \quad In the Laguerre case $f^{22}(x,s)$ is given by (\ref{lc1}),
while in the Jacobi case it is given by (\ref{jc1}). Substituting these
formulas as appropriate, and simplifying according to the strategy
of the proof of Proposition \ref{p10a} gives the stated formulas.
\hfill $\square$

\subsection{Summation formulas --- the odd-odd block}
We read off from (\ref{fs}) that
\begin{eqnarray}
f_{\rm oo}^{11}(x,x') & = & f_{\rm oo}^{22}(x,x') \: = \: 
- f_{\rm oe}^{21}(x',x) \nonumber \\
f_{\rm oo}^{12}(x,x') & = & - f_{\rm ee}^{22}(x,x') \nonumber \\
f_{\rm oo}^{21}(x,x') & = & \int_{-\infty}^x dt \, \kappa(x,t)
\int_{-\infty}^{x'} ds \, \kappa(x',s) f_{\rm oe}^{21}(t,s)
\nonumber \\
& = & f^{21}(x,x') + f_{\rm ee}^{12}(x,x') - f_{\rm oe}^{22}(x',x) 
- f_{\rm oe}^{22}(x,x').
\end{eqnarray}
The quantity $f_{\rm oe}^{21}$ is specified by (\ref{3.97a})
(with $f^{22}$ and $f_{\rm ee}^{11}$ therein having the explicit
forms (\ref{lc1}), (\ref{jc1}) and (\ref{ch.1}), (\ref{ch.2})
respectively); $f_{\rm ee}^{22}$ is given by the explicit forms
(\ref{lc3}), (\ref{jc3}); $f_{\rm ee}^{12}$ by (\ref{mf1}), (\ref{mf2});
$f_{\rm oe}^{22}$ by (\ref{3.97b}) (with $f_{\rm ee}^{12}$ given as noted,
and the integral by Proposition \ref{p11a}).

\section{Scaled form of the correlations}
\setcounter{equation}{0}

\subsection{Superimposed orthogonal ensembles with a parameter}
Consider first the Laguerre case. To leading order, the support of the
spectrum for the LOE is $[0,4n]$. As has been identified in previous
studies (see e.g.~\cite{Fo93a}), there are three distinct scaling regimes
in which different limiting forms of the correlations are obtained.
These are the hard edge, specified by the change of scale
\begin{equation}\label{sac1}
x_j \mapsto X_j/4n,
\end{equation}
the bulk of the spectrum, specified by
\begin{equation}\label{sac2}
x_j \mapsto c + \pi X_j/ \sqrt{n}, \qquad 0 < c < 4n \: \:  ({\rm fixed})
\end{equation}
and the soft edge, specified by
\begin{equation}\label{sac3}
x_j \mapsto 4n + 2(2n)^{1/3} X_j.
\end{equation}
In general, under the linear change of scale
$$
x_j = a(n) + b(n) X_j,
$$
the correlation functions transform to  correlation functions in the new
variables $\{X_j\}$ according to
\begin{equation}\label{sac4}
\rho_k(X_1,\dots,X_k) = (b(n))^k \rho_k(x_1,\dots,x_k).
\end{equation}
The significance of the particular scales (\ref{sac1})--(\ref{sac3}) is that
\begin{equation}\label{sac5}
\rho_k^{\rm scaled}(X_1,\dots,X_k) :=
\lim_{n \to \infty} (b(n))^k
\rho_k\Big ( a(n) + b(n) X_1, \dots, a(n) + b(n) X_k \Big )
\end{equation}
is well defined.

For the parameter dependent extension of the LOE (\ref{3.1}), and the parameter
dependent extension of the superimposed ensemble LOE$\, \cup \,$LOE
(\ref{4.1}), we expect the hard edge, bulk and soft edge scaled
limits to again all be well defined provided the scale of the parameter
$A$ is suitably chosen. The correct choice can be anticipated by the
requirement that the quantity $e^{A(x-y)/2}$ occurring in the
formulas (\ref{KL}) and 
(\ref{lc1})--(\ref{lc3}) be of order unity in the scaled limit. 
This is achieved by 
\begin{equation}\label{sac6}
A \mapsto \left \{ \begin{array}{ll} 4n \alpha, & {\rm hard \, edge} \\
\sqrt{n} \alpha / \pi, & {\rm bulk} \\
\alpha/2(2n)^{1/3}, & {\rm soft \, edge,} \end{array} \right.
\end{equation}
where $\alpha$ denotes the scaled parameter.

Let $c,\omega >0$ be otherwise
arbitrary fixed real numbers. Using the
asymptotic formulas \cite{Sz75}
\begin{equation}\label{ab1}
e^{-x/2}x^{a/2}L_n^a(x) = n'^{-a/2} {(n+a)! \over n!}
J_a(2(n'x)^{1/2}) + R_1,
\end{equation}
$$
n' = n + (a+1)/2, \qquad
R_1 = \left \{ \begin{array}{ll}
x^{5/4} {\rm O}(n^{a/2 - 3/4}), & cn^{-1} \le x \le \omega  \\
x^{a/2+2} {\rm O}(n^a), & 0 < x \le c n^{-1} \end{array} \right.
$$
where $J_a(z)$ denotes the Bessel function of order $a$,
\begin{equation}\label{ab2}
e^{-x/2}x^{a/2}L_n^a(x) = n^{a/2} 
{1 \over \pi^{1/2} (nx)^{1/4}} \bigg (
\cos ( 2 (nx)^{1/2} - a \pi / 2 - \pi / 4)
+ (nx)^{-1/2} {\rm O}(1) \bigg ), \quad cn^{-1} \le x \le \omega
\end{equation}
and \cite{Ol74}
\begin{equation}\label{ab3}
e^{-x/2}x^{a/2}L_n^a(x) =  n^{a/2} \bigg (
{(-1)^n \over 2^a (2n)^{1/3}} {\rm Ai}(t) + 
{\rm O}(e^{-t}){\rm o}(n^{-1/3}) \bigg )
\end{equation}
where $x = 4n + 2 + 2(2n)^{1/3} t$, $t \in [t_0, \infty)$, 
it is straightforward
to derive the well known formulas
\begin{eqnarray}\label{sap1}
 K^{\rm hard}(X,Y)  & := & \lim_{n \to \infty} {1 \over 4n}
K^L\Big ( {X \over 4n}, {Y \over 4n} \Big ) \nonumber \\
&
= & \chi_{X,Y>0}
{J_a(X^{1/2})Y^{1/2}J'_a(Y^{1/2}) - X^{1/2}J'_a(X^{1/2})J_a(Y^{1/2}) \over
2(X-Y)} \Big |_{a=0} \nonumber \\ &=&
 \chi_{X,Y>0}{1 \over 4} \int_0^1 J_a(\sqrt{Xt}) J_a(\sqrt{Yt}) \, dt
\Big |_{a=0}, \\
 K^{\rm bulk}(X,Y)  & := & \lim_{n \to \infty} {\pi \over \sqrt{n}}
K^L\Big ( c + {\pi X \over \sqrt{n}} ,  c + {\pi Y \over \sqrt{n}}
\Big ) \nonumber \\
 & = &{\sin \pi (X - Y) \over \pi (X - Y)} =
\int_0^1 \cos \pi (X - Y)t \, dt  \label{sap2} \\
 K^{\rm soft}(X,Y)  & := & \lim_{n \to \infty} 2 (2n)^{1/3}
K^L(4n+2(2n)^{1/3}X, 4n+2(2n)^{1/3}Y) \nonumber \\
 & = & { {\rm Ai}(X){\rm Ai}\,'(Y) - {\rm Ai}(Y){\rm Ai}\,'(X) \over X-Y}
=
\int_0^\infty {\rm Ai}(X+t) {\rm Ai}(Y+t) \, dt. \label{sap3}
\end{eqnarray}
We see from (\ref{KL}) that knowledge of (\ref{sap1})--(\ref{sap3})
immediately gives the scaled form of $K_{\rm ee}^L(y,y')$, since
we have
\begin{equation}\label{sap4}
K_{\rm ee}^{\rm scaled}(Y,Y') := \lim_{n \to \infty} b(n)
K_n^L(y,y') = K^{\rm scaled}(Y,Y').
\end{equation}
The scaled form of the remaining quantities in (\ref{KL}) is also obtained
by formally replacing $K_n^L$ by $K^{\rm scaled}$. We will see that the
derivation is straightforward in the cases of the hard edge and bulk limits.
For the soft edge limit the integrations in the formulas for
$K_{\rm eo}^L$ and $K_{\rm oo}^L$ are to leading order over the interval
$[0,4n]$ whereas (\ref{ab3}) applies to the interval $[4n,\infty)$. To
overcome this difficulty the following identity will be used.

\begin{lemma}
We have
\begin{eqnarray}\label{see0}
e^{Ax/2} \int_0^x e^{-Au/2} K_n^L(y,u) \, du & = &
\Big ( {A - 1 \over A + 1} \Big )^{n-1} e^{A(x-y)/2}
\int_y^\infty e^{-(1-A)u/2} \Big ( {d \over du} L_n(u) \Big ) \, du
\nonumber \\ && -
e^{Ax/2} \int_x^\infty e^{-A u/2} K_n^L(y,u) \, du.
\end{eqnarray}
\end{lemma}

\noindent
Proof. \quad It follows from the integration formula \cite{GR80}
$$
\int_0^\infty e^{-(1+A)t/2} L_k(t) \, dt = {2 \over 1 + A}
\Big ( {A - 1 \over A + 1} \Big )^k
$$
that
\begin{equation}\label{see1}
\int_0^\infty e^{-Au/2} K_n^L(y,u) \, du =
{2 \over 1 + A} e^{-y/2} \sum_{l=0}^{n-1}
\Big ( {A - 1 \over A + 1} \Big )^l L_l(y).
\end{equation}
But the sum in (\ref{see1}) can, according to (\ref{10.2}), be written as
an integral involving a single Laguerre polynomial, and (\ref{see0}) results.
\hfill $\square$

\begin{prop}\label{p14}
Let the term ``scaled'' refer to any of the hard edge, bulk or soft edge limits
as specified by (\ref{sac1})--(\ref{sac6}). Then in addition to
(\ref{sap4}) we have
\begin{eqnarray}
K_{\rm eo}^{\rm scaled}(Y,X) & := &
\lim_{n \to \infty} K_{\rm eo}^L(y,x) \: = \:
- e^{\alpha(X-Y)/2} \chi_{x > y} +
e^{\alpha X/2} \int_{-\infty}^X e^{-\alpha v/2}
K^{\rm scaled}(v,Y) \, dv \label{sam1} \\
K_{\rm oe}^{\rm scaled}(X,Y)  & := &
\lim_{n \to \infty} (b(n))^2 K_{\rm oe}^L(x,y) \: = \:
- e^{-\alpha X/2} {\partial \over \partial X} \Big \{
e^{\alpha X/2} K^{\rm scaled}(X,Y) \Big \} \label{sam2} \\
K_{\rm oo}^{\rm scaled}(X,X') & \! := \! &
\lim_{n \to \infty} b(n) K_{\rm oo}^L(x,x')  = 
-e^{\alpha (X-X') /2} 
{\partial \over \partial X} \Big \{ e^{\alpha X/2}
\int_{-\infty}^{X'} e^{-\alpha v/2}
K^{\rm scaled}(X,v) \, dv \Big \}, \nonumber \\\label{sam3}
\end{eqnarray}
valid for $\alpha \le 0$ in the hard edge and bulk cases, and for all
$\alpha$ in the soft edge case. 
\end{prop}

\noindent
Proof. \quad Using the differentiation formula
$$
{d \over dt} L_p^a(t) = - L_{p-1}^{a+1}(t)
$$
one can check the well known fact that the asymptotic formulas 
(\ref{ab1})--(\ref{ab3}) remain valid after differentiation. This,
together with (\ref{sap4}), then implies (\ref{sam2}). Consider next
$K_{\rm eo}^{\rm hard}$. The explicit form of the remainder term in
(\ref{ab1}) implies it does not contribute to the scaled limit 
(\ref{sac1}), rather the sole contribution comes from
(\ref{sap4}), and this implies (\ref{sam1}) in the hard edge case (note that
the lower terminal in (\ref{sam1}) can be replaced by 0 in this
case). For the bulk limit we note from (\ref{ab2}) and (\ref{ab3})
that we have
\begin{equation}\label{fa1}
K_{\rm eo}^{\rm bulk}(Y,X) =
- e^{\alpha(X-Y)/2} \chi_{x>y} + e^{\alpha X/2}
\int_{-\infty}^X e^{-\alpha v/2} K^{\rm scaled}(v,Y) \, dv + R,
\end{equation}
where
$$
R = \lim_{\epsilon \to 0} \lim_{n \to \infty}
e^{A \sqrt{n} c/ \pi} e^{\alpha X/2} {\pi \over \sqrt{n}}
\int_0^\epsilon e^{-A \sqrt{n} v /2} K^L_n(v,c) \, dv,
$$
and in which we are free to choose $\epsilon < c$. Now it follows from
(\ref{ab1}) that
$$
| K^L_n(v,c) | \le \sqrt{n} f(v), \qquad 0 \le v \le \epsilon
$$
where $f(v)$ is integrable. Thus with $A \le 0$
the remainder term  $R$ in
(\ref{fa1}) vanishes. 

The remaining case is the soft edge limit. We make use of the formula
(\ref{see0}) and the asymptotic expansion (\ref{ab3}), which together
imply
\begin{eqnarray*}
\lefteqn{ \lim_{n \to \infty \atop {\rm soft \, edge}} b(n)
e^{\alpha x/2} \int_0^x e^{-\alpha v /2} K_n^L(v,y) \, dy
} \nonumber  \\
&& = e^{-\alpha^3/24} e^{\alpha(X-Y)/2}
\int_Y^\infty e^{\alpha u/2} {\rm Ai}(u) \, du -
e^{\alpha X/2} \int_X^\infty e^{-\alpha u /2}
K^{\rm soft}(u,Y) \, du.
\end{eqnarray*}
The form (\ref{sam1}) now follows after making use of the easily
verified integration formula
\begin{equation}\label{e3}
\int_{-\infty}^\infty e^{\epsilon t} {\rm Ai}(t) \, dt =
e^{\epsilon^3/3}
\end{equation}
and the integral form of $K^{\rm soft}$ in (\ref{sap3}) to deduce that
$$
\int_{-\infty}^\infty e^{-\alpha u /2} K^{\rm soft}(u,Y) \, du =
e^{-\alpha^3/24} e^{-\alpha Y/2}
\int_Y^\infty {\rm Ai}(t) e^{\alpha t /2} \, dt.
$$

Finally, the result (\ref{sam3}) follows from the fact, already noted, that
the key asymptotic formulas (\ref{ab1})--(\ref{ab3}) remain valid
under differentiation, together with the method just used to
derive (\ref{sam1}). \hfill $\square$

We now turn our attention to the Jacobi case. Here there is a hard
edge scaling limit in the neighbourhood of both $x=-1$ and $x=1$, as well
as a bulk limit. From previous studies 
(see e.g.~\cite{NF95}) we know the appropriate scales are
\begin{equation}\label{sc1}
x \mapsto 1 - {X \over 2n^2}, \qquad x \mapsto -1 + {X \over 2n^2}
\end{equation}
for the hard edge at $x=1$, $x=-1$ respectively, and
\begin{equation}\label{sc2}
x \mapsto \cos \theta_0 - {X \over n} \sin \theta_0 \: \sim \:
\cos( \theta_0 + {X \over n}), \quad 0 < \theta_0 < \pi
\end{equation}
in the bulk. For the scaling of the parameter $A$ we choose
\begin{equation}\label{sc3}
A \mapsto \left \{ \begin{array}{ll} 4 n^2 \alpha, &
{\rm hard \, edge \, at \, } x = 1 \\
\alpha, & {\rm hard \, edge \, at \, } x = -1 \\
n \alpha (1 - \cos \theta_0)/ \sin \theta_0, & {\rm bulk} \end{array}
\right.
\end{equation}
This is suggested by the criterion that the term
$((1-x)/(1-y))^{-A/2}$ tend to a non-constant order one quantity.

A rigorous analysis of the scaling limits for the JOE and JSE has
recently been undertaken by Due\~nez \cite{Du01}. Following the
methodology therein, we make use of the asymptotic formulas
\begin{eqnarray}\label{du1}
&& P_n^{(a,b)}(\cos \theta) = (\pi n)^{-1/2}
\Big ( \sin {\theta \over 2} \Big )^{-a-1/2}
\Big ( \cos {\theta \over 2} \Big )^{-b-1/2}
\cos(n' \theta + \gamma) + E_1, \nonumber \\&&
\quad n' = n + (a+b+1)/2, \qquad \gamma = - {\pi \over 2}(a+1/2)
\end{eqnarray}
for $0 < \theta < \pi$, where
$$
E_1 = \theta^{-a-3/2} O(n^{-3/2}), \qquad
{\rm uniformly \, for } \, c/n \le \theta \le \pi - \epsilon
\: \: (c, \epsilon \ll 1)
$$
and
\begin{equation}\label{du2}
\Big ( \sin {\theta \over 2} \Big )^a
\Big ( \cos {\theta \over 2} \Big )^b
P_n^{(a,b)}(\cos \theta) = n^{-a}
{\Gamma (n+a + 1) \over n!}
\sqrt{ {\theta \over \sin \theta}} J_a(n' \theta) + E_2
\end{equation}
for $0 \le \theta < \pi$, where
$$
E_2 = \left \{ \begin{array}{ll} \theta^{1/2} O(n^{-3/2}), &
c/n \le \theta \le \pi - \epsilon \\
\theta^{a+2} O(n^a), & 0 < \theta \le c/n, \end{array} \right.
$$
again uniformly in $\theta$.
The asymptotic form (\ref{du1}) is dominant in the bulk, while (\ref{du2})
is dominant for the hard edge at $x=1$. To study the hard edge at
$x=-1$, we make use of the fact that
$$
P_n^{(a,b)}(-\cos \theta) = P_n^{(b,a)}(\cos \theta),
$$
and then  use  (\ref{du2}). The strategy of the proof of Proposition
\ref{p14} then yields the following forms for the matrix elements
determining the scaled
correlations.

\begin{prop}\label{p15}
The results of Proposition \ref{p14} hold for the scaled limit of the
matrix elements (\ref{KJ}) in the bulk and for the hard edge at $x=-1$.
For the hard edge at $x=1$ we have
\begin{eqnarray}
K_{\rm ee}^{\rm hard}(Y,Y') & = &
\Big ( {Y \over Y'} \Big )^{1/2} K^{\rm hard}(Y,Y') \nonumber \\
K_{\rm eo}^{\rm hard}(Y,X) & = & -
\Big ( {X \over Y} \Big )^{-\alpha/2} \chi_{x > y} +
{Y^{1/2} \over X^{\alpha/2}} \int_X^\infty v^{(\alpha - 1)/2}
K^{\rm hard}(v,Y) \, dv  \nonumber \\
K_{\rm oe}^{\rm hard}(X,Y) & = &  {X^{\alpha/2} \over Y^{1/2}}
{\partial \over \partial X} \Big \{ X^{(1-\alpha)/2}
K^{\rm hard}(X,Y) \Big \} \nonumber \\
K_{\rm oo}^{\rm hard}(X,X') & = & 
\Big ( {X \over X'} \Big )^{\alpha/2}
{\partial \over \partial X} \Big \{ X^{(1-\alpha)/2}
\int_{X'}^\infty v^{(\alpha - 1)/2}
K^{\rm hard}(X,v) \, dv \Big \}, 
\end{eqnarray}
valid for parameter values $\alpha < 3/2$. Here $K^{\rm hard}$ refers
to (\ref{sap1}), but without the restriction $a=0$.
\end{prop}

\subsection{Decimated orthogonal ensembles with a parameter}
The scaling limits introduced in the previous subsection in relation to
the superimposed orthogonal ensembles with a parameter all carry
over to the decimated orthogonal ensembles with a parameter.
Furthermore, the strategies used to derive the scaled limits in
Propositions \ref{p14} and \ref{p15} again suffice to derive the scaled
limits of the matrix elements determining the correlations for the
decimated Laguerre and Jacobi orthogonal ensembles with a parameter.
In the interest of economy of space we will restrict ourselves to
presenting the explicit form of the scaling limit for the parity blind
correlations only.

\begin{prop}\label{p16}
Let the term ``scaled'' refer to any of the hard edge, bulk or soft edge limits
for the parameter dependent Laguerre ensembles
as specified by (\ref{sac1})--(\ref{sac6}). Then the scaled form of the
matrix elements (\ref{lc1})--(\ref{lc3}) have the explicit form
\begin{eqnarray}
\lefteqn{f_{\rm scaled}^{22}(X,Y) := \lim_{n \to \infty}
b(n) f^{22}(x,y) \Big |_{2n \mapsto n}
= {1 \over 2} K^{\rm scaled}(X,Y) 
} \nonumber \\ && \quad - {1 \over 2}
{\partial \over \partial Y} \int_{-\infty}^X
e^{\alpha (X-t)/2} K^{\rm scaled}(t,Y) \, dt
- {\alpha \over 4} \int_{-\infty}^X dt \,
e^{\alpha (X - t)/2} \int_Y^\infty du \,
{\partial \over \partial t} K^{\rm scaled}(u,t) \label{wa1} \\
\lefteqn{f_{\rm scaled}^{12}(X,Y) := \lim_{n \to \infty}
(b(n))^2 f^{12}(x,y) \Big |_{2n \mapsto n} } \nonumber \\
&& \quad = {1 \over 4} \Big ( {\alpha \over 2} +
{\partial \over \partial X} \Big )
 \Big ( {\alpha \over 2} +
{\partial \over \partial Y }  \Big )
\Big \{ \int_X^\infty K^{\rm scaled}(Y,t) \, dt -
\int_Y^\infty K^{\rm scaled}(X,t) \, dt \Big \} \label{wa2} \\
\lefteqn{f_{\rm scaled}^{21}(X,Y) = \lim_{n \to \infty} 
f^{21}(x,y) \Big |_{2n \mapsto n} =
- e^{\alpha|X-Y|/2} {\rm sgn}(X-Y) } \nonumber \\
&& \quad - \Big \{
\int_{-\infty}^Y e^{\alpha(Y-t)/2} K^{\rm scaled}(X,t) \, dt -
\int_{-\infty}^X e^{\alpha(X-t)/2} K^{\rm scaled}(Y,t) \, dt
\Big \} \label{wa3}
\end{eqnarray}
\end{prop}

\noindent
Proof. \quad The only new feature of the form of the expressions in
(\ref{lc1})--(\ref{lc3}) relative to those in (\ref{KL}) is the need to
analyze
$$
\int_0^\infty K_{2n}^L(y,u) \, du.
$$
This is done by noting from (\ref{le4}) and (\ref{le4'}) that
\begin{eqnarray}\label{naf}
\int_0^\infty K_{2n}^L(y,u) \, du & = & {1 \over 2}
\int_y^\infty e^{-u/2} L_{2n}(u) \, du - e^{-y/2} L_{2n}(y)
\nonumber \\
 & = & 1 - {1 \over 2} \int_0^y e^{-u/2} L_{2n}(u) \, du - e^{-y/2} L_{2n}(y).
\end{eqnarray}
Apart from this, we follow the strategy used to deduce the scaled limits
in Proposition \ref{p14}. \hfill $\square$

In the Jacobi case the analogue of (\ref{naf}) is afforded by
(\ref{le5a})--(\ref{le5b}). Using the resulting identity, and employing
the strategies used to derive Propositions \ref{p14} and \ref{p15}
gives the following result.

\begin{prop}\label{p17}
The results of Proposition \ref{p16} hold for the scaled limit of
the matrix elements (\ref{jc1})--(\ref{jc3}) in the bulk and for the
hard edge at $x=-1$. For the hard edge at $x=1$ we have
\begin{eqnarray}
f_{\rm hard}^{22}(X,Y) & = & {1 \over 2} \Big ({X \over Y} \Big )^{1/2}
K^{\rm hard}(X,Y) + {1 \over 2} {\partial \over \partial Y} 
{Y^{1/2} \over X^{\alpha/2}}
\int_X^\infty t^{(\alpha -1)/2}
K^{\rm hard}(t,Y) \, dt \nonumber \\
&& - {\alpha \over 4 Y} \int_X^\infty dt \,
\Big ( {t \over X} \Big )^{\alpha/2} \int_0^Y du \, u^{-1/2}
{\partial \over \partial t} t^{1/2} K^{\rm hard}(u,t) \\
f_{\rm hard}^{12}(X,Y) & = & - {1 \over 4} \Big (
{\alpha \over 2 X} - {\partial \over \partial X} \Big )
\Big ( {\alpha \over 2 Y} - {\partial \over \partial Y} \Big )
\nonumber \\
&& \times \Big \{ Y^{1/2} \int_0^X t^{-1/2} K^{\rm hard}(Y,t) \, dt
- X^{1/2} \int_0^Y t^{-1/2} K^{\rm hard}(X,t) \, dt \Big \} 
\label{4.32} \\
f_{\rm hard}^{21}(X,Y) & = & - \Big ( {X \over Y} 
\Big )^{-\alpha{\rm sgn}(X-Y)/2}
{\rm sgn}(X-Y) \nonumber \\
&&+  \Big \{ {X^{1/2} \over Y^{\alpha/2}}
\int_Y^\infty t^{(\alpha - 1)/2} K^{\rm hard}(X,t) \, dt -
{Y^{1/2} \over X^{\alpha/2}} \int_X^\infty t^{(\alpha - 1)/2}
K^{\rm hard}(Y,t) \, dt \Big \}. \label{4.33}
\end{eqnarray}
valid for parameter values $\alpha < 3/2$, and where $K^{\rm hard}$ refers
to (\ref{sap1}), but without the restriction $a=0$ (both (\ref{4.32})
and (\ref{4.33}) have been multiplied by $-1$ --- an operation which
leaves qdet[$f_{\rm hard}$] unchanged --- so as to formally conserve
the relations (\ref{fsub2}) and (\ref{fsub3})).

\end{prop}

\section{Discussion}
\setcounter{equation}{0}
\subsection{The orthogonal symmetry limit}\label{s4.1}
In the orthogonal symmetry limit $A=0$ the matrix elements determining
the parity blind correlations are specified by (\ref{fsub})--(\ref{fsub3}).
We have already remarked that the value of $f^{11}(x,y)$ implied by
(\ref{fsub}) differs in structure to that known from previous
literature \cite{Wi99,FNH99,AFNV99}. To present the latter, which are
valid for general $a>-1$ in the LOE and general $a,b>-1$ in the JOE, we
introduce generalizations of (\ref{6.3.3}), (\ref{6.4.2}),
\begin{eqnarray}
K_{n,a}^L(x,y) & := & (xy)^{a/2} e^{-(x+y)/2} \sum_{l=0}^{n-1}
{1 \over h_{l,a}^L} L_l^a(x) L_l^a(y) \label{Kh} \\
K_{n,a,b}^J(x,y) & := & \Big ((1-x)(1-y) \Big )^{(a-1)/2}
 \Big ((1+x)(1+y) \Big )^{b/2} \sum_{l=0}^{n-1}
{1 \over h_{l,a,b}^J} P_l^{(a,b)}(x) P_l^{(a,b)}(y). \label{Kh1}
\end{eqnarray}
In (\ref{Kh}), $L_l^a$ denotes the Laguerre polynomial of degree $l$ with
orthogonality property
$$
\int_0^\infty t^a e^{-t} L_m^a(t) L_n^a(t) \, dt =
h_{n,a}^L \delta_{m,n}, \qquad  h_{n,a}^L = {\Gamma(a+n+1) \over
\Gamma(n+1)},
$$
while in (\ref{Kh1}), $P_l^{(a,b)}$ denotes the Jacobi polynomial of
degree $l$ with orthogonality property
\begin{eqnarray}
&&\int_{-1}^1(1-t)^a (1+t)^b P_m^{(a,b)}(t) P_n^{(a,b)}(t) \, dt =
h_{n,a,b}^J \delta_{m,n}, \nonumber \\ && \quad
h_{n,a,b}^J = {\Gamma(a+1+n) \Gamma(b+1+n) 2^{a+b+1} \over
\Gamma(n+1) \Gamma(a+b+1+n) (a+b+1+2n) }. \label{Kh1a}
\end{eqnarray}
The case $a=0$ of (\ref{Kh}) and $b=0$ of (\ref{Kh1}) reduces to
(\ref{6.3.3}) and (\ref{6.4.2}) respectively.

In terms of (\ref{Kh}) and (\ref{Kh1}), the results of \cite{AFNV99}
give that the $k$-point distribution for matrix ensembles
OE${}_n(x^{a/2} e^{-x/2})$ (LOE) and OE${}_n((1+x)^{b/2}
(1-x)^{(a-1)/2})$ (JOE), $n$ even,
 as specified by (\ref{4.3}) are given by
(\ref{6.3.2}) with the $2 \times 2$ matrix $f$ in (\ref{6.3.2})
having as its top left entry
\begin{equation}\label{Kh2}
f^{11}(x,y) = \Big ( {x \over y} \Big )^{1/2}
K_{n-1,a+1}^L(x,y) - {1 \over 4 h_{n-1,a}^L}
y^{a/2} e^{-y/2} L_{n-1}^{a+1}(y)
\int_0^\infty {\rm sgn}(x-u) L_{n-2}^{a+1}(u) u^{a/2} e^{-u/2} \, du
\end{equation}
in the Laguerre case, and
\begin{eqnarray}
f^{11}(x,y) & = & (1-x) \Big ( {1+x \over 1+y} \Big )^{1/2}
K_{n-1,a, b+1}^J(x,y) + {1 \over 4 h_{n-1,a,b}^J}
{ (a+b+n)(a-1+n) \over (a+b+2n-1)}
(1-y)^{(a-1)/2}  \nonumber \\
&& \times (1+y)^{b/2} P_{n-1}^{(a,b+1)}(y) \int_{-1}^1 {\rm sgn}(x-u)
P_{n-2}^{(a,b+1)}(u) (1-u)^{(a-1)/2} (1+u)^{b/2} \, du,
\label{Kh3}
\end{eqnarray}
in the Jacobi case. The other entries in the matrix $f$ are related to
$f^{11}$ by
\begin{equation}\label{Kh3h}
f^{22}(x,y) = f^{11}(y,x), \quad
f^{21}(x,y) = - {1 \over 2} {\rm sgn}(x-y) - \int_x^y f^{11}(x,u) \, du,
\quad
f^{12}(x,y) = {\partial \over \partial x} f^{11}(x,y),
\end{equation}
which are identical to the structure of the formulas obtained in
Section \ref{s3.4} provided we interchange $f^{11}$ and
$f^{22}$ (recall the first equality in (\ref{pr1}),
(\ref{fsub2}), (\ref{fsub3})).

We know from \cite{Wi99,FNH99,AFNV99} that (\ref{Kh2}) can be rewritten
to read
\begin{equation}\label{Kh3a}
f^{11}(x,y) = K_{n,a}^L(x,y) + F_1^L(y) F_2^L(x)
\end{equation}
where
\begin{eqnarray}
F_1^L(y) & = & {1 \over 4 h_{n-1,a}^L} y^{a/2} e^{-y/2} L_{n-1}^{a+1}(y)
\label{bh3a} \\
F_2^L(x) & = &  \Big ( -4 x^{a/2} e^{-x/2} L_{n-1}^a(x) -
\int_0^\infty {\rm sgn}(x-u) L_{n-2}^{a+1}(u) u^{a/2} e^{-u/2} \, du
\Big ) \nonumber \\
& = & - n \int_0^\infty {\rm sgn}(x-t) t^{a/2 -1} e^{-t/2} (
L_n^a(t) - L_{n-1}^a(t) ) \, dt. \label{bh3b}
\end{eqnarray}
The formula (\ref{Kh3}) in the Jacobi case can similarly be rewritten.

\begin{lemma}\label{lem18}
The formula (\ref{Kh3}) has the alternative form
\begin{equation}\label{Kh4}
f^{11}(x,y)  =  (1-x) K_{n,a,b}^J(x,y) + F_1^J(y) F_2^J(x)
\end{equation}
where
\begin{equation}
F_1^J(y)  =  {1 \over 4 h_{n-1,a,b}^J} \Big ( {a+b+n \over a+b-1+2n}
\Big ) (1-y)^{(a-1)/2} (1+y)^{b/2} P_{n-1}^{(a,b+1)}(y) 
\label{Kh4e}
\end{equation}
\begin{eqnarray} 
&&
F_2^J(x)  = 
\Big ( -4 (1-x)^{(a+1)/2} (1+x)^{b/2} P_{n-1}^{(a,b)}(x)  \nonumber
\\ && 
+ (a-1+n) \int_{-1}^1 {\rm sgn}(x-u) P_{n-2}^{(a,b+1)}(u)
(1-u)^{(a-1)/2} (1+u)^{b/2} \, du \Big ) 
\nonumber 
\\ &&
 =  {2n \over 2n + a + b}
\int_{-1}^1 {\rm sgn}(x-u) \Big ( (n+a+b) P_n^{(a,b)}(u)
+ (n+a) P_{n-1}^{(a,b)}(u) \Big ) (1-u)^{(a-1)/2} (1+u)^{b/2-1}du. 
\nonumber  \\ \label{Kh4q}
\end{eqnarray}
\end{lemma}

\noindent
Proof. \quad It follows from (\ref{Kh1a}) that
$$
{1 \over h_{l,a,b+1}^J} = {(a+b+1+l)(a+b+2+2l) \over 2 (b+1+l)
(a+b+1+2l)} {1 \over h_{l,a,b}^J}.
$$
Using this, and the Jacobi polynomial identity 
$$
(l + {a+b \over 2} + 1)(1+x) P_l^{(a,b+1)}(x) =
(l+b+1) P_l^{(a,b+1)}(x) + (l+1) P_{l+1}^{(a,b)}(x),
$$
we see that
\begin{eqnarray*}
\lefteqn{(1+x) \sum_{l=0}^{n-2} {1 \over h_{l,a,b+1}^J}
P_l^{(a,b+1)}(x) P_l^{(a,b+1)}(y)} \nonumber \\
&& = \sum_{l=0}^{n-1} {1 \over h_{l,a,b}^J}
\Big ( {1 \over a+b+1+2l} \Big ) P_l^{(a,b)}(x)
\Big \{ (a+b+1+l) P_l^{(a,b+1)}(y) + (a+l) P_{l-1}^{(a,b+1)}(y)
\Big \} \nonumber \\
&& - {1 \over h_{n-1,a,b}^J} \Big ( {a+b+n \over a+b-1+2n} \Big )
P_{n-1}^{(a,b)}(x) P_{n-1}^{(a,b+1)}(y).
\end{eqnarray*}
Simplifying the term in the brackets $\{ \, \}$ using the Jacobi
polynomial identity
$$
(a+b+1+l) P_l^{(a,b+1)}(x) + (a+l) P_{l-1}^{(a,b)}(x)
= (a+b+1+2l) P_l^{(a,b)}(x),
$$
recalling the definition (\ref{Kh1}), and substituting the result in
(\ref{Kh3}) we deduce (\ref{Kh4}) with $F_2^{J}(x)$ given by the first
equality in (\ref{Kh4q}).
The second equality in (\ref{Kh4q}) can be deduced from the first
by verifying that both expressions agree at $x=1$, and then verifying
that the derivative of both expressions agrees.
\hfill $\square$ 

We now equate the RHSs of the final equation in (\ref{Kh3h}), with the
substitutions (\ref{Kh3a}) and (\ref{Kh4}) (and $n \mapsto 2n$),
and (\ref{fsub2}), with the substitution (\ref{fsub}). In the
Laguerre case this gives
\begin{equation}\label{ph1}
{1 \over 2} \Big ( {\partial \over \partial x} +
{\partial \over \partial y} \Big ) K_{2n}^L(x,y) =
c_{2n,0}^L \phi_{2n,0}^L(x) \psi_{2n,0}^L(y)
\end{equation}
where 
\begin{eqnarray}\label{ph}
c_{n,a}^L & = & {n \over 2 h_{n-1,a}^L} \nonumber \\
\phi_{n,a}^L(x) & = & x^{a/2} e^{-x/2} L_{n-1}^{a+1}(x) 
 =  - x^{a/2 - 1} e^{-x/2} \Big ( nL_n^a(x) - (n+a) L_{n-1}^a(x)
\Big ) \nonumber \\
\psi_{n,a}^L(y) & = & y^{a/2 - 1} e^{-y/2} \Big (
L_n^a(y) - L_{n-1}^a(y) \Big )
\end{eqnarray}
(the parameter $a$ has been kept general in (\ref{ph})  for later
convenience). Note from (\ref{ph}) that the RHS of (\ref{ph1}) is
symmetric in $x$ and $y$, as is the LHS. The substitutions in the
Jacobi case give
\begin{equation}\label{ph3} 
{1 \over 2} \Big ( (1-x) {\partial \over \partial x} +
(1-y) {\partial \over \partial y} \Big ) 
(1-x) (1-y) K_{2n}^J(x,y) =
c_{2n,a,0}^J \phi_{2n,a,0}^J(x) \psi_{2n,a,0}^J(y)
\end{equation}
where
\begin{eqnarray}\label{ph4}
c_{n,a,b}^J & = & {1 \over 2 h_{n-1,a,b}^J} 
\Big ( {a+b+n \over a+b-1+2n} \Big ) 
\Big ( {2n \over 2n+a+b} \Big )
\nonumber \\
\phi_{n,a,b}^J(x) & = & (1-x)^{(a+1)/2}
P_{n-1}^{(a,b+1)}(x) \nonumber \\
& = & (1-x)^{(a+1)/2} (1+x)^{b/2-1}
{2 \over 2n+a+b} \Big ( (n+b) P_{n-1}^{(a,b)}(x) + n P_n^{(a,b)}(x) \Big ) 
\nonumber \\
\psi_{n,a,b}^J(y) & = & - (1-y)^{(a+1)/2} (1+y)^{b/2 - 1}
\Big ((n+a+b) P_n^{(a,b)}(y) + (n+a) P_{n-1}^{(a,b)}(y) \Big ).
\end{eqnarray}  
As with (\ref{ph1}), we see from (\ref{ph4}) that the RHS of (\ref{ph3}) is
symmetric in $x$ and $y$, as is the LHS.

The identity (\ref{ph1}) is the special case $a=0$ of the identity
\cite{Wi99}
\begin{equation}\label{w1}
\Big ( {\partial \over \partial x} +
{\partial \over \partial y} \Big ) K_{n,a}^L(x,y) =
c_{n,a}^L \Big ( \phi_{n,a}^L(x) \psi_{n,a}^L(y) +
\phi_{n,a}^L(y) \psi_{n,a}^L(x) \Big ).
\end{equation}
A simple
consequence of (\ref{w1}) is the integral representation \cite{Jo01}
\begin{equation}\label{ll1}
K_{n,a}^L(x,y) = c_{n,a}^L \int_0^\infty \Big (
\phi^L_{n,a}(x+t) \psi^L_{n,a}(y+t) + 
\phi^L_{n,a}(y+t) \psi^L_{n,a}(x+t) \Big ) \, dt.
\end{equation}
The result (\ref{ph3}) suggests an analogue of (\ref{w1}), and
(\ref{ll1}), in the Jacobi case.

\begin{prop}
With $K_{n,a,b}^J$ specified by (\ref{Kh1}), and $c_{n,a}^J,
\phi_{n,a,b}^J(x), \psi_{n,a,b}^J(y)$ by (\ref{ph4}), we have
\begin{equation}\label{w2}
\Big (  (1-x) {\partial \over \partial x} +
(1-y) {\partial \over \partial y} \Big ) 
(1-x)(1-y) K_{n,a,b}^J(x,y) =
c_{n,a,b}^J \Big ( \phi_{n,a,b}^J(x) \psi_{n,a,b}^J(y) +
\psi_{n,a,b}^J(x)  \phi_{n,a,b}^J(y) \Big ),
\end{equation}
and consequently, for $0< x,y < 1$,
\begin{equation}\label{w2a}
4 xy K_{n,a,b}^J(1-2x,1-2y) =
c_{n,a,b}^J \int_0^1 \Big (
\phi_{n,a,b}^J(1-xu) \psi_{n,a,b}^J(1-yu) +
\phi_{n,a,b}^J(1-yu) \psi_{n,a,b}^J(1-xu) \Big ) {du \over u}
\end{equation}
\end{prop}

\noindent
Proof. \quad We generally follow the strategy used in \cite{Wi99} in
the Laguerre case. But before doing so we note the general identity
$$
\Big ( (1-x) {\partial \over \partial x} +
(1-y) {\partial \over \partial y} \Big ) \Big (
{(1-x)^{1/2} (1-y)^{1/2} \over x - y} f \Big ) =
{(1-x)^{1/2} (1-y)^{1/2} \over x - y}
\Big ( (1-x) {\partial \over \partial x} +
(1-y) {\partial \over \partial y} \Big ) f.
$$
To make use of this identity we note from the Christoffel-Darboux
formula (see e.g.~\cite{Sz75}) that
\begin{eqnarray*}
K_{n,a,b}^J(x,y) & = & ((1-x)(1-y))^{(a-1)/2}
((1+x)(1+y))^{b/2} \bar{a}_n
{P_n^{(a,b)}(x) P_{n-1}^{(a,b)}(y) - 
P_n^{(a,b)}(y) P_{n-1}^{(a,b)}(x) \over x - y }
\end{eqnarray*}
where
$$
 \bar{a}_n = {n! \over 2^{a+b}} {\Gamma(n+a+b+1) \over
\Gamma(n+a) \Gamma(n+b) (2n+a+b)}.
$$
Thus
\begin{eqnarray}\label{wi1}
\lefteqn{
\Big ( (1-x) {\partial \over \partial x} +
(1-y) {\partial \over \partial y} \Big )
(1-x) (1-y)  K_{n,a,b}^J(x,y) } \nonumber \\
&& = {(1-x)^{1/2} (1-y)^{1/2} \over x - y} 
 \bar{a}_n
\Big ( (1-x) {\partial \over \partial x} +
(1-y) {\partial \over \partial y} \Big )
\Big ( q_n(x) q_{n-1}(y) - q_n(y) q_{n-1}(x) \Big )
\end{eqnarray}
where
$$
q_n(x) := (1-x)^{a/2} (1+x)^{b/2} P_n^{(a,b)}(x),
$$
so the task is to compute the action of the operator on the RHS of
(\ref{wi1}). 
 
For this purpose, we note from suitable Jacobi polynomial formulas that
\begin{eqnarray*}
(1-x^2) q_n'(x) & = & (\alpha_0 + \alpha_1 x) q_n(x) + \beta_0 q_{n-1}(x)
\\
(1-x^2) q_{n-1}'(x) & = & - \gamma_0  q_n(x) - ( \alpha_0 +
\alpha_1 x) q_{n-1}(x)
\end{eqnarray*}
with
\begin{eqnarray}\label{wi2}
&& \alpha_0 + \alpha_1 x  =  {b^2 - a^2 \over 2 (2n+a+b)} -
{(2n+a+b) \over 2} x \nonumber \\
&& \beta_0 = {2(n+a) (n+b) \over 2n+a+b}, \qquad
\gamma_0 = {2n(n+a+b)  \over 2n+a+b}.
\end{eqnarray}
Equivalently
\begin{equation}\label{wi3}
\left [ \begin{array}{c} (1-x) q_n'(x) \\ (1-x) q_{n-1}'(x) \end{array}
\right ] =
\left [ \begin{array}{cc} A(x) & B(x) \\ - C(x) & - A(x)
\end{array} \right ]
\left [ \begin{array}{c} q_n(x) \\ q_{n-1}(x) \end{array} \right ]
\end{equation}
where
$$
A(x) = {\alpha_0 + \alpha_1 x \over 1 + x}, \quad
B(x) = {\beta_0 \over 1 + x}, \quad
C(x) = {\gamma_0 \over 1 + x}.
$$
Introducing the matrix formulation 
$$
q_n(x) q_{n-1}(y) - q_n(y) q_{n-1}(x) =
  \begin{array}{cc} [ \, q_n(x) & q_{n-1}(x) \, ]  \\
& \end{array} 
\left [ \begin{array}{cc} 0 & 1 \\ -1 & 0 \end{array} \right ]
\left [ \begin{array}{c} q_n(y) \\ q_{n-1}(y) \end{array} \right ],
$$
a straightforward calculation using (\ref{wi3}) shows
\begin{eqnarray}\label{wi4}
\lefteqn{
{1 \over x - y}
\Big ( (1-x) {\partial \over \partial x} +
(1-y) {\partial \over \partial y} \Big )
\Big ( q_n(x) q_{n-1}(y) - q_n(y) q_{n-1}(x) \Big )} \nonumber \\
&& =
  \begin{array}{cc} [ \, q_n(x) & q_{n-1}(x) \, ] \\ & \end{array} 
\left [ \begin{array}{cc} \displaystyle
{C(x) - C(y) \over x - y} &  \displaystyle {A(x) - A(y) \over x - y} \\
\displaystyle {A(x) - A(y) \over x - y} & 
 \displaystyle {B(x) - B(y) \over x - y}
\end{array} \right ]
\left [ \begin{array}{c} q_n(y) \\ q_{n-1}(y) \end{array} \right ]
\nonumber \\
&& = - {1 \over (1+x) (1+y)}
  \begin{array}{cc} [\, q_n(x) & q_{n-1}(x) \, ]\\
& \end{array} 
\left [ \begin{array}{cc} \gamma_0 & \alpha_0 - \alpha_1 \\
\alpha_0 - \alpha_1 & \beta_0 \end{array} \right ]
\left [ \begin{array}{c} q_n(y) \\ q_{n-1}(y) \end{array} \right ]
\nonumber \\
&& = - {1 \over (1+x) (1+y)}
(1-x)^{a/2}(1+x)^{b/2} (1-y)^{a/2}(1+y)^{b/2} {2 \over 2n+a+b} \nonumber \\
&& \quad \times \Big [
n(n+a+b) P_n^{(a,b)}(x) P_n^{(a,b)}(y) + (n+a)(n+b)
P_{n-1}^{(a,b)}(x) P_{n-1}^{(a,b)}(y) \nonumber \\
&& \quad + {1 \over 4} (b^2 - a^2 + (2n+a+b)^2)
(P_n^{(a,b)}(x)  P_{n-1}^{(a,b)}(y) + P_n^{(a,b)}(y) P_{n-1}^{(a,b)}(x))
\Big ]
\end{eqnarray}
Substituting (\ref{wi4}) in (\ref{wi1}) gives (\ref{w2}).

With (\ref{w2}) established, we make the changes of variables
$$
x= 1 - 2e^{-s}, \qquad y = 1 - 2e^{-t}, \qquad \: 0 \le s,t < \infty
$$
which gives
\begin{eqnarray}\label{wi5}
&&
\Big ( {\partial \over \partial s} + {\partial \over \partial t}
\Big ) 4 e^{-(s+t)} K_{n,a,b}^J(1-2e^{-s}, 1-2e^{-t})
\nonumber \\ && \quad =
{c}_{n,a,b}^J \Big (
\phi_{n,a,b}^J(1 - 2e^{-s})
\psi_{n,a,b}^J(1 - 2e^{-t}) +
\phi_{n,a,b}^J(1 - 2e^{-t}) \psi_{n,a,b}^J(1 - 2e^{-s}) \Big ).
\end{eqnarray}
A simple calculation using integration by parts verifies that
\begin{eqnarray}\label{wi6}
\lefteqn{4 e^{-(s+t)} K_{n,a,b}^J(1-2e^{-s}, 1-2e^{-t})
= {c}_{n,a,b}^J}
\nonumber \\ &&
\times \int_0^\infty \Big (
\phi_{n,a,b}^J(1 - 2e^{-s-z})
\psi_{n,a,b}^J(1 - 2e^{-t-z}) +
\phi_{n,a,b}^J(1 - 2e^{-t-z}) \psi_{n,a,b}^J(1 - 2e^{-s-z}) \Big )
\, dz.
\end{eqnarray}
satisfies (\ref{wi5}). The general solution of (\ref{wi5}) is (\ref{wi6})
plus an arbitrary function of $s-t$. But we see from the definition of
$K_{n,a,b}^J$, $\phi_{n,a,b}^J$ and $\psi_{n,a,b}^J$ that both sides
of (\ref{wi6}) vanish in the limits $s,t \to \infty$, $s-t$ fixed, 
so this arbitrary function must be chosen to be the zero function.
Putting $e^{-s} = x$, $e^{-t} = y$ in (\ref{wi6}) and changing
variables $e^{-z} = u$ gives (\ref{w2a}). 
\hfill $\square$

The orthogonal symmetry limit of the scaled matrix elements of
Propositions \ref{p16} and \ref{p17} is obtained by setting
$\alpha = 0$. We see from the results of
Propositions \ref{p16} and \ref{p17} (with $f^{11}$ and $f^{22}$
interchanged for convenience) that the $k$-point distributions are
then given by (\ref{6.3.2}) with the $2 \times 2$ matrix $f$
having as its top left entry
\begin{equation}\label{ir1}
f_{\rm scaled}^{11}(X,Y) = {1 \over 2} K^{\rm scaled}(X,Y) -
{1 \over 2} {\partial \over \partial Y}
\int_{-\infty}^X  K^{\rm scaled}(t,Y) \, dt
\end{equation}
in the case of the soft edge and bulk
($ K^{\rm scaled}$ as given by (\ref{sap3}) and (\ref{sap2}) respectively),
and
\begin{equation}\label{ir2}
f_{\rm scaled}^{11}(X,Y) = {1 \over 2} \Big ( {X \over Y}
\Big )^{1/2} K^{\rm hard}(X,Y) +
{1 \over 2} {\partial \over \partial Y} Y^{1/2}
\int_{X}^\infty  t^{-1/2} K^{\rm hard}(t,Y) \, dt
\end{equation}
in the case of the hard edge with singularity $x^{(a-1)/2}$ as
$x \to 0$ ($K^{\rm hard}$ is given by (\ref{sap1}), without the
restriction $a=0$). The remaining entries in $f$ are related to $f^{11}$ as
in (\ref{Kh3h}).

The formula (\ref{ir2}) has been read off from Proposition \ref{p17},
deduced from the scaled limit of the Jacobi case. The hard edge singularity
$x^{(a-1)/2}  |_{a=1}$ as $x \to 0$ is also contained in the results
of Proposition \ref{p16} for the scaled Laguerre case. In this case we read
off from Proposition \ref{p16} the alternative formula
\begin{equation}\label{ir3}
f_{\rm scaled}^{11}(X,Y) = {1 \over 2} K^{\rm hard}(X,Y)  |_{a=0} -
{1 \over 2} {\partial \over \partial Y} \int_0^X
 K^{\rm hard}(t,Y)   |_{a=0} dt,
\end{equation}
which differs in form to (\ref{ir2}) with $a=1$. Our first discussion point
regarding the form of $f_{\rm scaled}^{11}$  exhibited by
(\ref{ir1}) and (\ref{ir2}) is to reconcile the two seemingly different
expressions in the case of the hard edge singularity
$x^{(a-1)/2}  |_{a=1}$ as $x \to 0$. For this purpose we make use of
the following recurrence.

\begin{lemma}\label{lemq}
Let $K^{\rm hard}(X,Y)$ be given by (\ref{sap1}), without the restriction
to $a=0$. We have
$$
\Big ( {X \over Y}
\Big )^{1/2} K^{\rm hard}(X,Y) \Big |_{a \to a+1} =
K^{\rm hard}(X,Y) - {1 \over 2} Y^{-1/2} J_a(X^{1/2}) J_{a+1}(Y^{1/2})
$$
where it is understood that $X,Y >0$.
\end{lemma}

\noindent
Proof. \quad We first make use of the Bessel function identity
$$
u J_a'(u) = a J_a(u) - u J_{a+1}(u)
$$
to rewrite the first equality in (\ref{sap1}), without the restriction
to $a=0$, as
$$
K^{\rm hard}(X,Y) =
{X^{1/2} J_{a+1}(X^{1/2}) J_a(Y^{1/2}) - Y^{1/2} J_{a+1}(Y^{1/2})
J_a(X^{1/2}) \over 2 (X - Y)}.
$$
The stated formula now follows by replacing $a$ by $a+1$, using the Bessel
function identity
$$
u J_{a+2}(u) = 2 (a+1) J_{a+1}(u) - u J_a(u),
$$
and simple manipulation. \hfill $\square$

Using Lemma \ref{lemq} in (\ref{ir2}) we obtain
\begin{eqnarray}\label{ir4}
f_{\rm scaled}^{11}(X,Y)  \Big |_{a \to a+1} & = &
{1 \over 2} K^{\rm hard}(X,Y) -
{1 \over 4} Y^{-1/2} J_a(X^{1/2}) J_{a+1}(Y^{1/2}) \nonumber \\
&& + {1 \over 2} {\partial \over \partial  Y} \int_X^\infty
 K^{\rm hard}(t,Y) \, dt - {1 \over 4}
{\partial \over \partial  Y} J_a(Y^{1/2})
 \int_X^\infty t^{-1/2} J_{a+1}(t^{1/2}) \, dt.
\end{eqnarray}
In the special case $a=0$ the last integral can be evaluated, and this
shows the final term cancels with the second term. We thus have agreement
with (\ref{ir3}) provided we can show
\begin{equation}\label{is}
{\partial \over \partial  Y} \int_0^\infty
 K^{\rm hard}(t,Y) \Big |_{a = 0} \, dt = 0.
\end{equation}
Now, using the integral representation from (\ref{sap1}) shows
$$
\int_0^\infty
 K^{\rm hard}(t,Y) \Big |_{a = 0}
\, dt = {1 \over 4} \int_0^\infty dt \int_0^1ds \,
J_0 (\sqrt{ts}) J_0(\sqrt{Ys}) =
{1 \over 4} \int_0^\infty dt J_0 (\sqrt{t})
 \int_0^1ds \, {1 \over s} J_0(\sqrt{Ys}),
$$
where the second equality follows by changing variables $t \mapsto t/s$.
But
$$
 \int_0^\infty  J_0 (\sqrt{t}) \, dt = 2 \int_0^\infty s J_0(s) \, ds
= 2  \int_0^\infty {d \over ds} (s J_1(s)) \, ds = 0
$$
so indeed (\ref{is}) holds true.

Let us now return to the consideration of $f_{\rm scaled}^{11}$ in general.
Previous studies have given formulas of a different form to (\ref{ir1})
and (\ref{ir2}). These read \cite{FNH99}
\begin{eqnarray}
f_{\rm bulk}^{11}(X,Y) & = & K^{\rm bulk}(X,Y) \label{fk1} \\
f_{\rm soft}^{11}(X,Y) & = & K^{\rm soft}(X,Y) + {1 \over 2}
{\rm Ai}(Y) \int_{-\infty}^X {\rm Ai}(t) \, dt \label{fk2} \\
f_{\rm hard}^{11}(X,Y) & = & K^{\rm hard}(X,Y) +
{J_{a+1}(\sqrt{Y}) \over 4 \sqrt{Y}}
\int_{\sqrt{X}}^\infty J_{a-1}(u) \, du \label{fk3}
\end{eqnarray}
where here $f_{\rm hard}^{11}(X,Y)$ is for the scaled hard edge with 
singularity
$x^{a/2}$ as $x \to 0^+$. Agreement between (\ref{ir1}) and
(\ref{fk1}), (\ref{fk2}) is immediate upon substituting the integral
formulas from (\ref{sap2}), (\ref{sap3}) in (\ref{ir1}) and integrating
by parts. It remains to show that the RHS of (\ref{fk3}) agrees with the
RHS of (\ref{ir2}) with the replacement $a \mapsto a + 1$. The verification
is done by making use of Lemma \ref{lemq} in (\ref{ir2}) to obtain
\begin{eqnarray}\label{ir5}
f_{\rm scaled}^{11}(X,Y)  \Big |_{a \to a+1} & = &
{1 \over 2} K^{\rm hard}(X,Y) -
{1 \over 4 \sqrt{Y}} J_a( \sqrt{X}) J_{a+1}(\sqrt{Y}) 
 + {1 \over 2} {\partial \over \partial  Y} \int_X^\infty
{Y \over t} K^{\rm hard}(t,Y) \, dt 
\nonumber \\&&
- {1 \over 4}
{\partial \over \partial  Y} \sqrt{Y} J_{a+1}(\sqrt{Y})
 \int_X^\infty t^{-1} J_{a}( \sqrt{t}) \, dt
\end{eqnarray}
(c.f.~(\ref{ir4})). Using the integral formula from (\ref{sap1})
(without the restriction $a=0$), a straightforward calculation shows 
$$
{\partial \over \partial  Y} \int_X^\infty
{Y \over t} K^{\rm hard}(t,Y) \, dt =
 K^{\rm hard}(X,Y) + {J_a(\sqrt{Y}) \over 4}
 \int_X^\infty t^{-1} J_{a}( \sqrt{t}) \, dt.
$$
Substituting this in (\ref{ir5}) gives agreement with (\ref{fk3}) provided
$$
- {1 \over 4 \sqrt{Y}} J_a(\sqrt{X}) J_{a+1}(\sqrt{Y}) +
{J_a(\sqrt{Y}) \over 8} \int_X^\infty  t^{-1} J_{a}(\sqrt{t}) \, dt -
 {1 \over 4} {\partial \over \partial Y} \sqrt{Y}
J_{a+1}(\sqrt{Y}) \int_X^\infty  t^{-1} J_{a}(\sqrt{t}) \, dt
$$
$$
= {J_{a+1}(\sqrt{Y}) \over 4 \sqrt{Y}}
\int_{\sqrt{X}}^\infty J_{a-1}(u) \, du.
$$
Since both sides vanish as $X \to \infty$, it suffices to check that the
derivative with respect to $X$ of both sides agrees. This is easily verified
using suitable Bessel function identities.

\subsection{The symplectic symmetry limit}\label{s5.2}
We know from (\ref{3.12a}) and (\ref{3.12b}) that in the limit $A \to - \infty$
the parameter dependent Laguerre and Jacobi ensembles tend to the LSE
with weight $e^{-x}$ and the JSE with weight $(1-x)^{a+1}$ respectively.
The results of the present study give that the corresponding $k$-point
distribution is given by (\ref{6.3.2}) with matrix elements specified by
(\ref{fsubu})--(\ref{fsubub}). As with the orthogonal symmetry limit, previous
studies \cite{Wi99,FNH99,AFNV99} 
have obtained the $k$-point distribution for the LSE
with general $a>-1$, and the JSE with general $a,b>-1$. In particular, in
terms of (\ref{Kh}) and (\ref{Kh1}), the results of \cite{AFNV99} give that the
$k$-point distribution for the matrix ensembles SE${}_n(x^a e^{-x})$
(LSE) and SE${}_n((1+x)^b(1-x)^{a+1})$ (JSE), as specified by
(\ref{hs}), are given by (\ref{6.3.2}) with the $2 \times 2$ matrix $f$ in
(\ref{6.3.2}) having as its top left entry
\begin{equation}\label{dra}
2f^{11}(x,y) = \Big ( {x \over y} \Big )^{1/2} K_{2n,a-1}^L(x,y)
+ {2(2n+a-1) \over h_{2n,a-1}} y^{(a-2)/2} e^{-y/2} L_{2n}^{a-1}(y)
\int_x^\infty t^{(a-2)/2} e^{-t/2} L_{2n-1}^{a-1}(t) \, dt
\end{equation}
in the Laguerre case, and
\begin{eqnarray}\label{drb}
2f^{11}(x,y) & = & (1-x) \Big ( {1+x \over 1+y} \Big )^{1/2}
K_{2n,a,b-1}^J(x,y) - {2n(2n+b-2) \over h_{2n-1,a,b-1}^J (2n + {1 \over 2}
(a+b-2))}(1-y)^{(a-1)/2}  \nonumber \\&&
\times (1+y)^{(b-2)/2}
P_{2n}^{(a,b-1)}(y) \int_x^1 (1-t)^{(a-1)/2} (1+t)^{(b-2)/2}
P_{2n-1}^{(a,b-1)}(t) \, dt 
\end{eqnarray}
in the Jacobi case (we have taken the transpose of the qdet formula in
\cite{AFNV99}, and thus have interchanged $x$ and $y$ in $f^{11}$ relative
to the expression in \cite{AFNV99}). The other entries in the matrix $f$ are
related to $f^{11}$ by
\begin{equation}\label{dr1}
f^{22}(x,y) = f^{11}(y,x), \quad
f^{12}(x,y) = - \int_y^x f^{11}(y,t) \, dt, \quad
f^{21}(x,y) = {\partial \over \partial y} f^{11}(y,x).
\end{equation}
We see from the first equality in (\ref{pr1}), 
(\ref{fsubua}) and (\ref{fsubub}) that
the equations (\ref{dr1}) are identical to those obtained in Section
\ref{s3.4} for the symplectic limit in the Laguerre case. Furthermore, in
the Jacobi case, defining
\begin{equation}\label{dr2}
f^{22}(x,y) = {1-y \over 1-x} f^{22}(x,y) =
{1 \over 2} (1-y) K_{2n}^J(x,y) + {1 \over 2}
\Big ( (1-x) {\partial \over \partial x} - 1 \Big )
\int_y^1 K_{2n}^J(t,x) \, dt
\end{equation}
the first equality in (\ref{pr1}) and (\ref{fsubu})--(\ref{fsubub}) together
with simple scaling invariances of the quaternion determinant imply that
the $k$-point correlation can be written as qdet$[\tilde{f}]$ with the
elements of the $2 \times 2$ matrix $\tilde{f}$ given as in (\ref{dr1})
but with each $f^{ss'}$ replaced by $\tilde{f}^{ss'}$. Thus it remains to show
agreement between the first equality in (\ref{fsubu}), substituted in the
first equation of (\ref{dr1}), with (\ref{dra}) in the case $a=0$,
and (\ref{dr2}), substituted in the 
first equation of (\ref{dr1}), with (\ref{dr1}) in the case $b=0$.

Consider first the Laguerre case. We know from \cite{Wi99,FNH99} 
that (\ref{dra})
can be rewritten to read
\begin{equation}\label{dr3}
2f^{11}(x,y) = K_{2n,a}^L(x,y) + G_1^L(y) G_2^L(x)
\end{equation}
where
\begin{eqnarray*}
G_1^L(y) & = & {1 \over 2 h_{2n,a-1}^L} y^{a/2-1} e^{-y/2} L_{2n}^{a-1}(y)
\\
G_1^L(x) & = & \Big ( 2 x^{a/2} e^{-x/2} L_{2n-1}^a(x) +
(2n+a-1) \int_x^\infty t^{a/2 - 1} e^{-t/2} L_{2n-1}^{a-1}(t) \, dt
\Big ) \\
 & = & \int_0^x t^{a/2 - 1} e^{-t/2} \Big ( 2n L_{2n}^a(x) -
(2n+a) L_{2n-1}^a(x) \Big ).
\end{eqnarray*}
For the RHS of (\ref{fsubu}) to equal the RHS of (\ref{dr3}) substituted in the
first equality of (\ref{dr1}) in the case $a=0$, by setting $y=0$ in
both expressions we see that a necessary condition is that
$$
{1 \over 2} K_{2n}^L(x,0) + {1 \over 2} {\partial \over \partial x}
\int_0^\infty K_{2n}^L(t,x) \, dx = K_{2n,0}^L(x,0),
$$
which is obtained by setting $y=0$ in both expressions. This is just the
previously established identity (\ref{3.87'}). Knowing that both sides
agree at $y=0$, to show they are equal for all $y$
it suffices to show that their partial derivatives
with respect to $y$  agree. In fact equating the partial derivatives
gives the previously established identity (\ref{ph1}).

To show agreement in the Jacobi case, we require the analogue of
(\ref{dr3}).

\begin{lemma}
The formula (\ref{drb}) has the alternative form
\begin{equation}\label{dr5}
2 f^{11}(x,y) = (1-x) K_{2n,a,b}^J(x,y) + G_1^J(y) G_2^J(x)
\end{equation}
where
\begin{eqnarray*}
G_1(y) & = & {4n \over 4n + a+b-1} {1 \over h_{2n-1,a,b}^J}
(1-y)^{(a-1)/2} (1+y)^{b/2-1} P_{2n}^{(a,b-1)}(y) \nonumber \\
G_2(y)  & = & - \Big \{ P_{2n-1}^{(a,b)}(x) (1-x)^{(a+1)/2}
(1+x)^{b/2} \nonumber \\
&& + (b+2n-1) \int_x^1 dt \, (1-t)^{(a-1)/2} (1+t)^{(b-2)/2}
P_{2n-1}^{(a,b-1)}(t) \Big \}  \nonumber \\
& = & {a+b+2n \over 4} \int_{-1}^x (1-t)^{(a-1)/2}
(1+t)^{b/2} P_{2n-1}^{(a,b+1)}(t) \, dt.
\end{eqnarray*}
\end{lemma}

\noindent
Proof. \quad This is derived in a similar fashion to the result of
Lemma \ref{lem18}. \hfill $\square$

With (\ref{dr5}) substituted in the first equality of (\ref{dr1}),
agreement of the resulting expression with (\ref{dr2}) in the case
$b=0$ can be established by first checking that both expressions
coincide at $y=-1$ (this follows from (\ref{le2})), and then showing
both expressions have the same partial derivative with respect to
$y$ (this follows from (\ref{ph3})).

Next we consider the form of the scaled matrix elements of
Propositions \ref{p16} and \ref{p17} as $\alpha \to - \infty$, which
corresponds to the symplectic limit. Integration by parts shows
\begin{eqnarray}
{Y \over X} f_{\rm scaled}^{22}(X,Y) & \! \sim & \!
\tilde{f}_{{\rm scaled}
}^{22} (X,Y) \: = \: 
{1 \over 2} \sqrt{{Y \over X}} 
K^{\rm scaled}(X,Y) + {1 \over 2} \int_0^Y {1 \over \sqrt{u}}
{\partial \over \partial X}
K^{\rm scaled}(X,t) \label{tf1} \\
{8 \over \alpha^2} f_{\rm scaled}^{12}(X,Y) & \sim &
\int_X^\infty \tilde{f}_{{\rm scaled} 
}^{22} (t,Y) \, dt \nonumber
\\
{\alpha^2 \over 8} f_{{\rm scaled} 
}^{21} (X,Y) & \sim &
{\partial \over \partial Y} 
 \tilde{f}_{{\rm scaled} 
}^{22} (X,Y)
\nonumber
\end{eqnarray}
in the case of the soft edge and the bulk, and
\begin{eqnarray}\label{tf2}
&&{Y \over X} f_{\rm hard}^{22}(X,Y)  \sim   \tilde{f}_{{\rm hard} 
}^{22} (X,Y)  = 
{1 \over 2} \sqrt{Y \over X}
K^{\rm hard}(X,Y) + {1 \over 2} \int_0^Y {1 \over \sqrt{u}}
{\partial \over \partial X} \sqrt{X} K^{\rm hard}(X,t) \, dt \nonumber \\
\\
&&{8 \over \alpha^2} XY f_{\rm scaled}^{12}(X,Y)  \sim 
- \int_0^X 
f_{{\rm hard} 
}^{22} (t,Y) \, dt \nonumber
\\
&&{\alpha^2 \over 8} {1 \over XY} f_{\rm hard}^{21} 
(X,Y)  \sim 
{\partial \over \partial Y} 
 \tilde{f}_{{\rm hard} \atop
\alpha = -\infty}^{22} (X,Y)
\nonumber
\end{eqnarray}
in the case of the hard edge with singularity $x^{a+1}$ as $x \to 0$.

Previous studies have given formulas of a different form to (\ref{tf1})
and (\ref{tf2}), these being \cite{FNH99}
\begin{eqnarray*}
f_{\rm bulk}^{22}(X,Y) & = & K^{\rm bulk}(X,Y) \\
f_{\rm soft}^{22}(X,Y) & = & K^{\rm soft}(X,Y) -
{1 \over 2} {\rm Ai}(X) \int_Y^\infty {\rm Ai}(s) \, ds \\
f_{\rm hard}^{22}(X,Y) & = & K^{\rm hard}(X,Y) -
{1 \over 4 \sqrt{X}} J_{a-1}(\sqrt{X}) \int_0^{\sqrt{Y}}
J_{a+1}(s) \, ds,
\end{eqnarray*}
where here $f_{\rm hard}^{22}$ relates to the scaled hard edge with
singularity $x^a$ as $x \to 0^+$. The verification that these formulas are
identical to (\ref{tf1}) or (\ref{tf2}) is done in the same way as
reconciling (\ref{ir1})--(\ref{ir2}) with (\ref{fk1}) and (\ref{fk3}).

\subsection{Bulk structure function for the decimated orthogonal ensemble
with a parameter}
In general the 2-point correlation implied by (\ref{6.3.2}) 
is given in terms of 
the corresponding 1-point correlation (the density) and the matrix elements
$f^{ss'}$ $(s,s'=1,2)$ of $f$ according to
\begin{equation}\label{fc}
\rho_2(x_1,x_2) = \rho_1(x_1) \rho_1(x_2) - \Big ( f^{11}(x_1,x_2)
f^{22}(x_1,x_2) - f^{12}(x_1,x_2) f^{21}(x_1,x_2) \Big ).
\end{equation}
Now the truncated quantity
\begin{equation}\label{fc1'}
\rho_2^T(x_1,x_2) := \rho_2(x_1,x_2) - \rho_1(x_1) \rho_1(x_2)
\end{equation}
decays for large $|x_1 - x_2|$. Furthermore, in the bulk it is a function
only of the difference $|x_1 - x_2|$. Thus in the bulk the Fourier transform
of $\rho_2^T(x_1-x_2,0)$ is a well defined quantity. We seek its
evaluation in the case of the parameter dependent orthogonal ensemble
with a parameter, when the matrix elements are specified by
(\ref{wa1})--(\ref{wa3}). Explicitly, we will 
compute the quantity
\begin{eqnarray}\label{fc1}
S(k) & = & 1 + \int_{-\infty}^\infty \rho_2^T(x) e^{ikx} \, dx
\nonumber \\
& = & 1 - \int_{-\infty}^\infty ( f^{11}(x) f^{22}(-x) +
f^{12}(x) f^{21}(-x) ) e^{ikx} \, dx
\end{eqnarray}
where to obtain the second equality we have substituted (\ref{fc})
and made use of the facts, apparant from (\ref{wa1})--(\ref{wa3}), that
$f^{11}(x), f^{22}(x)$ are even functions of $x$, while
$f^{12}, f^{21}$ are odd functions of $x$. Following \cite{FJM00}
we know that the most efficient way to compute such a Fourier
transform is to make use of the general formula
\begin{equation}\label{fc1a}
\int_{-\infty}^\infty f(x) f(-x) e^{ikx} \, dx =
{1 \over 2 \pi} \int_{-\infty}^\infty
\hat{f}(l) \hat{f}(l-k) \, dl, \quad
\hat{f}(l) := \int_{-\infty}^\infty f(x) e^{ilx} \, dx.
\end{equation}
Thus we must first compute the Fourier transform of the individual matrix
elements.

The calculation of the Fourier transform of the individual matrix
elements is simplified by first noting, making use of the fact that
$K^{\rm bulk}(X,Y)$ is a function of $X-Y$, that the expression
(\ref{wa1}) for $f_{\rm bulk}^{22}(X,Y)$ can be simplified to read
$$
f^{22}_{\rm bulk}(X,Y) = K^{\rm bulk}(X,Y),
$$
independent of the parameter $\alpha$. Recalling the first equality in
(\ref{pr1}), and the integral representation of (\ref{sap2}), we thus have
$$
\hat{f}_{\rm bulk}^{22}(l) = \hat{f}_{\rm bulk}^{11}(l) =
\chi_{|l| < \pi}.
$$
Furthermore, substituting the integral representation of (\ref{sap2})
in (\ref{wa2})--(\ref{wa3}) allows us to compute
$$
\hat{f}^{12}_{\rm bulk}(l) = {(\alpha/2)^2 + l^2 \over 2i l} \chi_{|l| < \pi},
\qquad \hat{f}^{21}_{\rm bulk}(l) = - {2il \over
(\alpha/2)^2 + l^2}  \chi_{|l| > \pi}.
$$
Thus
\begin{eqnarray}
\int_{-\infty}^\infty \hat{f}^{11}_{\rm bulk}(l) 
\hat{f}_{\rm bulk}^{22}(l-k) \, dl & = & \int_{-\pi}^\pi
\chi_{|l-k| < \pi} dl =
\left \{ \begin{array}{ll} 2 \pi - |k|, & |k| < 2 \pi \\
0, & |k| \ge 2 \pi \end{array} \right. \label{zf1} \\
\int_{-\infty}^\infty \hat{f}^{12}(l)
\hat{f}^{21}(l-k) \, dl & = &
- \int^{{\rm min}(\pi - |k|, - \pi)}_{- \pi - |k|}
{(\alpha/2)^2 + (l+|k|)^2 \over l + |k|}
{l \over (\alpha/2)^2 + l^2} \, dl. \label{zf2}
\end{eqnarray} 
The integral (\ref{zf2}) can be evaluated in terms of elementary functions,
with there being a different functional form for $|k| < 2 \pi$,
$|k| \ge 2 \pi$. Substituting this and (\ref{zf1}) in (\ref{fc1})
(appropriately rewritten using (\ref{fc1a})), we find that for
$|k| < 2 \pi$
\begin{eqnarray}\label{zf3}
S(k) & = & {|k| \over \pi} +
{|k| \over 4 \pi ((\alpha/2)^2 + k^2)} \Big [ \alpha |k|
\Big ( \arctan {2 \pi \over \alpha} - \arctan {2 |k| + 2 \pi \over \alpha}
\Big ) \nonumber \\
&& - ( {\alpha^2 \over 2} + k^2 ) \log \Big (
{(\alpha/2)^2 + (|k| + \pi)^2 \over (\alpha/2)^2 + \pi^2} \Big )
- {\alpha^2 \over 2} \log | 1 - |k|/ \pi | \Big ],
\end{eqnarray}
while for $|k| \ge 2 \pi$
\begin{eqnarray}\label{zf4}
S(k)  & = & 2 +
{\alpha k^2 \arctan(2(|k| - \pi)/\alpha) \over 4 \pi ((\alpha/2)^2 + k^2)}
- {\alpha k^2 \arctan(2(|k| + \pi)/\alpha) \over 4 \pi ((\alpha/2)^2 + k^2)} 
\nonumber \\
&& - {(2(\alpha/2)^2|k| + |k|^3) \over 4 \pi ((\alpha/2)^2 + k^2)}
\log {(\alpha/2)^2 + (|k|+\pi)^2 \over (\alpha/2)^2 +
(|k| - \pi)^2}.
\end{eqnarray} 
Of particular interest is the small $|k|$ expansion of (\ref{zf3}). We
find
\begin{equation}\label{zf5}
S(k) = {|k| \over \pi} + {1 \over 2 \pi^2}
{(\alpha/2)^2 - \pi^2 \over (\alpha/2)^2 + \pi^2} k^2 +
{((\alpha/2)^2 - \pi^2)^2 \over 4 \pi^3  ((\alpha/2)^2 + \pi^2)^2}
|k|^3 + {\rm O}(k^4).
\end{equation} 

We see from (\ref{zf5}) that the coefficient of the leading order term,
proportional to $|k|$,
in the small $|k|$ expansion of $S(k)$ is independent of the
parameter $\alpha$, and has the value $1/\pi$. This is to anticipated from
the interpretation of (\ref{3.1}) as a one-component log-potential
Coulomb system with coupling $\beta = 1$. The coupling within 
pairs can be regarded as a short range potential which should not affect
properties determined by the long-ranged logarithmic potential. One
such property is the behaviour
$$
S(k) \mathop{\sim}\limits_{|k| \to 0}
{|k| \over \pi \beta}
$$
for a one-component log-potential system with coupling $\beta$ (see
e.g.~\cite{FJM00}), thus implying the leading behaviour seen in (\ref{zf5}).

\subsection{Distribution of odd labelled coordinates for a special
parameter}
In the Introduction, attention was drawn to the properties
(\ref{3.1b}), (\ref{1ma}) of the parameter dependent ensembles
(\ref{3.1}), (\ref{6.1}) relating to the distribution of the even
labelled coordinates. Similarly, we noted the properties
(\ref{4.2}), (\ref{1m'}) of the even labelled
coordinates in the superimposed parameter dependent ensembles
(\ref{4.1}), (\ref{6.2}). For the special value of the parameter 
for which the one body factor for the even labelled coordinates reduces
to a constant,
it turns out
that the odd labelled coordinates also have a distribution which
coincides with that of other matrix ensembles. This follows from the
following integration formulas.
\begin{lemma}
Let $x_1,x_2,\dots,x_{2n}$ be ordered as in (\ref{3.1a}), and label
this ordering $X$. We have
\begin{equation}\label{xq1}
\int_X dx_2 dx_4 \cdots dx_{2n} \prod_{1 \le j < k \le 2n}
(x_j - x_k) = {1 \over (2n)!}
\prod_{j=1}^n x_{2j-1}^2 \prod_{1 \le j < k \le n}
(x_{2j-1} - x_{2k-1})^4
\end{equation}
and
\begin{equation}\label{xq2}
\int_X dx_2 dx_4 \cdots dx_{2n} \prod_{1 \le j < k \le n}
(x_{2j} - x_{2k}) = {1 \over n!}
\prod_{j=1}^n x_{2j-1} \prod_{1 \le j < k \le n}
(x_{2j-1} - x_{2k-1}).
\end{equation}
\end{lemma}

\noindent
Proof. \quad To derive (\ref{xq1}), we use the Vandermonde determinant
formula to write
$$
\prod_{1 \le j < k \le 2n} (x_j - x_k) =
\det [ x_{2n+1-j}^{k-1} ]_{j,k=1,\dots,2n}.
$$
The method of integration over alternative variables gives
\begin{eqnarray*}
&&
\int_X dx_2 dx_4 \cdots dx_{2n} \prod_{1 \le j < k \le 2n}
(x_j - x_k) = \det \left [ \begin{array}{c} {1 \over k} x_{2n+1-2j}^k \\
x_{2n+1-2j}^{k-1} \end{array} \right ]_{j=1,\dots,n \atop
k=1,\dots,2n} \\
&& \quad = {1 \over (2n)!} \prod_{j=1}^n x_{2j-1}
\det \left [ \begin{array}{c}  x_{2n+1-2j}^{k-1} \\
k x_{2n+1-2j}^{k-1} \end{array} \right ]_{j=1,\dots,n \atop
k=1,\dots,2n} = {1 \over (2n)!} \prod_{j=1}^n x_{2j-1}^2
\det \left [ \begin{array}{c}  x_{2n+1-2j}^{k-1} \\
(k-1) x_{2n+1-2j}^{k-2} \end{array} \right ]_{j=1,\dots,n \atop
k=1,\dots,2n}.
\end{eqnarray*}
This final determinant is well known to be equal to the product of differences
to the fourth power, and thus (\ref{xq1}) follows.

A similar, even simpler, computation gives (\ref{xq2}). \hfill $\square$

\noindent
We remark that (\ref{xq1}) and (\ref{xq2}) are equivalent to the special case
of (\ref{5.1}) and (\ref{1.13a}) in which $(f,g)$ is given by the
Jacobi weight in (\ref{5.2}) with $a=1$. 

For the ensembles (\ref{3.1}), (\ref{6.1}), it follows immediately
from (\ref{xq1}) that
\begin{equation}\label{wp1}
{\rm odd}({\rm OE}_{2n}(f_{\rm o}, f_{\rm e})) = {\rm SE}_n(h)
\end{equation}
with
\begin{equation}\label{wi}
(f_{\rm o}, f_{\rm e}, h) = \left \{
\begin{array}{l} (e^{-x/2} e^{Ax/2}, e^{-x/2} e^{-Ax/2}, x^2 e^{-x})
\Big |_{A=-1} \: \: (x>0) \\[.1cm]
((1-x)^{(a-A-1)/2}, (1-x)^{(a+A-1)/2}, (1+x)^2 (1-x)^{a-1})
\Big |_{A=1-a} \: \: (-1 < x < 1) \end{array} \right.
\end{equation}
in the Laguerre and Jacobi cases respectively. Similarly, for the ensembles 
(\ref{4.1}), (\ref{6.2}), it follows immediately
from (\ref{xq2}) that
\begin{equation}\label{wp2}
{\rm odd}({\rm OE}_{n}(f_{\rm o}, f_{\rm e})
\cup {\rm OE}_{n}(f_{\rm o}, f_{\rm e})) = {\rm UE}_n(\tilde{h})
\end{equation}
where $f_{\rm o}$, $f_{\rm e}$ are as in (\ref{wi}), while
\begin{equation}
\tilde{h} = \left \{\begin{array}{ll} x e^{-x}, & {\rm Laguerre \: case} \\
(1+x) (1 - x)^{a-1}, & {\rm Jacobi \: case} \end{array} \right.
\end{equation} 

A consequence of (\ref{wp1}) and (\ref{wp2}) is that the $k$-point odd-odd
correlation for the ensemble on the LHS must coincide with the
$k$-point correlation for the ensemble on the RHS. This is simple to
explicitly verify for the relation (\ref{wp2}). Then the $k$-point
distribution on the RHS is given by
\begin{equation}\label{5.59}
\det [K_{n,1}^L(x_j,x_l)]_{j,l=1,\dots,k}, \qquad
\det [(1-x_j) K_{n,a-1,1}^J(x_j,x_l)]_{j,l=1,\dots,k}
\end{equation}
in the Laguerre and Jacobi cases respectively. On the LHS the odd-odd
$k$-point correlation is given by
\begin{equation}\label{5.60}
\det [K_{\rm oo}^L(x_j,x_l)\Big |_{A=-1}]_{j,l=1,\dots,k}, \qquad
\det [(1-x_j) K_{\rm oo}^J(x_j,x_l)]_{j,l=1,\dots,k},
\end{equation}
where $K_{\rm oo}^L$, $K_{\rm oo}^J$ are given by (\ref{KL}) and
(\ref{KJ}) respectively. Using appropriate Laguerre and Jacobi
polynomial formulas to evaluate
$$
\int_0^{x'} L_l(x) \, dx, \quad
{d \over dx} e^{-x} L_l(x), \quad
\int_0^{x'} P_l^{(a,0)}(x) \, dx, \quad
{d \over dx} (1-x)^a P_l^{(a,0)}(x)
$$
it is readily seen that (\ref{5.60}) can be reduced to (\ref{5.59}).

\subsection{Gap probabilities and eigenvalue distributions}
\setcounter{equation}{0}
Observable quantities for eigenvalue sequences closely related to
correlation functions are gap probabilities and distribution
functions of individual eigenvalues. The  gap probability specifies the
probability --- to be denoted $E(p;I;{\rm ME})$ --- that an interval $I$
(the gap) in a given matrix ensemble ME contains precisely $p$
eigenvalues. With the eigenvalues ordered as in (\ref{3.1a}), we
may choose to  observe eigenvalues of a definite parity,
even or odd labelled. In this case we denote the gap probability as
$E^{(\cdot)}(p;I;{\rm ME})$, $(\cdot) = $ (e)ven, (o)dd, where $p$
now refers to eigenvalues of parity $(\cdot)$ only. For $I=(s,\infty)$
with $s$ inside the support of ME, it follows from the ordering
(\ref{3.1a}) that
\begin{eqnarray}\label{beg.1}
E^{(\rm o)}(p;(s,\infty);{\rm ME}) & = & \Big (
E(2p-1;(s,\infty);{\rm ME}) +
E(2p;(s,\infty);{\rm ME}) \Big ) \nonumber \\
E^{(\rm e)}(p;(s,\infty);{\rm ME}) & = & \Big (
E(2p;(s,\infty);{\rm ME}) +
E(2p+1;(s,\infty);{\rm ME}) \Big )
\end{eqnarray}
where $E(-1;I;{\rm ME}) := 0$. 
Consequently we can express $E(p;(s,\infty);{\rm ME})$ in terms of
$\{E^{(\cdot)}(p;(s,\infty);{\rm ME})\}$,
\begin{eqnarray}\label{beg.1a}
E(2p;(s,\infty);{\rm ME}) & = & 
\sum_{j=0}^p E^{(\rm o)}(j;(s,\infty);{\rm ME}) -
\sum_{j=0}^{p-1} E^{(\rm e)}(j;(s,\infty);{\rm ME})  \nonumber \\
E(2p+1;(s,\infty);{\rm ME}) & = & 
\sum_{j=0}^p E^{(\rm e)}(j;(s,\infty);{\rm ME}) -
\sum_{j=0}^{p} E^{(\rm o)}(j;(s,\infty);{\rm ME}).
\end{eqnarray}
We will denote the PDF for the distribution
function of the $k$th eigenvalue $x_k$ (with the ordering (\ref{3.1a})) by
$p(k-1;s;{\rm ME})$ (here $k-1$ is the number of eigenvalues greater than
$x_k$). A standard formula (see e.g.~\cite{Fo02}) gives that
\begin{equation}\label{beg.2}
p(k-1;s;{\rm ME}) = {d \over ds} E(k-1;(s,\infty);{\rm ME}) +
p(k-2;s;{\rm ME}), \qquad k \ge 1
\end{equation}
where $p(-1;s;{\rm ME}) := 0$, so knowledge of $\{E(p;(s,\infty);{\rm ME})
\}_{p=0,\dots,k-1}$ suffices to compute $p(k-1;s;{\rm ME})$.

Let us consider first the gap probabilities and eigenvalue
distributions for ${\rm ME} = ({\rm LOE}_n \cup {\rm LOE}_n)^A$ and
${\rm ME} = ({\rm JOE}_n \cup {\rm JOE}_n)^A$ (i.e.~the PDFs (\ref{4.1})
and (\ref{6.1})). The identities (\ref{5.1}) and (\ref{1m}) tell us
that
\begin{eqnarray}\label{beg.3}
 E^{(\rm e)}(p;(s,\infty);({\rm LOE}_n \cup {\rm LOE}_n)^A)
& = & E(p;(s,\infty);{\rm LUE}_n |_{a=0}), \nonumber \\
E^{(\rm e)}(p;(s,1);({\rm JOE}_n \cup {\rm JOE}_n)^A)
& = & E(p;(s,1);{\rm JUE}_n |_{b=0}) \nonumber \\
 p(2k-1;s;({\rm LOE}_n \cup {\rm LOE}_n)^A) & = &
p(k-1;s;{\rm LUE}_n |_{a=0}), \nonumber \\
p(2k-1;s;({\rm JOE}_n \cup {\rm JOE}_n)^A) & = &
p(k-1;s;{\rm JUE}_n |_{b=0}),
\end{eqnarray}
which we can check are consistent with (\ref{beg.2}). Note in particular
that each quantity in (\ref{beg.3}) is independent of the parameter $A$,
and we remark too that each has a known Painlev\'e transcendent
evaluation \cite{TW94c,FW01b,FW02}. To specify $E^{(\rm o)}$ and thus
$p(2k;s;{\rm ME}), E$ for these matrix ensembles we make use of the standard
formula relating the gap probability to the correlation functions,
\begin{equation}\label{erh}
E^{(\rm o)}(p;I;{\rm ME}) = {(-1)^p \over p!}
{\partial^p \over \partial \xi^p} \Big (1 +
\sum_{k=1}^n {(-\xi)^k \over k!} \int_I dx_1 \cdots \int_I dx_k \,
\rho_{(k)}^{(\rm o)}(x_1,\dots,x_k) \Big ) \Big |_{\xi = 1}.
\end{equation}
We know from (\ref{1.17}) that $\rho_{(k)}^{(\rm o)}$ is the $k \times k$
determinant $\det [K_{\rm oo}(x_i,x_j)]_{i,j=1,\dots,k}$. In this
circumstance the expression in brackets is just the expansion
\cite{WW65} of the Fredholm determinant of the integral operator
$K_{\rm oo}$ with kernel $K_{\rm oo}(x,y)$ supported on $I$,
\begin{equation}\label{erh1}
E^{(\rm o)}(p;I;{\rm ME}) = {(-1)^p \over p!}
{\partial^p \over \partial \xi^p} \det ( {\bf 1} - \xi K_{\rm oo})
|_{\xi = 1}
\end{equation}
(here ${\bf 1}$ denotes the identity operator). When $p=0$ this reads
\begin{equation}\label{erh2}
E^{(\rm o)}(0;I;{\rm ME}) = E(0;I;{\rm ME}) = \det ({\bf 1} - K_{\rm oo})
\end{equation}
where the first equality follows from the first equation of
(\ref{beg.1}). Using the fact that with $A=0$,
$({\rm LOE}_n \cup {\rm LOE}_n)^A$ and
$({\rm JOE}_n \cup {\rm JOE}_n)^A$ reduce to ${\rm LOE}_n \cup {\rm LOE}_n$
and ${\rm JOE}_n \cup {\rm JOE}_n$ respectively, it follows that
$E(0;I;{\rm ME} |_{A = 0}) =
(E(0;I;{\rm OE}))^2 $ and thus we deduce from (\ref{erh2}) that
\begin{equation}\label{erh2a}
\Big ( E(0;(s,\infty);{\rm LOE}|_{a=0}) \Big )^2 =
\det ( {\bf 1} - K_{\rm oo}^L |_{A = 0}), \quad
\Big ( E(0;(s,\infty);{\rm JOE}|_{a\mapsto (a-1)/2 \atop b=0}) \Big )^2 =
\det ( {\bf 1} - K_{\rm oo}^J |_{A = 0}).
\end{equation}
We remark that $E(0;(s,\infty);{\rm LOE}|_{a=0})$ has recently been
evaluated in terms of Painlev\'e transcendents \cite{Ba02,FW02}.

Let us now consider the gap probabilities and eigenvalue distributions
for ${\rm ME} = ({\rm LOE}_{2n})^A$ and ${\rm ME} = ({\rm JOE}_{2n})^A$ 
(i.e.~the PDFs (\ref{3.1}) and (\ref{6.1})). The identities (\ref{1.13a})
and (\ref{1m'}) tell us that
\begin{eqnarray}\label{beg.3'}
E^{(\rm e)}(p;(s,\infty);({\rm LOE}_{2n})^A)
& = & E(p;(s,\infty);{\rm LSE}_n |_{a=0}) \nonumber \\
E^{(\rm e)}(p;(s,1);({\rm JOE}_{2n})^A)
& = & E(p;(s,1);{\rm JSE}_n |_{a \mapsto a+1 \atop b=0}) \nonumber \\
p(2k-1;s;({\rm LOE}_{2n})^A) & = &
p(k-1;s;{\rm LSE}_n |_{a=0}), \nonumber \\ 
p(2k-1;s;({\rm JOE}_n \cup {\rm JOE}_n)^A) & = &
p(k-1;s;{\rm JSE}_n |_{a  \mapsto a+1 \atop b=0 }),
\end{eqnarray}
(c.f.~(\ref{beg.3})). Of these quantities $E(0;(s,\infty);{\rm LSE}_n |_{a=0})$
and $p(0;s;{\rm LSE}_n |_{a=0})$ are known in terms of 
Painlev\'e transcendents \cite{Ba02,FW02}. To specify $E^{(\rm o)}$ we again
make use of (\ref{erh}), this time noting from (\ref{6.2.1}) that
$\rho_{(k)}^{(\rm o)}$ is the $k \times k$ quaternion determinant
${\rm qdet} [ f_{\rm oo}(x_i, x_j) ]_{i,j=1,\dots, k}$ where
$f_{\rm oo}$ is the $2 \times 2$ matrix representation of a particular
real quanternion, which in turn implies
$$
E^{(\rm o)}(p:I;{\rm ME}) = {(-1)^p \over p!}
{\partial^p \over \partial \xi^p} \rm qdet ( {\bf 1} - \xi f_{\rm oo})
|_{\xi = 1}
$$
(with $\{ \lambda_j \}$ denoting the distinct eigenvalues of $f_{\rm oo}$,
${\rm qdet}( {\bf 1} - \xi f_{\rm oo}) = \prod_j(1 - \xi \lambda_j)$).
We remark that for general $A$, $E^{(\rm o)}(0;(s,\infty),
({\rm LOE})_{2n}^A)$ has recently been evaluated in terms of Painlev\'e
transcendents \cite{Ba02}.

All the above formulas have well defined scaled limits. In particular
\begin{eqnarray*}
\lefteqn{ \lim_{n \to \infty} E^{(\rm o)}(p;(4n + 2(2n)^{1/3}s, \infty);
({\rm LOE}_n \cup {\rm LOE}_n)^{A = \alpha/2(2n)^{1/3}})} \nonumber \\
&& := E^{(\rm o) \, soft}(p;(s,\infty); ({\rm OE} \cup {\rm OE})^\alpha)
= {(-1)^p \over p!} {\partial^p \over \partial \xi^p}
\det ( {\bf 1} - \xi K_{\rm oo}^{\rm soft}) \Big |_{\xi = 1}
\end{eqnarray*}
\begin{eqnarray*}
\lefteqn{ \lim_{n \to \infty} E^{(\rm o)}(p;(1 - {s^2 \over 2n^2},1);
({\rm JOE}_n \cup {\rm JOE}_n)^{A = 4 n^2 \alpha})} \nonumber \\
&& := E^{(\rm o) \, hard}(p;(0,s); ({\rm OE} \cup {\rm OE})^{\alpha,a})
= {(-1)^p \over p!} {\partial^p \over \partial \xi^p}
\det ( {\bf 1} - \xi K_{\rm oo}^{\rm hard}) \Big |_{\xi = 1}
\end{eqnarray*} 
(for the justification of the limiting processes see \cite{BF02}).
An evaluation of $E^{(\rm o) \, soft}(p;(s,\infty); ({\rm OE} \cup 
{\rm OE})^\alpha)$ in terms of a Riemann-Hilbert problem and
Painlev\'e II transcendents has been given in \cite{BR01b}.

\subsection*{Acknowledgements}
We thank J.~Baik for encouraging us to take up the problem of
calculating the correlation functions for (\ref{3.1}).  The work of
PJF was supported by the Australian Research Council.


\end{document}